\pdfoutput=1

\documentclass[a4paper, 11pt, abstract=true, abstraction, titlepage, oneside]{scrartcl}
\makeatletter

\usepackage{amsmath,amssymb,amsfonts,array}

\usepackage[T1]{fontenc}
\usepackage[utf8]{inputenc}
\usepackage[onehalfspacing]{setspace}
\usepackage[headsepline]{scrlayer-scrpage}
\clearscrheadfoot %
\usepackage{layout}
\usepackage[english]{babel}
\usepackage{geometry}
\AtBeginDocument{
	\newgeometry{
		textheight=9in,
		textwidth=6.5in,
		top=1in,
		headheight=17pt,
		headsep=25pt,
		footskip=30pt
	}
}

\usepackage{amsmath,amssymb}
\usepackage{adjustbox}
\usepackage{eurosym}
\usepackage{lscape}
\usepackage{nicefrac}
\usepackage{array}
\usepackage{paralist}
\usepackage{authblk}
\usepackage{graphicx}
\usepackage{booktabs}
\usepackage{placeins}
\usepackage{mathtools}
\usepackage{bm}
\usepackage{color}
\usepackage[acronym]{glossaries}
\usepackage{pgfplots}
\usepackage[normal]{threeparttable}
\usepackage{ulem}
\usepackage{natbib}
\usepackage[titletoc,title]{appendix}
\usepackage[hyphens]{url} %
\usepackage{ellipsis}
\usepackage{breqn}

\usepackage{chngcntr}  
\usepackage{enumitem}
\usepackage{theorem}
\usepackage{etoolbox}
\usepackage{multirow}
\usepackage{colortbl}
\usepackage{subcaption}

\usepackage{tikz}
\usetikzlibrary{arrows.meta}
\usepgfplotslibrary{groupplots}
\usepgfplotslibrary{dateplot}
\usepackage[utf8]{inputenc}
\ifdefined\DeclareUnicodeCharacter
\DeclareUnicodeCharacter{2212}{-}
\fi

\let\proof\relax 
\let\endproof\relax
\let\theoremstyle\relax 

\usepackage{amsthm}

\ifdefined\mathleft
\else
\newcommand{\mathleft}{}
\fi	
\ifdefined\mathright
\else
\newcommand{\mathright}{}
\fi
\ifdefined\mathcenter
\else
\newcommand{\mathcenter}{}
\fi

\DeclareMathOperator*{\argmin}{arg\,min}

\makeatletter
\providecommand\phantomcaption{\caption@refstepcounter\@captype}
\makeatother

\definecolor{lightblue}{rgb}{0.60784,0.76078,0.90196}
\definecolor{darkblue}{rgb}{0.26667,0.44706,0.76863}
\definecolor{lightgreen}{rgb}{0.66275,0.81569,0.55686}
\definecolor{darkgreen}{rgb}{0.43922,0.67843,0.27843}
\definecolor{orange}{rgb}{0.92941,0.49020,0.19216}
\definecolor{yellow}{rgb}{1.00000,0.75294,0.00000}
\definecolor{grey}{rgb}{0.64706,0.64706,0.64706}
\definecolor{purple}{rgb}{0.51373,0.23529,0.04706}

\newsavebox{\abstractbox}
\renewenvironment{abstract}
{\begin{lrbox}{0}\begin{minipage}{\textwidth}
			\begin{center}\normalfont\sectfont\abstractname\end{center}\quotation}
		{\endquotation\end{minipage}\end{lrbox}%
	\global\setbox\abstractbox=\box0 }

\makeatletter
\expandafter\patchcmd\csname\string\maketitle\endcsname
{\vskip\z@\@plus3fill}
{\vskip\z@\@plus2fill\box\abstractbox\vskip\z@\@plus1fill}
{}{}
\makeatother

\addtokomafont{captionlabel}{\footnotesize\bfseries} %
\addtokomafont{caption}{\footnotesize\bfseries} %

\setlist[enumerate]{topsep=1pt, itemsep=-1ex,partopsep=1ex,parsep=1ex}

\AtBeginDocument{%
	\abovedisplayskip=7.5pt plus 0pt minus 0pt
	\abovedisplayshortskip=7.5pt plus 0pt minus 0pt
	\belowdisplayskip=7.5pt plus 0pt minus 0pt
	\belowdisplayshortskip=7.5pt plus 0pt minus 0pt
}

\theoremstyle{definition}
\newtheorem{definition}{Definition}[section]

\theoremstyle{definition}

\theoremstyle{definition}

\usepackage[normal]{threeparttable}
\usepackage{longtable}
\usepackage{threeparttablex}
\usepackage{booktabs}
\usepackage{tabularx}
\usepackage{twoopt}

\usepackage{lmodern}

\usepackage{subcaption}

\usepackage{multicol}

\usepackage[ruled, noend, linesnumbered]{algorithm2e}
\SetKw{Continue}{continue}

\usepackage[author=Patrick]{pdfcomment}

\usepackage{xparse}

\usepackage{mathtools}

\usepackage{makecell}

\usepackage{bm}

\usepackage{adjustbox}

\usepackage{tikz}
\usetikzlibrary{shapes}
\usetikzlibrary{math}
\usetikzlibrary{intersections}
\usetikzlibrary{calc}

\usepackage{eurosym}

\usepackage{thmtools, thm-restate}

\newcommand{\Revision}[1]{#1}

\newcommand{\ipref}[1]{IP~\ref{#1}}
\newcommand{\mipref}[1]{MIP~\ref{#1}}

\def\checkmark{\tikz\fill[scale=0.4](0,.35) -- (.25,0) -- (1,.7) -- (.25,.15) -- cycle;}

\newenvironment{paragraphs}
{\begin{list}
		{}
		{
			\setlength{\topsep}{0pt}
			\setlength{\parskip}{0pt}
		}
	}
	{\end{list}}

\renewcommand{\emph}[1]{\textit{#1}}

\usepackage{url}
\usepackage{floatrow}
\usepackage{tikzscale}

\newacronym{abk:alns}{ALNS}{adaptive large neighbourhood search}
\newacronym{abk:ampereHours}{Ah}{ampere hours}
\newacronym{abk:afv}{AFV}{alternate fuel vehicle}
\newacronym{abk:acc}{ACC}{average cycle count}
\newacronym{abk:awc}{AWC}{average wear cost}
\newacronym{abk:bss}{BSS}{battery swapping station}
\newacronym{abk:bnb}{B\&B}{branch and bound}
\newacronym{abk:bnp}{B\&P}{branch and price}
\newacronym{abk:bsp}{BSP}{bicriteria shortest path}
\newacronym{abk:cng}{CNG}{compressed natural gas}
\newacronym{abk:cc-cv}{CC-CV}{{constant current - constant voltage}}
\newacronym{abk:cc}{CC}{constant current}
\newacronym{abk:cv}{CV}{constant voltage}
\newacronym{abk:csgs}{CSGS}{charging schedule generation scheme}
\newacronym{abk:cplex}{CPLEX}{IBM CPLEX Optimizer}
\newacronym{abk:cfp}{CFP}{charging function propagation}
\newacronym{abk:cg}{CG}{column generation}
\newacronym{abk:csp}{CSP}{charge scheduling problem}
\newacronym{abk:dod}{DoD}{depth of discharge}
\newacronym{abk:dod-acc}{DOD-ACC}{depth of discharge - achievable cycle count}
\newacronym{abk:dsl}{DSL}{domain specific language}
\newacronym{abk:ecv}{ECV}{electric commercial vehicle}
\newacronym{abk:evrp}{EVRP}{electric vehicle routing problem}
\newacronym{abk:evrpcs}{EVRCSP}{electric vehicle routing and charge scheduling problem}
\newacronym{abk:evrpjcs}{EVRPJCS}{\gls{abk:evrp} with joint charge scheduling}
\newacronym{abk:evrptw}{EVRP-TW}{electric vehicle routing problem with time windows}
\newacronym{abk:elrp}{ELRP}{electric location routing problem}
\newacronym{abk:evrpnl}{EVRP-NL}{electric vehicle routing problem with non-linear charging function}
\newacronym{abk:evrpnlc}{EVRP-NL-C}{E-VRP-NL with capacitated charging stations}
\newacronym{abk:evsp}{EVSP}{electric vehicle scheduling problem}
\newacronym{abk:evscp}{EVSCP}{electric vehicle scheduling and charging problem}
\newacronym{abk:evse}{EVSE}{electric vehicle supply equipment}
\newacronym{abk:efvcsp}{EFV-CSP}{electric freight vehicle charge scheduling problem}
\newacronym{abk:ea}{EA}{evolutionary algorithm}
\newacronym{abk:espp}{ESSP}{elementary shortest path problem}
\newacronym{abk:frvcp}{FRVCP}{fixed-route vehicle charging problem}
\newacronym{abk:gvrp}{GVRP}{green vehicle routing problem}
\newacronym{abk:gvrppr}{GVRP-MTPR}{green vehicle routing problem with multiple technologies and partial recharging}
\newacronym{abk:ga}{GA}{genetic algorithm}
\newacronym{abk:ghg}{GHG}{greenhouse gas}
\newacronym{abk:hc}{HC}{heuristic concatenation}
\newacronym{abk:icev}{ICEV}{internal combustion engine vehicle}
\newacronym{abk:ilp}{ILP}{integer linear program}
\newacronym{abk:ip}{IP}{integer program}
\newacronym{abk:isecp}{ISECP}{fixed interval scheduling with energy constraints}
\newacronym{abk:ils}{ILS}{iterated local search}
\newacronym{abk:ide}{IDE}{integrated desktop environment}
\newacronym{abk:kiloWattHours}{kWh}{kilowatt hours}
\newacronym{abk:lns}{LNS}{large neighbourhood search}
\newacronym{abk:lhio}{Li-Ion}{lithium-ion}
\newacronym{abk:lsp}{LSP}{logistics service provider}
\newacronym{abk:lp}{LP}{Linear Program}
\newacronym{abk:mip}{MIP}{mixed integer program}
\newacronym{abk:mwcp}{MWCP}{maximum weight clique problem}
\newacronym{abk:myproblem}{\textcolor{teal}{\acrshort{abk:efvcsp}}}{\textcolor{teal}{\acrlong{abk:efvcsp}}}
\newacronym{abk:myfrvcp}{FRVCP-TW-EP}{Fixed Route Vehicle Charging Problem with Time Windows and Energy Prices}
\newacronym{abk:mp}{MP}{master problem}
\newacronym{abk:opl}{OPL}{optimization programming language}
\newacronym{abk:pr}{PR}{path relinking}
\newacronym{abk:pcsgs}{P-CSGS}{parallel charging schedule generation scheme}
\newacronym{abk:pyomo}{Pyomo}{Python Optimization Modeling Objects}
\newacronym{abk:pwl}{PWL}{piecewise-linear}

\newacronym{abk:rvrp}{RVRP}{recharging vehicle routing problem}
\newacronym{abk:rmp}{RMP}{restricted master problem}
\newacronym{abk:ref}{REF}{resource extension function}
\newacronym{abk:soc}{SoC}{state of charge}
\newacronym{abk:sos2}{SOS2}{Special-Ordered Set of Type 2}
\newacronym[longplural={shortest path problems with resource constraints}]{abk:spprc}{SPPRC}{shortest path problem with resource constraints}
\newacronym{abk:scsgs}{S-CSGS}{sequential charging schedule generation scheme}
\newacronym{abk:spp}{SPP}{shortest path problem}
\newacronym{abk:sdvrp}{SDVRP}{site-dependent vehicle routing problem}
\newacronym{abk:ts}{TS}{tabu search}
\newacronym{abk:tsp}{TSP}{traveling salesman problem}
\newacronym{abk:tco}{TCO}{total cost of ownership}
\newacronym{abk:tou}{TOU}{time-of-use}

\newacronym{abk:vrp}{VRP}{vehicle routing problem}
\newacronym{abk:vsp}{VSP}{vehicle scheduling problem}
\newacronym{abk:vrptw}{VRP-TW}{vehicle routing problem with time windows}
\newacronym{abk:vns}{VNS}{variable neighborhood search}
\newacronym{abk:vrpis}{VRPIS}{vehicle routing problems with intermediate stops}
\newacronym{abk:wdf}{WDF}{(cumulative) wear density function}

\begin{document}
	\emergencystretch 3em

\newcommand{\Edge}{\OriginVertex, \TargetVertex}
\newcommand{\AllEdges}{(\Edge) \in \Edges}

\newcommand{\SetOfEdges}{\mathcal{E}^{\Vehicle}}
\newcommand{\SetOfChargingEdges}{\SetOfEdges_{\mathcal{F}}}
\newcommand{\SetOfChargingArcs}{\SetOfEdges_{\mathcal{F}}}
\newcommand{\SetOfServiceArcs}[1][{\SetOfTours}]{\SetOfEdges_{{#1}}}
\newcommand{\SetOfSourceArcs}{\SetOfEdges_{\SourceNode}}
\newcommand{\SetOfIdleArcs}{\SetOfEdges_{I}}

\newcommand{\PeriodOfVertex}[1][\OriginVertex]{p(#1)}
\newcommand{\ChargerOfVertex}[1][\OriginVertex]{f(#1)}
\newcommand{\DummyCharger}{\Charger_{0}}

\newcommand{\SetOfStationNodes}{\SetOfVertices_{\mathcal{F}}}
\newcommand{\SetOfTourNodes}{\SetOfVertices_{\mathcal{T}}}

\newcommand{\ArcConsumption}[1][\Edge]{q_{#1}}
\newcommand{\ArcCost}[1][\Edge]{c_{#1}}

\newcommand{\ArcDuration}[1][\Edge]{\tau_{#1}}

\newcommand{\Arc}{\Edge}
\newcommand{\ChargingArc}[1][\TargetVertex]{(\OriginVertex, #1)}
\newcommand{\ServiceArc}[1][\TargetVertex]{(\OriginVertex, #1)}
\newcommand{\SourceArc}[1][\TargetVertex]{(\OriginVertex, #1)}
\newcommand{\IdleArc}[1][\TargetVertex]{(\OriginVertex, #1)}

\newcommand{\GarageNode}[1][\Period]{v_{#1,f_0}}
\newcommand{\SetOfGarageNodes}{\mathcal{V}_x}

\newcommand{\LabelFixCost}[1][\Label]{c^{fix}_{#1}}
\newcommand{\LabelFixSoC}[1][\Label]{q^{fix}_{#1}}
\newcommand{\LabelCharger}[1][\Label]{\Charger_{#1}}
\newcommand{\LabelVisitedTours}[1][\Label]{O_{#1}}
\newcommand{\LabelCostProfile}[1][\Label]{\Psi\langle#1\rangle}

\newcommand{\CostProfile}[1][\Label]{\psi_{{#1}}}
\newcommand{\InvCostProfile}[1][\Label]{\psi_{#1}^{-1}}

\newcommand{\StationProfile}[1][\Arc]{\CostProfile[{#1}]}
\newcommand{\InvStationProfile}[1][\Arc]{\InvCostProfile[{#1}]}

\newcommand{\ShiftedStationProfile}{\StationProfile^{\rightarrow}}
\newcommand{\InvShiftedStationProfile}{(\ShiftedStationProfile)^{-1}}

\newcommandtwoopt{\CostProfileBreakpoints}[2][\Label][p]{\mathcal{B}_{\CostProfile[#1][#2]}}
\newcommand{\CostProfileMinCost}[1][\Label]{c_{\min}(\CostProfile[#1])}
\newcommand{\CostProfileMaxCost}[1][\Label]{c_{\max}(\CostProfile[#1])}
\newcommand{\CostProfileMinSoC}[1][\Label]{q_{\min}(\CostProfile[#1])}
\newcommand{\CostProfileMaxSoC}[1][\Label]{q_{\max}(\CostProfile[#1])}
\newcommandtwoopt{\SoCProfileMinTime}[2][\Label][p]{\tau_{\min}(\SoCProfile[#1][#2])}
\newcommandtwoopt{\SoCProfileMaxTime}[2][\Label][p]{\tau_{\max}(\SoCProfile[#1][#2])}
\newcommandtwoopt{\SoCProfileMinSoC}[2][\Label][p]{q_{\min}(\SoCProfile[#1][#2])}
\newcommandtwoopt{\SoCProfileMaxSoC}[2][\Label][p]{q_{\max}(\SoCProfile[#1][#2])}

\newcommand{\ChargingTime}[1][v]{\Delta\tau_{#1}}
\newcommandtwoopt{\SoCProfile}[2][\Label][\PartialPath]{\Gamma_{#2}\langle #1 \rangle}
\newcommandtwoopt{\InverseSoCProfile}[2][\Label][\PartialPath]{\Gamma^{-1}_{#2}\langle #1 \rangle}
\newcommandtwoopt{\SoCProfileSlope}[2][\Label][\PartialPath]{\dot{\Gamma}_{#2}\langle #1 \rangle}
\newcommandtwoopt{\CostProfileSlope}[2][\Label][\PartialPath]{\dot{\psi}_{#2}\langle #1 \rangle}
\newcommandtwoopt{\CostProfileTime}[2][\Label][\PartialPath]{T_{#2}\langle #1 \rangle}
\newcommandtwoopt{\CostProfileTimeSlope}[2][\Label][\PartialPath]{\dot{T}_{#2}\langle #1 \rangle}

\newcommandtwoopt{\PWLSlope}[2][f][x]{\frac{\partial #1}{\partial #2}}

\newcommandtwoopt{\PropagateRegularArc}[2][\Label][(\Edge)]{#1 \underset{#2}{\gets} /}

\newcommandtwoopt{\PropagateAndReplace}[2][\Label][c]{#1 \overunderset{\ChargerOfVertex}{(\Edge)}{\gets} #2}
\newcommandtwoopt{\PropagateAndCharge}[2][\Label][\tau]{#1 \underset{(\Edge)}{\gets} #2}

\newcommandtwoopt{\PropagateAndReplaceNoC}[2][\Label][(\Edge)]{#1 \overunderset{\ChargerOfVertex}{#2}{\gets}}
\newcommandtwoopt{\PropagateAndChargeNoTau}[2][\Label][(\Edge)]{#1 \underset{#2}{\gets}}

\newcommandtwoopt{\PropagateAndReplaceAlone}[2][\ChargerOfVertex][(\Edge)]{\overunderset{#1}{#2}{\gets}}
\newcommand{\PrevStationVertex}{\OriginVertex'}
\newcommand{\NewStationVertex}{\OriginVertex}
\newcommand{\PrevLabel}{\Label}
\newcommand{\NewLabel}{\Label'}
\newcommand{\PrevCostProfile}{\CostProfile[\PrevLabel]}
\newcommand{\InvPrevCostProfile}{\InvCostProfile[\PrevLabel]}
\newcommand{\NewCostProfile}{\CostProfile[\NewLabel]}
\newcommand{\InvNewCostProfile}{\InvCostProfile[\NewLabel]}
\newcommand{\PrevTrackedPeriod}{\PeriodOfVertex[\PrevStationVertex]}
\newcommand{\NewTrackedPeriod}{\PeriodOfVertex[\NewStationVertex]}
\newcommand{\NewStationProfile}{\CostProfile[\OriginVertex, \TargetVertex]}
\newcommand{\InvNewStationProfile}{\InvCostProfile[\OriginVertex, \TargetVertex]}
\newcommand{\NewPhi}{\ChargingFunction[{\ChargerOfVertex}]}
\newcommand{\BreakpointSwitchingSequence}{\chi}
\newcommand{\SupersetBreakpointSwitchingSequence}{\tilde{\chi}}

\newcommand{\SetOfVehicleSchedules}[1][\Vehicle]{\mathcal{A}_{#1}}
\newcommand{\ReducedSetOfVehicleSchedules}[1][\Vehicle]{\tilde{\SetOfVehicleSchedules[#1]}}
\newcommand{\SetOfSchedules}{\mathcal{A}}
\newcommand{\ReducedSetOfSchedules}{\tilde{\SetOfSchedules}}
\newcommand{\ScheduleMatrix}[1][\VehicleSchedule]{\bm{A}^{#1}}
\newcommand{\ChargeMatrix}[1][\VehicleSchedule]{\bm{B}^{#1}}
\newcommand{\ChargerCapacity}[1][\Charger]{C_{#1}}
\newcommand{\SoC}{\texttt{SoC}}
\newcommand{\MinSoC}{\SoC_{\min}}
\newcommand{\MaxSoC}{\SoC_{\max}}
\newcommand{\TerminalVoltage}[1][\texttt{SoC}]{V_{term}(#1)}
\newcommand{\OpenCircuitVoltage}[1][\texttt{SoC}]{V_{OC}(#1)}
\newcommand{\MaxTerminalVoltage}{V_{term}^{\max}}
\newcommand{\EarliestTourDeparture}[1][\Tour]{a_{#1}}
\newcommand{\Breakpoint}{b}
\newcommand{\ChargerSegment}{b}
\newcommand{\AllSegments}[1][\Charger]{\ChargerSegment \in \SetOfChargerSegments[#1]}
\newcommand{\TourCost}[1][\Tour]{c_{#1}}
\newcommand{\PricingObjectiveValue}[1][\Vehicle]{c^*_{\Vehicle}}
\newcommand{\VertexCost}[1][\Vertex]{c_{#1}}
\newcommand{\ScheduleCost}[1][\Schedule]{c(#1)}
\newcommand{\DegradationCostBreakpoint}{c_{\WearDensityFunction, \Breakpoint}}
\newcommand{\LatestTourDeparture}[1][\Tour]{d_{#1}}
\newcommand{\VertexEarliestDeparture}[1][\Vertex]{\underline{d}_{#1}}
\newcommand{\VertexLatestDeparture}[1][\Vertex]{\overline{d}_{#1}}
\newcommand{\PeriodEnergyCost}[1][\Period]{e_{#1}}
\newcommand{\PeriodEnergyPrice}[1][\Period]{e_{#1}}
\newcommand{\VertexEnergyCost}[1][\Vertex]{e_{#1}}
\newcommand{\Charger}{f}
\newcommand{\BaseCharger}{\hat{f}}
\newcommand{\AllChargers}{\Charger \in \SetOfChargers}
\newcommand{\DegradationSoCIntervalUB}[1][\DegradationSegment]{\overline{g_{#1}}}
\newcommand{\DegradationSoCIntervalLB}[1][\DegradationSegment]{\underline{g_{#1}}}
\newcommand{\VertexPotential}{h}
\newcommand{\OriginVertex}{i}
\newcommand{\TargetVertex}{j}
\newcommand{\Vehicle}{k}
\newcommand{\AllVehicles}{k \in \SetOfVehicles}
\newcommand{\DegradationCost}[1][\DegradationSegment]{o_{#1}}
\newcommand{\Period}{p}
\newcommand{\ChargerSoCIntervalUB}[1][\Charger,\ChargerSegment]{\overline{q_{#1}}}
\newcommand{\ChargerSoCIntervalLB}[1][\Charger,\ChargerSegment]{\underline{q_{#1}}}
\newcommand{\ChargerSoCBreakpoint}[1][\Charger,\ChargerSegment]{q_{#1}}
\newcommand{\DegradationSoCBreakpoint}[1][\WearDensityFunction, \Breakpoint]{q_{#1}}
\newcommand{\SourceNode}{s^{-}}
\newcommand{\SinkNode}{s^{+}}
\newcommand{\ChargerTimeIntervalUB}[1][\Charger, \ChargerSegment]{\overline{t_{#1}}}
\newcommand{\ChargerTimeIntervalLB}[1][\Charger, \ChargerSegment]{\underline{t_{#1}}}
\newcommand{\ChargerTimeBreakpoint}[1][\Charger, \ChargerSegment]{t_{#1}}
\newcommandtwoopt{\VehicleChargesOnDegSeg}[2][\Vehicle][\Tour, \DegradationSegment]{u^{#1}_{#2}}
\newcommandtwoopt{\DegradationSegmentSelected}[2][\Vehicle][\Arc, \DegradationSegment]{u^{#1}_{#2}}
\newcommandtwoopt{\VehicleUsesChargerInPeriod}[2][\Vehicle][\Period, \Charger]{v^{#1}_{#2}}
\newcommand{\Vertex}{v}
\newcommand{\DegradationSegment}{w}
\newcommand{\AllDegSegments}{\DegradationSegment \in \SetOfDegradationSegments}
\newcommandtwoopt{\VehicleChargesInPeriod}[2][\Vehicle][\Period]{x^{#1}_{#2}}
\newcommandtwoopt{\ScheduleUsed}[2][\Vehicle][\VehicleSchedule]{x^{#1}_{#2}}
\newcommand{\TravelsEdge}[1][\Edge]{x^{\Vehicle}_{#1}}
\newcommandtwoopt{\VehicleUsesCharger}[2][\Vehicle, \Period][\Charger, \ChargerSegment]{y^{#1}_{#2}}
\newcommandtwoopt{\VehicleUsesChargerEntry}[2][\Vehicle, \Period][\Charger, \ChargerSegment]{\overrightarrow{y^{#1}_{#2}}}
\newcommandtwoopt{\VehicleUsesChargerExit}[2][\Vehicle, \Period][\Charger, \ChargerSegment]{\overleftarrow{y^{#1}_{#2}}}
\newcommand{\ChargerUsedSchedule}[1][\Schedule, \Period, \Charger]{y_{#1}}
\newcommand{\VehicleUnplugEvent}[1][\Period]{z^{\Vehicle}_{#1}}

\newcommand{\RelaxedRestrictedMasterProblemFunctionValue}{\tilde{z}_{RMP}}
\newcommand{\RestrictedMasterProblemFunctionValue}{z_{RMP}}
\newcommand{\RelaxedMasterProblemFunctionValue}{\tilde{z}_{MP}}
\newcommand{\RelaxedMasterProblemOptimalFunctionValue}{\tilde{z}^*_{MP}}
\newcommand{\CurRMPSol}{\bar{z}_{RMP}}
\newcommand{\RestrictedMasterProblemOptimalFunctionValue}{z^*_{RMP}}
\newcommand{\MasterProblemFunctionValue}{z_{MP}}
\newcommand{\MasterProblemOptimalFunctionValue}{z^*_{MP}}
\newcommand{\OptimalSol}{z^*}
\newcommand{\OptimalMPSol}{z_{MP}^*}
\newcommand{\SetOfChargerSegments}[1][\Charger]{\mathcal{B}({#1})}
\newcommand{\SetOfBreakpoints}[1][\Charger]{\mathcal{B}({#1})}
\newcommand{\SetOfDegradationSegments}{\SetOfBreakpoints[\WearDensityFunction]}
\newcommand{\SetOfDays}{\mathcal{D}}
\newcommand{\Edges}{\mathcal{E}}
\newcommand{\SetOfChargers}{\mathcal{F}}
\newcommand{\SetOfWDFIntervals}{\mathcal{W}}
\newcommand{\SetOfVehicles}{\mathcal{K}}
\newcommand{\Label}{\ell}
\newcommand{\SourceLabel}{\Label_{\SourceNode}}
\newcommand{\SetOfLabels}{\mathcal{L}}
\newcommand{\SetOfUnsettledLabels}[1][\OriginVertex]{\SetOfLabels^{uns}_{#1}}
\newcommand{\SetOfSettledLabels}[1][\OriginVertex]{\SetOfLabels^{set}_{#1}}
\newcommand{\SetOfGeneratedLabels}{\SetOfLabels_{new}}
\newcommand{\SetOfPeriods}{\mathcal{P}}
\newcommand{\AllPeriods}{\Period \in \SetOfPeriods}
\newcommand{\SetOfDepotPeriodsPerVehicle}[1][\Vehicle]{\SetOfPeriods_{#1}}
\newcommand{\AllDepotPeriods}{\Period \in \SetOfDepotPeriodsPerVehicle}
\newcommand{\NumberOfPeriods}{|\SetOfPeriods|}
\newcommand{\NodeQueue}{\mathcal{Q}}

\newcommand{\SetOfVertices}{\mathcal{V}^{\Vehicle}}
\newcommand{\Vertices}[1][{\Vehicle}]{\mathcal{V}^{#1}}
\newcommand{\TourVertices}[1][\Vehicle]{\Vertices_{\SetOfTours[#1]}}
\newcommand{\TourVerticesWithSink}[1][\Vehicle]{\Vertices[\Vehicle, +]_{\SetOfTours[#1]}}
\newcommand{\TourVerticesWithSource}[1][\Vehicle]{\Vertices[\Vehicle, -]_{\SetOfTours[#1]}}
\newcommand{\TourVerticesWithDummys}[1][\Vehicle]{\Vertices^{\circ}_{\SetOfTours[#1]}}
\newcommand{\ChargerVertices}{\Vertices_{\SetOfChargers}}
\newcommand{\VerticesOfCharger}{\Vertices_{\Charger}}
\newcommand{\AllVertices}{\Vertex \in \SetOfVertices}

\newcommand{\TourConsumption}[1][\Tour]{\Delta\texttt{SoC}_{#1}}
\newcommand{\TourDuration}[1][\Tour]{\Delta\tau_{#1}}
\newcommand{\VertexConsumption}[1][\Arc]{\Delta\gls{abk:soc}_{#1}}
\newcommand{\SetOfTours}[1][\Vehicle]{\Theta_{#1}}
\newcommand{\AllToursDay}{\Tour \in \SetOfTours}
\newcommand{\AllTours}{\Tour \in \SetOfTours[]}
\newcommand{\WearDensityFunction}{\Upsilon}
\newcommand{\ChargeFunctionSymbol}{\Phi}
\newcommand{\ChargeFunction}[1][\Charger]{\ChargeFunctionSymbol_{#1}}
\newcommand{\InvChargeFunction}[1][\Charger]{\ChargeFunctionSymbol^{-1}_{#1}}
\newcommand{\PWLChargeFunction}[1][\Charger]{\overline{\ChargeFunction[#1]}}
\newcommand{\ChargingFunction}[1][\Charger]{\ChargeFunction[#1]}
\newcommand{\PWLChargingFunction}[1][\Charger]{\PWLChargeFunction[#1]}
\newcommand{\FleetSchedule}{\Omega}
\newcommand{\VehicleTourDepartureTime}[1][\Tour]{\alpha_{#1}}
\newcommandtwoopt{\VehicleSoCAtBegOfPeriod}[2][\Vehicle][\Period]{\beta^{#1}_{#2}}
\newcommand{\VehicleArrivalSoC}[1][\Vertex]{\beta^{\Vehicle}_{#1}}
\newcommandtwoopt{\VehicleRechargeInPeriod}[2][\Vehicle][\Period]{\gamma^{#1}_{#2}}
\newcommand{\VehicleReplenishedCharge}[1][\Vertex]{\gamma^{\Vehicle}_{#1}}

\newcommand{\ScheduleChargingDecisions}[1][\VehicleSchedule]{\bar{\gamma}_{#1}}
\newcommand{\ChargingDecisions}{\bar{\gamma}}
\newcommand{\PeriodChargingDecision}[1][\Period]{\tau_{#1}}
\newcommand{\ArcChargingDecision}[1][\Edge]{\gamma_{#1}}
\newcommand{\VertexChargingDecision}[1][\Vertex]{\gamma_{#1}}
\newcommand{\PathChargingDecisions}[1][\PartialPath]{\bar{\gamma}_{#1}}
\newcommand{\OutgoingArcs}[1][\OriginVertex]{\delta^{-}(#1)}
\newcommand{\IncomingArcs}[1][\TargetVertex]{\delta^{+}(#1)}
\newcommand{\Tour}{\vartheta}
\newcommand{\DummyTourStart}{\Tour_0}
\newcommand{\DummyTourEnd}{\Tour_{n+1}}
\newcommand{\TourPrecedence}{\gg}
\newcommand{\ArcOperation}[1][\Edge]{\Tour_{#1}}
\newcommand{\ConvexMultEntrySoC}[1][\Vertex, \ChargerSegment]{\lambda^{k, in}_{#1}}
\newcommand{\ConvexMultExitSoC}[1][\Vertex, \ChargerSegment]{\lambda^{k, out}_{#1}}
\newcommand{\PredecessorPeriodSymbol}[1][\Tour]{\mu}
\newcommand{\PredecessorPeriod}[1][\Tour]{\mu_{#1}}
\newcommand{\AccumulatedSoC}[1][\Vertex]{\mu^{\Vehicle}_{#1}}
\newcommand{\ConvexMultEntrySoCDeg}{\mu^{\Vehicle, in}_{\Vertex, \Breakpoint}}
\newcommand{\ConvexMultExitSoCDeg}{\mu^{\Vehicle, out}_{\Vertex, \Breakpoint}}
\newcommand{\PartialPricingScheduleCount}{\nu}
\newcommand{\PeriodDuration}[1][\Period]{\xi}
\newcommand{\DualScheduleUsed}[1][\Vehicle]{\pi^{\scriptsize\eqref{MP-mip:oneSchedulePerVehicle}}_{#1}}
\newcommand{\DualChargerCapacity}[1][\Period, \Charger]{\pi^{\scriptsize\eqref{MP-mip:stationCapacityRespected}}_{#1}}
\newcommandtwoopt{\VehicleRechargeOnDegSegment}[2][\Vehicle][\Tour, \DegradationSegment]{\rho^{#1}_{#2}}
\newcommandtwoopt{\DegradationSegmentSoC}[2][\Vehicle][\Arc, \DegradationSegment]{\rho^{#1}_{#2}}
\newcommand{\DegradationOfChargingArc}{\rho^{\Vehicle}_{\Vertex}}
\newcommand{\PartialPath}{\rho}
\newcommand{\VehicleArrivalPeriod}[1][\Vertex]{\tau^{\Vehicle}_{#1}}

\newcommand{\VehicleSchedule}{\omega}
\newcommand{\Schedule}{\VehicleSchedule}
\newcommandtwoopt{\DeltaChargingCost}[2][q'][\Delta q]{\Delta c_{\PeriodOfVertex, \ChargerOfVertex}\langle #1 \rangle(#2)}
\newcommandtwoopt{\InverseDeltaChargingCost}[2][q'][\Delta q]{\Delta c_{\PeriodOfVertex, \ChargerOfVertex}^{-1}\langle #1 \rangle(#2)}
\newcommand{\AssumedCost}{c}

\newcommand{\OriginalLabel}{\Label_{i}}
\newcommand{\NoChargeLabel}{\Label_{j}}
\newcommand{\ToBeDominatedLabel}{{\Label'}}
\newcommand{\DominatingLabel}{{\Label''}}

\newcommand{\InverseOriginalProfile}{\InvCostProfile[\OriginalLabel]}
\newcommand{\OriginalProfile}{\CostProfile[\OriginalLabel]}
\newcommand{\InverseNoChargeProfile}{\InvCostProfile[\NoChargeLabel]}
\newcommand{\NoChargeProfile}{\CostProfile[\NoChargeLabel]}
\newcommand{\InverseToBeDominatedProfile}{{\InvCostProfile[\ToBeDominatedLabel]}}
\newcommand{\ToBeDominatedProfile}{{\CostProfile[\ToBeDominatedLabel]}}
\newcommand{\DominatingProfile}{{\CostProfile[\DominatingLabel]}}
\newcommand{\InverseDominatingProfile}{{\InvCostProfile[\DominatingLabel]}}

\newcommand{\EntryCostDominatedStation}{c_{j}'}
\newcommand{\EntrySoCDominatedStation}{q'}

\newcommand{\EntryCostDominating}{c_{j}''}
\newcommand{\EntrySoCDominating}{q''}

\title{\large Electric vehicle charge scheduling with flexible service operations}

\author[1]{\normalsize Patrick Sean Klein}
\author[2]{\normalsize Maximilian Schiffer}
\affil{\small 
	TUM School of Management, Technical University of Munich, 80333 Munich, Germany
	
	\scriptsize patrick.sean.klein@tum.de
	
	\small
	\textsuperscript{2}TUM School of Management \& Munich Data Science Institute,
	
	Technical University of Munich, 80333 Munich, Germany
	
	\scriptsize schiffer@tum.de}

\date{}

\lehead{\pagemark}
\rohead{\pagemark}

\begin{abstract}
	\begin{singlespace}
		{\small\noindent Operators who deploy large fleets of electric vehicles often face a challenging charge scheduling problem. Specifically, time-ineffective recharging operations limit the profitability of charging during service operations such that operators recharge vehicles off-duty at a central depot. Here, high investment cost and grid capacity limit available charging infrastructure such that operators need to schedule charging operations to keep the fleet operational. In this context, flexible service operations, i.e. allowing to delay or expedite vehicle departures, can potentially increase charger utilization. Beyond this, jointly scheduling charging and service operations promises operational cost savings through better utilization of \acrlong{abk:tou} energy tariffs and carefully crafted charging schedules designed to minimize battery wear.
Against this background, we study the resulting joint charging and service operations scheduling problem accounting for battery degradation, non-linear charging, and \acrlong{abk:tou} energy tariffs. We propose an exact Branch \& Price algorithm, leveraging a custom branching rule and a primal heuristic to remain efficient during the Branch \& Bound phase. Moreover, we develop an exact labeling algorithm for our pricing problem, constituting a resource-constrained shortest path problem that considers variable energy prices and non-linear charging operations.
We benchmark our algorithm in a comprehensive numerical study and show that it can solve problem instances of realistic size with computational times below one hour, thus enabling its application in practice. Additionally, we analyze the benefit of jointly scheduling charging and service operations. We find that our integrated approach lowers the amount of charging infrastructure required by up to $57\%$ besides enabling operational cost savings of up to $5\%$.\\
			\smallskip}
	\end{singlespace}
\end{abstract}

\maketitle
\section{Introduction}
\label{sec:introduction}
\pagenumbering{arabic}

Increasing societal and political environmental awareness resulting from climate change and local and global emission problems call for a paradigm change towards sustainable transportation systems. Herein, \glspl{abk:ecv} are seen as a promising alternative to \glspl{abk:icev}, allowing up to 20\% reduction in life-cycle greenhouse gas emissions when considering the current European energy mix \citep[cf.][]{Agency2018}. Moreover, \glspl{abk:ecv} may provide an economic advantage due to lower operational costs \citep{Taefi2016, SchifferKleinEtAl2020}.
Accordingly, major players in the freight and passenger transportation sectors started to electrify their fleets. Seminal examples of this development include the Deutsche Post DHL Group (DPDHL), UPS,  FedEx, General Electric, Hertz, and Amazon in the freight transportation sector, as well as Uber, Lyft, and Addison Lee in the passenger transportation sector \citep[cf.][]{DPDHL2017,Clark2019, JuanMendezEtAl2016, RodriguezHildermeierEtAl2020, Lyft2021, AddisonLee2021}.

A central challenge in all of these applications is the efficient scheduling of charging operations, which are often conducted during off-service periods using private charging infrastructure installed at a central depot to avoid inefficient use of drivers' time. Here, grid constraints and high investment costs limit the availability of dedicated charging infrastructure, such that there are generally fewer (fast) chargers than vehicles. Accordingly, operators must synchronize the fleet's charging operations to avoid charger capacity bottlenecks.
Moreover, \gls{abk:tou} energy tariffs, which charge different prices depending on the time of consumption, further complicate this scheduling problem: with on-peak prices up to three times as high as off-peak prices \citep{OpenEI2022}, it becomes economically worthwhile to consider energy prices when planning charging operations.
Generally, charger capacity and \gls{abk:tou} pricing favor schedules where a vehicle's \gls{abk:soc} peaks at certain times, e.g., when a fast charger becomes available or energy is cheap \citep[cf.][]{PelletierJabaliEtAl2018}. However, these schedules impose considerable stress on an \gls{abk:ecv}'s battery, such that the long-term effects of battery degradation may mitigate short-term energy cost savings (see Appendix~\ref{app:technical-fundamentals}).
Operators who want to utilize this trade-off between charger utilization, off-peak energy prices, and battery degradation must consider an accurate (non-linear) charging model as simple (e.g., linear) approximations may over- or underestimate charging rates, which potentially distorts the cost savings attainable through the trade-off mentioned above \citep[cf.][]{MontoyaGueretEtAl2017, PelletierJabaliEtAl2018}.

In practice, operators often determine service schedules in an upstream planning problem, e.g., by solving a respective \gls{abk:vrp} or \gls{abk:vsp}, and schedule charging operations for the resulting fixed service schedule and vehicle assignment subsequently.
This hierarchical decomposition often stems from applications with complex rostering constraints, when operators value consistent service \citep{Stavropoulou2022}, or when compatibility dependencies between service operations and vehicles or drivers exist \citep{BatsynBatsynaEtAl2021}. Examples of such applications are abundant: in city logistics, narrow or particularly congested roads may limit vehicle length, height, or weight. In law enforcement and military applications, access clearance may constrain the driver pool. Maintenance problems may place requirements on vehicle equipment or crew skill, while continuity of care may be a hard constraint in health care applications.
These predetermined service schedules often have some (unavoidable) slack due to, e.g., driver service regulations and restrictive time windows \citep[cf.][]{KokHansEtAl2011}, such that individual service operations are \textit{flexible}, i.e., can be shifted in time to a limited extent without violating upstream scheduling constraints. Operators may benefit from this flexibility and delay or expedite a service operation to allow charging at a slower charger, e.g., to make a faster charger available to another vehicle of the fleet, to charge during cheap off-peak periods, or to balance charging operations across the planning horizon to avoid charging patterns with high impact on battery health.

Concluding, operators who deploy (large) fleets of \glspl{abk:ecv} face an inherently complex planning problem comprising decisions on charging and service operation schedules, which may significantly impact the viability and practicability of an \gls{abk:ecv} fleet. Here, they need to account for \textit{i)} capacity restrictions of available charging infrastructure, \textit{ii)} battery degradation effects, \textit{iii)} \gls{abk:tou} energy tariffs, \textit{iv)} non-linear battery behavior, and \textit{v)} flexible service operations.
We study the resulting planning problem in the remainder of this paper. In the following, we first review related work in Section~\ref{sec:state-of-the-art} before we state our aims and scope in Section~\ref{sec:aims-and-scope} and outline the paper's structure in Section~\ref{sec:outline}.

\subsection{State-of-the-Art}
\label{sec:state-of-the-art}
\glsreset{abk:evrp}\glsreset{abk:vsp}\glsreset{abk:csp}
We concisely review the state-of-the-art of related research areas, namely \glspl{abk:evrp}, \glspl{abk:vsp}, and \glspl{abk:csp}. For in-depth reviews of these research fields, we refer to \citet{SchifferSchneiderEtAl2018b} and \citet{Olsen2020}. 

Most publications in the context of \glspl{abk:evrp} focus on routing decisions and simplify charging-related issues such as battery degradation, variable energy prices, and non-linear battery behavior.
In fact, apart from a few recent publications \citep{MontoyaGueretEtAl2017, Lee2020b, LiangDabiaEtAl2021, LamDesaulniersEtAl2022}, charging operations were either modeled as a fixed time penalty \citep{ConradFigliozzi2011, ErdoganMiller-Hooks2012} or were considered to be linear with respect to time and residual battery capacity \citep{SchneiderStengerEtAl2014, DesaulniersErricoEtAl2016, SchifferWalther2018}. Moreover, with the exception of \citet{LinGhaddarEtAl2021}, no publications in the realm of \glspl{abk:evrp} considered variable energy prices. 
Similarly, capacity constraints at charging stations have so far, to the best of our knowledge, only been considered in \cite{FrogerJabaliEtAl2022}, \citet{BruglieriManciniEtAl2019}, and \cite{LamDesaulniersEtAl2022}.

\glspl{abk:vsp}, which focus on assigning a set of (fixed) trips to a fleet of vehicles, have been limited similarly. Here, most publications focused on conventional vehicles and did not consider charging operations. \Glspl{abk:evsp} assumed either instantaneous \citep{AdlerMirchandani2017,YaoLiuEtAl2020}, (partial) linear \citep{WenLindeEtAl2016, AlvoAnguloEtAl2021, ParmentierMartinelliEtAl2021}, or (discretized) non-linear charging \citep{KootenNiekerkAkkerEtAl2017}.

\glspl{abk:csp} differ from \glspl{abk:evsp} by assuming a fixed assignment of trips to vehicles, which reduces the problem's complexity to scheduling charging operations in-between trips. In the realm of \glspl{abk:csp}, early publications have assumed linear charging operations and did not consider station capacity constraints or heterogeneous chargers \citep{SassiOulamara2014,SassiOulamara2017}.
More recent work on charge-scheduling problems alleviated some of these shortcomings. Specifically, \citet{AbdelwahedBergEtAl2020} consider station capacity and heterogeneous chargers but do not account for non-linear battery charging and degradation. They derive and compare discrete-time and discrete-event mixed-integer formulations using a commercial solver. \citet{PelletierJabaliEtAl2018} contribute a \gls{abk:mip} that models realistic battery behavior, accounting for non-linear battery degradation and charging. They conduct an extensive case study using a commercial solver to assess the influence of both cyclic and calendric battery aging, energy price, and grid restrictions in several city logistics scenarios.

Concluding, Table~\ref{table:literature} categorizes the most-related publications in the realm of \glspl{abk:evsp} and \glspl{abk:csp}. As can be seen, related publications do not consider the charging process in sufficient detail, particularly with respect to non-linear charging and, with the exception of \citet{KootenNiekerkAkkerEtAl2017}, charger capacity constraints. The work of \citet{PelletierJabaliEtAl2018} and \citet{AbdelwahedBergEtAl2020} does not consider service scheduling and relies on a standard \gls{abk:mip}, solved with commercial solvers, such that it remains limited in its computational scalability. To the best of our knowledge, a comprehensive, integrated approach for joint charging and service operation scheduling of \glspl{abk:ecv} has not been studied so far.
\begin{table}[b!]
	\newcolumntype{Y}{>{\centering\arraybackslash}X}
	\centering
	\caption{Related publications.\label{table:literature}}
	\begin{threeparttable}
		\begin{tabularx}{\linewidth}{l *{7}{Y}}
			\toprule
			 & [1] & [2] & [3] & [4] & [5] & [6] & \small Our work\\
			\toprule
			Service scheduling &  &  & & & & & $\checkmark$ \\
			Vehicle assignment & $\checkmark$ & $\checkmark$ & $\checkmark$ & & $\checkmark$ & & \\
			Continuous charging & $\checkmark$ & $\checkmark$ & & $\checkmark$ & $\checkmark$ & $\checkmark$ & $\checkmark$ \\
			Non-linear charging &  &  &  & $\checkmark$ &  & & $\checkmark$ \\
			Battery degradation &  &  & $\checkmark$ & $\checkmark$ &  & & $\checkmark$ \\
			Energy price & $\checkmark$ & $\checkmark$ & $\checkmark$ & $\checkmark$ & & & $\checkmark$ \\
			Heterogeneous chargers &  & $\checkmark$ & $\checkmark$ & $\checkmark$ & $\checkmark$ & $\checkmark$ & $\checkmark$ \\
			Station capacity &  &  & $\checkmark$ & $\checkmark$ &  & $\checkmark$ & $\checkmark$ \\
			\midrule
			Exact & $\checkmark$ & $\checkmark$ &  & $\checkmark$ & $\checkmark$ & $\checkmark$ & $\checkmark$ \\
			Scalable & $\checkmark$ & & $\checkmark$ &  & $\checkmark$ & & $\checkmark$ \\
			\bottomrule
		\end{tabularx}
		\begin{tablenotes}[flushleft]
			\item\smaller Indices {[1]} to {[6]} signify publications as follows: {[1]} \cite{SassiOulamara2017}, {[2]} \cite{SassiOulamara2014}, {[3]} \cite{KootenNiekerkAkkerEtAl2017}, {[4]} \cite{PelletierJabaliEtAl2018}, {[5]} \cite{ParmentierMartinelliEtAl2021}, {[6]} \cite{AbdelwahedBergEtAl2020}.
		\end{tablenotes}
	\end{threeparttable}
\end{table}
\subsection{Contribution}
\label{sec:aims-and-scope}
This paper proposes a joint charging and service operation scheduling problem that accounts for a realistic battery behavior model with \textit{i)} limited charger capacity, \textit{ii)} non-linear charging operations, \textit{iii)} battery degradation, and \textit{iv)} variable energy prices.
Here, our contribution is twofold: 

from a methodological perspective, we develop an efficient \gls{abk:bnp} algorithm that significantly outperforms commercial solvers and allows to solve problem sizes encountered in practice. 
This algorithm relies on a problem-specific branching rule, a primal heuristic, and partial pricing to remain scaleable. Our pricing problem constitutes a so-far unconsidered extension to the \gls{abk:frvcp} \citep[cf.][]{BaumDibbeltEtAl2019, FrogerMendozaEtAl2019, KullmanFrogerEtAl2020}. Specifically, we consider time-constrained charging operations. We develop a label-setting algorithm that utilizes a continuous label representation and relies on a set-based dominance rule to solve this pricing subproblem efficiently. 
A comprehensive numerical study shows the efficiency of our approach and asserts its scalability to large problem sizes.

From a managerial perspective, we analyze the impact of jointly scheduling charging and service operations on the amount of charging infrastructure required and the benefit of accounting for variable energy prices and battery degradation. Specifically, our computational study shows that integrated planning of charging and service operations improves the utilization of variable energy prices, lowers the cost incurred from battery degradation, and allows charging infrastructure savings. Specifically, our integrated approach lowers the amount of charging infrastructure required by up to $57\%$ and reduces operational costs by up to $5\%$. We further reveal that both the degree of service schedule flexibility and the energy price distribution significantly impact these savings.

\subsection{Outline}
\label{sec:outline}
The remainder of this paper is as follows.
Section~\ref{sec:problem-definition} provides a formal definition of our problem setting and derives a set-covering-based \gls{abk:ip} before Section~\ref{sec:column-generation} develops a column generation procedure. Section~\ref{sec:branch-and-price} embeds this column generation procedure into a \gls{abk:bnb} algorithm, leveraging a primal heuristic and problem-specific branching- and node-selection rules.
Section~\ref{sec:instances} details the design of our computational and managerial studies, the results of which we discuss in Section~\ref{sec:results}. Finally, Section~\ref{sec:conclusion} concludes this paper with a summary and an outlook on future research.
\section{Problem definition}
\label{sec:problem-definition}
\label{subsec:problem-definition}
We consider a set of vehicles $\Vehicle \in \SetOfVehicles$, each required to service a set of operations $\Tour \in \SetOfTours$, starting and ending at a central depot. Servicing an operation $\Tour \in \SetOfTours[], \SetOfTours[] = \cup_{\Vehicle \in \SetOfVehicles} \SetOfTours$, consumes a certain amount of energy $\TourConsumption$, takes a certain amount of time $\TourDuration$, and is restricted to an operation-specific time window, such that vehicle $\Vehicle$ must depart between $\EarliestTourDeparture$ and $\LatestTourDeparture$ to serve $\Tour$. Two comments on this setting are in order. First, we assume a fixed assignment of vehicles to service operations, determined in an upstream planning problem. This assumption is realistic in applications where compatibility constraints between service operations and vehicles or drivers exist, or where operators value consistent service \citep[cf.][]{BatsynBatsynaEtAl2021, Stavropoulou2022}.
Second, we specify that we do not assume any ordering of service operations such that any two operations $\Tour_i, \Tour_j \in \SetOfTours$ assigned to the same vehicle $\Vehicle \in \SetOfVehicles$ may be served in arbitrary order if their departure time windows allow. We note that our solution methodology can be extended straightforwardly to account for such precedence constraints.

We schedule operations on a finite time horizon, discretized with a time step width of $\PeriodDuration$ minutes, given as an ordered set $\SetOfPeriods$. This discretization is conservative with respect to operation time windows as it shifts departure and arrival times to the beginning and end of the respective periods. We denote the $i\textsuperscript{th}$ period as $\SetOfPeriods_i$ and assign to each period $\Period \in \SetOfPeriods$ an energy price~$\PeriodEnergyPrice$.

Vehicles can be recharged using a set of (heterogeneous) charging stations $\Charger \in \SetOfChargers$ available at the depot, each capable of simultaneously charging up to $\ChargerCapacity$ vehicles. To avoid charging more energy than necessary, we allow partial charging operations independent of the time discretization, such that charging may be started or interrupted at any time. 
However, our discretization remains conservative concerning charger capacity, such that vehicles charging in some period $\Period \in \SetOfPeriods$, occupy the respective charger for the entire period. In contrast to \citet{PelletierJabaliEtAl2018} and \citet{SassiOulamara2014}, we do not limit the number of uninterrupted charging operations but note that including such constraints is straightforward in our solution methodology.
We model non-linear charging behavior with charger-specific piecewise linear charging functions $\ChargeFunction: \text{Time} \mapsto \SoC$, which capture a vehicle's \gls{abk:soc} evolution over time when charging with an initially empty battery (cf. Figure~\ref{fig:non-linear-charging-pwl}). We assume convex charging functions $\ChargeFunction$ in line with \citet{MontoyaGueretEtAl2017} and \citet{PelletierJabaliEtAl2017}. We further define bivariate ${\ChargeFunction(\beta, \tau) \coloneqq \ChargeFunction(\InvChargeFunction(\beta) + \tau)}$ for the sake of conciseness. These give the \gls{abk:soc} after charging for time $\tau$ with an initial \gls{abk:soc} of $\beta$.
\begin{figure}[b!]
	\captionsetup[subfigure]{width=0.9\textwidth}
	\centering
	\begin{subfigure}[T]{0.45\textwidth}
		\subcaption{Piecewise linearization of $\ChargeFunction$.\label{fig:non-linear-charging-pwl}}
		\centering
		\resizebox{0.666\textwidth}{!}{
			\begin{tikzpicture}[scale=0.65]

\tikzset{
	bp/.style={
		circle, draw, thick, solid, inner sep=1.5pt
	},
	cost_profile/.style={
		black, very thick
	},
	label_profile/.style={
		cost_profile, dashed
	},
	dual/.style={
		black, dashed, bend left,->
	},
	dummy/.style={
		diamond, draw, inner sep=0pt, minimum size=0.7cm
	},
	station/.style={
		regular polygon, regular polygon sides = 3, draw, inner sep=0pt, minimum size=0.7cm
	},
	charging_arc/.style={
		black, solid, thick, ->
	},
	idle/.style={
		rectangle, draw, minimum size=0.5cm
	},
	arc_cost/.style={
		above
	},
	arc_consumption/.style={
		below
	},
}

\clip(-1, -1) rectangle (8.3,8.7);

\draw[step=1cm, gray, very thin] (0, -0) grid (8.3, 8.3);
\draw[->, thick] (0,0) -- (8.3, 0) node[above=0.25cm, left=0.25cm] {Time};
\draw[->, thick] (0,-0) -- (0, 8.3) node[right] {SoC};

\draw[thick] (2mm, 0) -- (-2mm, 0) node[left] {$q_0$};
\draw[thick] (2mm, 5.35) -- (-2mm, 5.35) node[left] {$q_1$};
\draw[thick] (2mm, 7) -- (-2mm, 7) node[left] {$q_2$};
\draw[thick] (2mm, 8) -- (-2mm, 8) node[left] {$q_3$};

\draw[thick] (-2mm, 8) -- (2mm, 8);

\draw[thick] (0, 2mm) -- (0, -2mm) node[below] {$t_0$};
\draw[thick] (3, 2mm) -- (3, -2mm) node[below] {$t_1$};
\draw[thick] (5, 2mm) -- (5, -2mm) node[below] {$t_2$};
\draw[thick] (8, 2mm) -- (8, -2mm) node[below] {$t_3$};

\draw[dotted, thick] (0, 5.35) -- ++(3, 0) -- ++ (0, -5.35);
\draw[dotted, thick] (0, 7) -- ++(5, 0) -- ++(0, -7);
\draw[dotted, thick] (0, 8) -- ++(8, 0) -- ++(0, -8);
\draw[thick] (8, 8.32) -- (8, 8.28) node[left]  {$\MaxSoC$};

\draw[cost_profile] (0, 0) node[bp]{} -- ++(3,5.35) node[bp]{} -- node[below, right=2mm] {$\ChargingFunction(\tau)$} ++(2, 1.65) node[bp]{} -- (8, 8) node[bp]{};

\end{tikzpicture}
		}
		\begin{minipage}{\textwidth}
			\centering \smaller \textit{Note.} The set of breakpoints corresponds to ${\SetOfBreakpoints[{\ChargeFunction}] = \{t_i \mid i \in [0, 3]\}}$.
		\end{minipage}
	\end{subfigure}
	\begin{subfigure}[T]{0.45\textwidth}
		\subcaption{An example \acrshort{abk:wdf} $(\WearDensityFunction)$. \label{fig:discrete-wdf}}
		\centering
		\resizebox{0.666\textwidth}{!}{
			\begin{tikzpicture}[scale=0.65]

\tikzset{
	bp/.style={
		circle, draw, thick, solid, inner sep=1.5pt
	},
	cost_profile/.style={
		black, very thick
	},
	label_profile/.style={
		cost_profile, dashed
	},
	dual/.style={
		black, dashed, bend left,->
	},
	dummy/.style={
		diamond, draw, inner sep=0pt, minimum size=0.7cm
	},
	station/.style={
		regular polygon, regular polygon sides = 3, draw, inner sep=0pt, minimum size=0.7cm
	},
	charging_arc/.style={
		black, solid, thick, ->
	},
	idle/.style={
		rectangle, draw, minimum size=0.5cm
	},
	arc_cost/.style={
		above
	},
	arc_consumption/.style={
		below
	},
}

\clip(-1, -1) rectangle (8.3,8.7);

\draw[step=1cm, gray, very thin] (0, -0) grid (8.3, 8.3);
\draw[->, thick] (0,0) -- (8.3, 0) node[above=0.25cm, left=0.25cm] {SoC};
\draw[->, thick] (0,-0) -- (0, 8.3) node[right] {Cost};

\draw[thick] (2mm, 0) -- (-2mm, 0) node[left] {$c_0$};

\draw[thick] (2mm, 1) -- (-2mm, 1) node[left] {$c_1$};
\draw[thick] (2mm, 5) -- (-2mm, 5) node[left] {$c_2$};
\draw[thick] (2mm, 8) -- (-2mm, 8) node[left] {$c_3$};

\draw[thick] (8, 8.32) -- (8, 8.28) node[left]  {$\MaxSoC$};
\draw[dotted, thick] (8, 2mm) -- (8, 8.3);

\draw[thick] (0, 2mm) -- (0, -2mm) node[below] {$q_0$};
\draw[thick] (2, 2mm) -- (2, -2mm) node[below] {$q_1$};
\draw[thick] (6, 2mm) -- (6, -2mm) node[below] {$q_2$};
\draw[thick] (8, 2mm) -- (8, -2mm) node[below] {$q_3$};

\draw[cost_profile] (0, 0) node[bp]{} -- ++(2,1) node[bp]{} -- node[below, right=2mm]{$\WearDensityFunction(q)$} ++(4, 4) node[bp]{} -- ++(2, 3) node[bp]{};

\draw[dotted, thick] (0, 1) -- ++(2, 0) -- ++ (0, -1);
\draw[dotted, thick] (0, 5) -- ++(6, 0) -- ++(0, -5);
\draw[dotted, thick] (0, 8) -- ++(8, 0) -- ++(0, -8);

\end{tikzpicture}
		}
		\begin{minipage}{\textwidth}
			\centering \smaller \textit{Note.} Here, ${\SetOfBreakpoints[\WearDensityFunction] = \{q_i \mid i \in [0, 3]\}}$ gives the set of breakpoints.
		\end{minipage}
	\end{subfigure}
	\caption{Piecewise linear approximation of non-linear battery behavior.\label{fig:battery-behavior-models}}
\end{figure}

We quantify the charging cost attributed to battery deterioration in a \emph{\acrlong{abk:wdf}}\glsunset{abk:wdf}, denoted by ${\WearDensityFunction: \SoC \mapsto \texttt{cost}}$. This function is piecewise linear and convex on the battery's operational range, $[\MinSoC, \MaxSoC]$, and maps \gls{abk:soc} $q$ to the total cost of charging an initially empty battery up to $q$ (cf. Figure~\ref{fig:discrete-wdf}). We define a bivariate ${\WearDensityFunction(q, q_t) \coloneqq \WearDensityFunction(q_t) - \WearDensityFunction(q)}$, which describes the battery degradation related cost of charging from an initial \gls{abk:soc} $q$ to a target \gls{abk:soc} $q_t$. We refer to Appendix~\ref{app:technical-fundamentals} for a formal definition of our charging functions $\ChargingFunction$ and \gls{abk:wdf} $\WearDensityFunction$.

Finally, we denote the set of breakpoints of some piecewise linear function $g$ by $\SetOfBreakpoints[g]$ (cf. Figure~\ref{fig:battery-behavior-models}) and use $\PWLSlope[g][x](x)$ to refer to its right derivative.

With this setting and notation, we state the objective of our optimization problem: Given a fleet of vehicles $\Vehicle \in \SetOfVehicles$, each assigned operations $\Tour \in \SetOfTours[\Vehicle]$ with energy consumption $\TourConsumption$, duration $\TourDuration$, and departure time window $[\EarliestTourDeparture, \LatestTourDeparture]$, we aim to find a cost-minimal, feasible \emph{fleet schedule}. We consider a fleet schedule feasible if it respects charger capacity constraints in each period and each \emph{vehicle schedule} satisfies the following constraints:
\begin{enumerate}
	\item the vehicle's \gls{abk:soc} remains within its operational limits $[\MinSoC, \MaxSoC]$ at all times,
	\item the vehicle provides service to all assigned operations,
	\item the vehicle meets each operation's departure time window.
\end{enumerate}
Formally, we represent a fleet schedule ${\FleetSchedule \coloneqq \{\VehicleSchedule_{1}, \dots, \VehicleSchedule_{|\SetOfVehicles|}\}}$ as a set of vehicle schedules $\VehicleSchedule_i$. A vehicle schedule $\VehicleSchedule_i$ captures the vehicle's actions at each point in time, utilizing
\begin{description}
	\item[$\ScheduleMatrix \in \mathbb{B}^{|\SetOfPeriods| \times (\SetOfChargers \cup \SetOfTours[\Vehicle])}$:] a binary matrix indicating scheduled operations. Here, $\ScheduleMatrix_{i, \eta}$, $\eta \in \SetOfChargers \cup \SetOfTours[\Vehicle]$ indicates that the vehicle uses period $\SetOfPeriods_{i}$ to charge at charger $\Charger \in \SetOfChargers$ if $\ScheduleMatrix_{i, \Charger} = 1$, services operation $\Tour \in \SetOfTours[\Vehicle]$ if $\ScheduleMatrix_{i, \Tour} = 1$, or remains idle if ${\ScheduleMatrix_{i, \eta} = 0}$ ${\forall \eta \in \SetOfChargers \cup \SetOfTours}$. We note that feasible schedules satisfy ${\sum_{\forall \eta \in \SetOfChargers \cup \SetOfTours} \ScheduleMatrix_{i, \eta} \leq 1\; \forall i \in [1, |\SetOfPeriods|]}$.
	\item[$\ChargeMatrix \in \mathbb{R}^{|\SetOfPeriods|}$:] a vector that holds the amount of charge replenished in each period $\Period \in \SetOfPeriods$. Here, negative values $\ChargeMatrix_{i} < 0$ indicate consumption. We allocate the consumption of operations that span multiple periods to the departure period.
\end{description}
With this notation, we state the cost of a vehicle schedule $\VehicleSchedule$ as the sum of energy and battery degradation related costs incurred in each period:
\begin{equation}
	\label{eq:vehicle-schedule-cost}
	c(\VehicleSchedule) \coloneqq \sum_{i=1}^{|\SetOfPeriods|} \max(0, \PeriodEnergyPrice[\SetOfPeriods_i] \cdot \ChargeMatrix_{i} + \WearDensityFunction(\sum_{j=1}^{i-1} \ChargeMatrix_{j}, \sum_{j=1}^{i-1} \ChargeMatrix_{j} + \ChargeMatrix_{i})).
\end{equation}
Here, energy costs result from the amount of charge replenished in each period multiplied by the respective period's energy price. Degradation costs result from the \gls{abk:wdf}. Note that periods $\SetOfPeriods_{i}$ that consume energy, i.e., where $\ChargeMatrix_{i} < 0$, do not incur cost as $\WearDensityFunction$ is increasing.
The cost of a fleet schedule, $c(\FleetSchedule)$, then corresponds to the sum of it's vehicle schedule costs:
\begin{equation}
	\label{eq:fleet-schedule-cost}
	c(\FleetSchedule) \coloneqq \sum_{\VehicleSchedule \in \FleetSchedule} c(\VehicleSchedule).
\end{equation}
We model this optimization problem as a set-covering problem over the set of vehicle schedules. For this purpose, we refer to the set of feasible schedules for vehicle $\Vehicle \in \SetOfVehicles$ as $\SetOfVehicleSchedules$ and let $\SetOfSchedules = \cup_{\Vehicle \in \SetOfVehicles} \SetOfVehicleSchedules$ denote the set of all feasible vehicle schedules.
Using binary variables $\ScheduleUsed$, which indicate the inclusion of a schedule in the final solution ($\ScheduleUsed = 1$), we propose the following \gls{abk:ip}.
\begin{subequations}
	\label{mips:MP-mip}
	\setlength{\abovedisplayskip}{0pt}
	\setlength{\belowdisplayskip}{0pt}
	\setlength{\abovedisplayshortskip}{0pt}
	\setlength{\belowdisplayshortskip}{0pt}
	\begin{multline}
		\hfill 
		\min \sum_{\AllVehicles} \sum_{\Schedule \in \SetOfVehicleSchedules} \ScheduleUsed \ScheduleCost
		\hfill
		\label{MP-mip:objective}
	\end{multline}
	\begin{multline}
		\quad \quad \sum_{\AllVehicles} \sum_{\Schedule \in \SetOfVehicleSchedules} \ScheduleUsed \cdot \ScheduleMatrix_{\Period, \Charger} \leq \ChargerCapacity
		\hfill \AllChargers, \AllPeriods
		\label{MP-mip:stationCapacityRespected}
	\end{multline}
	\begin{multline}
		\quad \quad \sum_{\Schedule \in \SetOfVehicleSchedules} \ScheduleUsed \geq 1
		\hfill \AllVehicles
		\label{MP-mip:oneSchedulePerVehicle}
	\end{multline}
	\begin{multline}
		\quad \quad \ScheduleUsed \in \{0, 1\}
		\hfill \Schedule \in \SetOfVehicleSchedules
		\label{MP-mip:bounds}
	\end{multline}
\end{subequations}
\noindent Objective~\eqref{MP-mip:objective} minimizes overall scheduling costs. 
\emph{Linking Constraints}~\eqref{MP-mip:stationCapacityRespected} enforce charger capacity limitations, while \emph{Convexity Constraints}~\eqref{MP-mip:oneSchedulePerVehicle} ensure that each vehicle is assigned a schedule. Finally, Constraints~\eqref{MP-mip:bounds} state our decision variables' domain.
\section{Column generation}
\glsreset{abk:spprc}
\glsreset{abk:frvcp}
\glsreset{abk:myfrvcp}
\label{sec:column-generation}

\newcommand{\TimeExpandedNetwork}{G^{\Vehicle}}
\newcommand{\SetOfArcs}{\SetOfEdges}
\newcommand{\ReducedCost}{rc}

The proposed set covering formulation comprises a variable for each feasible schedule, such that solving \ipref{mips:MP-mip} using standard integer programming techniques remains intractable even for small instances. As a remedy, we approach \ipref{mips:MP-mip} with \gls{abk:cg}. The fundamental idea of \gls{abk:cg} is to consider only a small subset of variables (\emph{columns}) in a \gls{abk:rmp}, which we obtain by relaxing \ipref{mips:MP-mip} and substituting each $\SetOfVehicleSchedules$ with some subset $\ReducedSetOfVehicleSchedules \subseteq \SetOfVehicleSchedules$.
The \gls{abk:cg} procedure then iteratively extends $\ReducedSetOfVehicleSchedules$ with schedules that improve the \gls{abk:rmp}'s objective value until no such schedules can be identified. For this purpose, \gls{abk:cg} solves the current \gls{abk:rmp} to obtain dual prices $\DualChargerCapacity$ and $\DualScheduleUsed$ of Constraints~\eqref{MP-mip:stationCapacityRespected} and \eqref{MP-mip:oneSchedulePerVehicle}, respectively. With these, we state the reduced cost of a schedule $\Schedule \in \SetOfVehicleSchedules$ as:
\begin{equation}
	\label{eq:pricing-problem}
	rc(\Schedule) \coloneqq \ScheduleCost[\Schedule] - \DualScheduleUsed - \sum_{\AllPeriods} \sum_{\AllChargers} \ScheduleMatrix_{\Period, \Charger} \cdot \DualChargerCapacity.
\end{equation}
Then, improving columns correspond to schedules $\VehicleSchedule \in \SetOfSchedules$ with negative reduced costs, which we generate in a so-called \emph{pricing problem} by solving $|\SetOfVehicles|$ \glspl{abk:spprc} on vehicle specific time-expanded networks~$\TimeExpandedNetwork$. 

In the following, we first detail the construction of these time-expanded networks in Section~\ref{sec:pricing-network} and give an overview of our labeling algorithm in Section~\ref{sec:labeling-algorithm}.
We then detail each central algorithmic component, namely label representation (\ref{subsec:label-representation}), label dominance (\ref{subsec:dominance}), label propagation (\ref{sec:label-propagation}), and non-dominated charging decisions (\ref{sec:committing-charging-operations}) in separate sections (\ref{subsec:label-representation}-\ref{sec:committing-charging-operations}). 
Section \ref{subsec:propagation-example} provides a detailed example that applies our algorithm to a simplified pricing network.
Finally, we outline speedup techniques used to establish the computational efficiency of our labeling algorithm in Section~\ref{subsec:speedup-techniques}.

\subsection{Pricing networks}
\label{sec:pricing-network}
\label{sec:network-design}

For each vehicle $\Vehicle \in \SetOfVehicles$, we model the pricing problem as a \gls{abk:spprc} defined on a time-expanded network $\TimeExpandedNetwork = (\SetOfVertices, \SetOfArcs)$. Here, vertices $\Vertex \in \SetOfVertices$ represent the vehicle's location in time and space. More precisely, $\TimeExpandedNetwork$ comprises \textit{station vertices} and a \textit{garage vertex} for each period $\Period \in \SetOfPeriods$. Station vertices correspond to the chargers available in $\Period \in \SetOfPeriods$, while garage vertices represent locations where vehicles can idle without occupying a charger.
Additionally, we add (dummy) \textit{source} and \textit{sink} vertices $\SourceNode$ and $\SinkNode$, which serve as network entry and exit points to \textit{virtual} periods $\SetOfPeriods_0$ and $\SetOfPeriods_{n+1}$, respectively. These dummy periods correspond intuitively to the first period before and after the planning horizon. We use functions $\PeriodOfVertex[\Vertex]$ and $\ChargerOfVertex[\Vertex]$ to denote the period and charger associated with some vertex $\Vertex \in \SetOfVertices$.

Arcs $(\Arc) \in \SetOfArcs$ correspond to actions that a vehicle performs in period $\PeriodOfVertex$ and allow a vehicle to move in time and space. Traversing an arc incurs a certain (fixed) cost $\ArcCost$ and consumes $\ArcConsumption$ units of energy. Our network comprises three types of arcs:
\begin{paragraphs}
\item[\textit{Charging arcs}] $\ChargingArc \in \SetOfChargingEdges$ represent charging at $\ChargerOfVertex[\OriginVertex]$ and incur fixed costs according to the station's dual multiplier ($\ArcCost \coloneqq \DualChargerCapacity[{\PeriodOfVertex[\OriginVertex], \ChargerOfVertex[\OriginVertex]}]$) without consuming energy ($\ArcConsumption \coloneqq 0$). Charging with arrival \gls{abk:soc}~$q$ replenishes $\Delta q$ \gls{abk:soc} at a price of $c(q, \Delta q) \coloneqq \PeriodEnergyPrice[{\PeriodOfVertex[\OriginVertex]}] \cdot \Delta q + \WearDensityFunction(q, q + \Delta q)$, such that traversing a charging arc incurs a total cost of $\ArcCost + c(q, \Delta q)$. Here, $\Delta q$ is variable and bounded implicitly by the period length $\PeriodDuration$, such that $0 < \Delta q \leq \ChargingFunction[ {\ChargerOfVertex[{\OriginVertex}]} ](q, \PeriodDuration)$.
Charging arcs connect station vertices to all vertices of the following period.
\item[\textit{Idle arcs}] $\IdleArc \in \SetOfIdleArcs$ model a vehicle idling at the depot. Idling is possible at zero cost ($\ArcCost \coloneqq 0$) and does not consume energy ($\ArcConsumption \coloneqq 0$). Analogous to charging arcs, idling arcs connect \textit{garage vertices} to all vertices of the following period.
\item[\textit{Service arcs}] $\ServiceArc \in \SetOfServiceArcs$ model a vehicle's departure in period $\PeriodOfVertex[\OriginVertex]$ to service an operation $\Tour \in \SetOfTours$. This consumes energy according to the operation's consumption ($\ArcConsumption \coloneqq \TourConsumption$) but incurs no additional costs $(\ArcCost \coloneqq 0)$. For each $\Tour \in \SetOfTours$, service arcs connect all garage vertices that lie within the departure time window of $\Tour$ to garage vertices in the respective arrival period. We denote the operation serviced by arc $(\Edge)$ with $\ArcOperation \coloneqq \{\Tour\}$, and let $\ArcOperation \coloneqq \emptyset$ for non-service arcs.
\end{paragraphs}

\begin{figure}[b!]
	\resizebox{\textwidth}{!}{
		\begin{tikzpicture} 

\tikzset{
	every node/.style={
		inner sep=0pt, minimum size=0.7cm
	},
	every edge/.append style={
		thick,->
	},
	legend/.style = {
		minimum size=3mm
	},
legend_arc/.style = {
		thin, ->
	},
	dummy/.style={
		diamond, draw
	},
	station/.style={
		regular polygon, regular polygon sides = 3, draw
	},
	idle/.style={
		rectangle, draw, minimum size=0.5cm
	},
	charging_arc/.style={
		black, solid
	},
	idle_arc/.style={
		green!50!black, dashed
	},
	service_arc/.style={
		blue, dotted, bend left = 25
	}
}

	\node[] (period_label0) at (0, 2.65) {\small $\mathcal{P}_{0}$};
\foreach \x in {1,...,8} {
	\tikzmath{\xscaled = \x * 1.5 - 0.25;}
	\node[] (period_label\x) at (\xscaled, 2.65) {\small $\mathcal{P}_{\x}$};
}
\node[] at (13.25, 2.65) {\small $\mathcal{P}_{n+1}$};

\foreach \x in {0.5, 2, ..., 13.5} {
	\draw[solid, gray!50, very thin] (\x, 3) -- (\x, -0.5);
}

\node[draw, dummy] (source) at (0, 1) {$s^-$};
\node[draw, dummy] (sink) at (13.25, 1) {$s^+$};

\foreach \x in {1, ..., 8} {
	\tikzmath{\xscaled = \x * 1.5 - 0.25;}
	\node[draw, station] (station\x) at (\xscaled, 0) {$f_1$};
	\node[draw, idle] (idle\x) at (\xscaled,2) {};
}

\foreach \x in {1, ..., 7} {
	\pgfmathtruncatemacro{\xplusone}{\x + 1}
	
	\path[idle_arc] (idle\x) edge (idle\xplusone);
	\path[idle_arc] (idle\x) edge node[fill=white, inner sep=0, minimum size=4mm] {} (station\xplusone);	

	\path[charging_arc] (station\x) edge node[above=-4mm,rotate=55] {\tiny $\pi_{\mathcal{P}_{\x}, f_{1}}$} (idle\xplusone);
	\path[charging_arc] (station\x) edge node[above=-2.25mm] {\tiny $\pi_{\mathcal{P}_{\x}, f_{1}}$} (station\xplusone);
}
\foreach \x in {2, ..., 5} {
	\pgfmathtruncatemacro{\target}{\x + 2}
	\path[service_arc] (idle\x) edge (idle\target);
}
\path[->] (source) edge node[above=-.75mm, right=-3.5mm, rotate=40] {\tiny $\pi$} (idle1);
\path[->] (source) edge node [below=.75mm, left=-3.5mm, rotate=-40] {\tiny $\pi$} (station1);

\path[->] (station8) edge node[above=3.5mm, left=-2mm, rotate=40] {\tiny $\pi_{\mathcal{P}_{8}, f_{1}}$} (sink);
\path[->, idle_arc] (idle8) edge (sink);

\path (14.5, 2.25) node[dummy, legend] {} -- node[legend,anchor=west] {\tiny Source/Sink} +(6mm, 0) -- ++(0, -0.5) node[idle, legend] {} -- node[legend,anchor=west] {\tiny Garage} +(6mm, 0) -- ++(0, -0.5) node[station, legend] {} -- node[legend,anchor=west] {\tiny Station} +(6mm, 0) --  ++(-2mm, -0.5) edge[legend_arc, charging_arc] +(4mm, 0) -- node[legend,anchor=west] {\tiny Charging} +(10mm, 0) -- ++(0, -0.5) edge[legend_arc, idle_arc] +(4mm,0) -- node[legend,anchor=west] {\tiny Idling} +(10mm, 0) -- ++(0, -0.5) edge[legend_arc, service_arc] +(4mm,0) -- node[legend,anchor=west] {\tiny Service} +(10mm, 0);

\draw[black, thin] (14.125, 2.5) -- ++(2.25, 0) -- ++(0, -3) -- ++(-2.25, 0) -- ++(0, 3);

\end{tikzpicture}
	}
	\caption{\centering An example time-expanded network.\label{fig:example-network}}
	\begin{minipage}{\textwidth}
		\smaller \textit{Note.} This example time-expanded network models the scheduling decisions of some vehicle $\Vehicle \in \SetOfVehicles$ that is assigned a single service operation with a departure time window of $[\SetOfPeriods_2, \SetOfPeriods_5]$ and duration $\TourDuration[] = 2$. Solid triangles and black squares represent station and garage vertices, respectively. Diamonds illustrate source and sink vertices. Solid black, dashed green, and dotted blue arcs represent charging, idle, and service arcs. We denote $\DualChargerCapacity$ and $\DualScheduleUsed$ as $\pi_{\Period, \Charger}$~and~$\pi$, respectively.
	\end{minipage}
\end{figure}
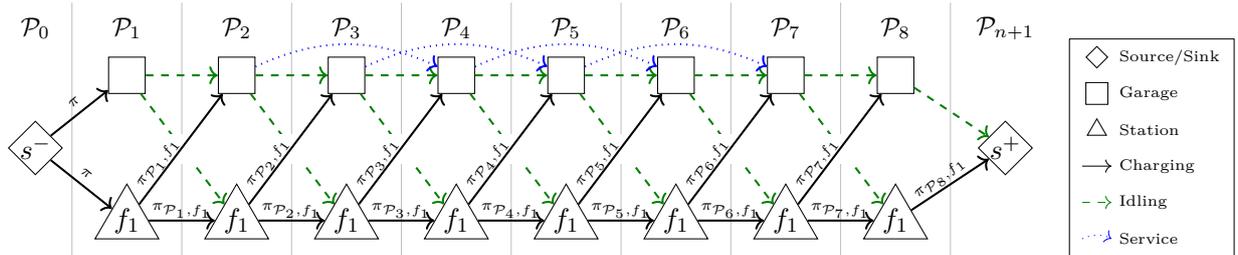

Finally, we add dummy arcs $(\Arc)$ connecting the source vertex with all vertices of the first period at a cost according to the coverage dual ($\ArcCost \coloneqq \DualScheduleUsed$). Figure~\ref{fig:example-network} shows an example of a time-expanded network with a single charger ($\Charger_1$) and a single service operation.
\subsection{Labeling algorithm}
\label{sec:labeling-algorithm}
\newcommand{\PropagateLabel}[1][u]{\rightarrow #1}

By construction, each source-sink path $\PartialPath$ in the pricing network $\TimeExpandedNetwork$ corresponds to a vehicle schedule~$\VehicleSchedule$. Moreover, the cost of $\PartialPath$ in $\TimeExpandedNetwork$ matches the reduced cost of schedule $\VehicleSchedule$. Consequently, by defining resource constraints that establish the feasibility of the corresponding schedule $\VehicleSchedule$ (cf. Section~\ref{subsec:problem-definition}), we can express the pricing subproblem of vehicle $\Vehicle \in \SetOfVehicles$ as a \gls{abk:spprc} on $\TimeExpandedNetwork$. Specifically, we treat \gls{abk:soc} and serviced operations as constrained resources, i.e., restrict the \gls{abk:soc} at non-source vertices to $[\MinSoC, \MaxSoC]$ (\emph{energy~feasibility}), and require that all operations have been serviced when reaching the sink (\emph{service~feasibility}).

We obtain a solution to these \glspl{abk:spprc} using a problem-specific label-setting algorithm. 
The design of this algorithm remains challenging as continuous (non-linear) charging raises a trade-off between cost and \gls{abk:soc} at charging stations. Specifically, charging any ${0 < \Delta q \leq \ChargingFunction(q, \PeriodDuration)}$ with arrival \gls{abk:soc}~$q$ results in a potentially optimal pair of cost and \gls{abk:soc}, such that visits to charging stations may generate an unbounded number of labels. While this is so far unconsidered in previous work on \gls{abk:evsp} and \gls{abk:csp}, e.g., \citet{KootenNiekerkAkkerEtAl2017}, which discretize charging operations,
we note that recent work on related problem settings, specifically time-minimizing shortest paths for \glspl{abk:ecv} \citep{BaumDibbeltEtAl2019, FrogerMendozaEtAl2019}, faced a similar challenge: here, charging operations raise a trade-off between arrival time and \gls{abk:soc}. To resolve this issue, these works proposed a function-based label representation, which allows capturing all time-\gls{abk:soc} trade-offs at the last visited station in a single label. This effectively delays the charging decision at the respective station until a finite subset of potentially optimal charging decisions can be identified.
Although related, the methodology developed in these works is insufficient for our problem setting. The reason for this relates to variable energy prices in combination with implicit bounds on the \gls{abk:soc} rechargeable at station vertices imposed by the time discretization. 
This violates a central assumption of the methodology developed in \citet{BaumDibbeltEtAl2019} and \citet{FrogerMendozaEtAl2019}, such that their algorithms fail to find an optimal solution if energy prices vary across periods. We give proof to this issue in Appendix~\ref{app:counterexample-baum} and note that these additional challenges are not unique to charge scheduling but may occur in other problem settings. For example, in time-dependent routing, travel times vary over time analogously to how energy prices vary in our \gls{abk:csp} and can hence raise a similar issue.
Accordingly, we contribute to the state of the art by presenting a new label-setting algorithm that accounts for such bounded charging operations. Moreover, we present speed-up and preprocessing techniques specifically tailored to the challenges of this new label-setting algorithm.

The following outlines our algorithm and introduces core algorithmic components before we detail each component separately in Sections~\ref{subsec:label-representation}-\ref{subsec:speedup-techniques}. We refer to Appendix~\ref{app:algorithm-details} for an in-depth technical description of our algorithm.

Our algorithm iteratively explores pricing networks $\TimeExpandedNetwork$, starting with an empty path at the source vertex $\SourceNode$.
We extract the, w.r.t. cost, currently cheapest path in each iteration and extend it to all neighboring vertices, thus creating new paths.
The algorithm terminates when extracting a path at the sink or when no \emph{unsettled} paths remain.
We capture path resources (operations serviced, \gls{abk:soc}, and cost) in labels $\Label \in \SetOfLabels_{\Vertex}$ to ensure that paths $\rho \coloneqq (\SourceNode, \dots, \Vertex)$ are feasible. As in \citet{BaumDibbeltEtAl2019}, and \citet{DabiaRopkeEtAl2013}, we use a function-based label representation for this purpose. Specifically, our labels store a function that maps the \gls{abk:soc} on arrival at the last node of the corresponding path to the total path cost. In contrast to \citet{BaumDibbeltEtAl2019}, our labels are not limited to capturing only the last visited charging station but instead capture charging decisions at several previously visited charging stations instead (see Section~\ref{subsec:label-representation}).
	
When extending paths along arc $(\Arc)$, we thus need to create new labels at $\TargetVertex$ to capture the extended path's state. We derive these in so-called \emph{propagation functions}, which generate a set of new labels at vertex $\TargetVertex$ according to arc resources and charging decisions based on the label of the current path $\rho \coloneqq (\SourceNode, \dots, \OriginVertex)$ and the newly extended arc $(\OriginVertex, \TargetVertex)$ (Section~\ref{sec:label-propagation}). Here, our algorithm considers already generated labels at vertex $\TargetVertex$ to avoid exploring \emph{dominated} paths. Specifically, it discards labels that do not improve on the set of labels already developed at $\TargetVertex$ using a set-based dominance criterion (Section~\ref{subsec:dominance}).
\subsection{Label representation}
\label{subsec:label-representation}
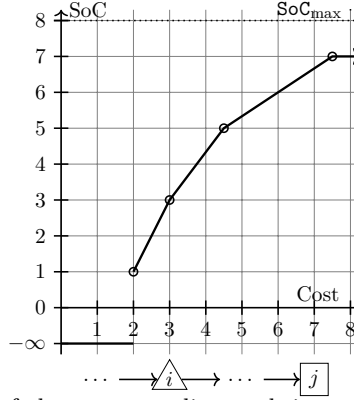
\begin{figure}
	\centering
	\resizebox{0.3\textwidth}{!}{%
		\begin{tikzpicture}[scale=0.65]

\tikzset{
	bp/.style={
		circle, draw, thick, solid, inner sep=1.5pt
	},
	cost_profile/.style={
		black, very thick
	},
	label_profile/.style={
		cost_profile, dashed
	},
	dual/.style={
		black, dashed, bend left,->
	},
	dummy/.style={
		diamond, draw, inner sep=0pt, minimum size=0.7cm
	},
	station/.style={
		regular polygon, regular polygon sides = 3, draw, inner sep=0pt, minimum size=0.7cm
	},
	charging_arc/.style={
		black, solid, thick, ->
	},
	idle/.style={
		rectangle, draw, minimum size=0.5cm
	},
	arc_cost/.style={
		above
	},
	arc_consumption/.style={
		below
	},
}

\draw[step=1cm, gray, very thin] (0, -1.3) grid (8.3, 8.3);
\draw[->, thick] (0,0) -- (8.3, 0) node[above=0.25cm, left=0.25cm] {Cost};
\draw[->, thick] (0,-1.3) -- (0, 8.3) node[right] {SoC};

\draw[thick] (2mm, -1) -- (-2mm, -1) node[left] {$-\infty$};
\foreach \y in {0, ..., 8} {
	\draw[thick] (2mm, \y) -- (-2mm, \y) node[left] {$\y$};
}
\draw[thick] (2mm, 8) -- (-2mm, 8);
\draw[dotted, thick] (2mm, 8) -- (8.3, 8);
\draw[thick] (8, 8.32) -- (8, 8.28) node[left]  {$\MaxSoC$};

\foreach \x in {1, ..., 8} {
	\draw[thick] (\x, 2mm) -- (\x, -2mm) node[below] {$\x$};
}

\draw[cost_profile] (0, -1) -- (2, -1);
\draw[->,cost_profile] (2, 1) node[bp]{} -- ++(1,2) node[bp]{} -- ++(1.5, 2) node[bp]{} -- ++(3, 2) node[bp]{} -- (8.3, 7);

\node (source) at (1, -2) {$\dots$};
\node[draw, station] (first_station) at (3, -2) {$i$};
\node (rest) at (5, -2) {$\dots$};
\node[draw, idle] (target_vertex) at (7, -2) {$j$};

\path[charging_arc] (source) edge node[arc_cost]{} node[arc_consumption] {} (first_station);
\path[charging_arc] (first_station) edge node[arc_cost]{} node[arc_consumption] {} (rest);
\path[charging_arc] (rest) edge node[arc_cost]{} node[arc_consumption] {} (target_vertex);

\end{tikzpicture}	
	}
	\caption{Example of a cost profile tracking the cost-\gls{abk:soc} trade-off of path $\PartialPath = (\SourceNode,  \dots, i, \dots, j)$. \label{fig:cost-profile}\vspace{-0.25cm}}
	\begin{minipage}{\textwidth}
		\smaller \textit{Note.} In this example, the lowest cost of the corresponding path is $c = 2$, at which we reach $v$ with a \gls{abk:soc} of $1$. Spending less than $c = 2$ is infeasible, which we indicate with a \gls{abk:soc} of $-\infty$. We reach a \gls{abk:soc} of $5$ at a cost of $4.5$, i.e., when spending ${\Delta c \coloneqq 4.5 - 2 = 2.5}$ to charge at $i$. The maximum reachable \gls{abk:soc} is $7$, which corresponds to spending ${\Delta c \coloneqq 7.5 - 2 = 5.5}$ at $j$. Any $c > 7.5$ does not increase the \gls{abk:soc} further. This upper bound on the \gls{abk:soc} can be caused by either the battery capacity or the period length, which limits the time spent charging at the tracked station. The example illustrates the latter case.%
	\end{minipage}
\end{figure}

Our labels $\Label \in \SetOfLabels_{\TargetVertex}$ store path cost and \gls{abk:soc} in \emph{cost profiles} $\CostProfile(c)$, which map the total cost $c$ of the labeled path to the resulting (arrival) \gls{abk:soc} at vertex $\TargetVertex$ (cf. Figure~\ref{fig:cost-profile}).
\noindent Formally, we represent labels as tuples with components:
\begin{description}[itemsep=-2ex,topsep=1pt,left=\parindent]
\item[$\CostProfile(c): \mathbb{R} \mapsto \lbrack\MinSoC, \MaxSoC \rbrack \cup \{-\infty\}$] denoting the vehicle's \gls{abk:soc} at a total cost of $c$, and
\item[$\LabelVisitedTours \in \mathbb{P}(\SetOfTours)$] denoting the operations serviced so far. 
\end{description}
We call a label energy feasible if there exists some $c \in \mathbb{R}$ such that $\CostProfile(c) \neq -\infty$ and denote the first and last breakpoints of $\CostProfile$ with $(\CostProfileMinCost,\CostProfileMinSoC)$ and $(\CostProfileMaxCost, \CostProfileMaxSoC)$, respectively. 
We further note that $\CostProfile(c)$ is not well defined on $\mathbb{R}$ and hence define $\InvCostProfile(q)$, preserving piecewise-linearity, as follows:
\begin{equation}
	\InvCostProfile(q) \coloneqq \begin{cases}
		-\infty &\text{if } q \leq \CostProfileMinSoC\\
		\CostProfileMaxCost &\text{if } q \geq \CostProfileMaxSoC\\
		\displaystyle\argmin_{c \in \mathbb{R}}\;\{c \mid \CostProfile(c) = q\} &\text{otherwise.}\\
	\end{cases}
\end{equation}
Note that this extended label representation remains efficient as $\CostProfile$ is piecewise-linear and can thus be represented as a finite sequence of \emph{breakpoints}.
Lastly, we define the label corresponding to the empty path used to initialize the algorithm at the source as $\SourceLabel \coloneqq (\CostProfile[\SourceLabel](c), \emptyset)$, with $\CostProfile[{\SourceLabel}](c)$ as follows:
\begin{equation}
	\CostProfile[\SourceLabel](c) \coloneqq \begin{cases}
		0 &\text{if } c \geq 0\\
		-\infty &\text{otherwise.}
	\end{cases}
\end{equation}
Intuitively, the source represents a (dummy) station with a charging rate $0$.
\subsection{Dominance}
\label{subsec:dominance}
\newcommand{\LabelDominatesLabel}{\succeq}
\newcommand{\SetDominatesLabel}{\succeq}

We use two dominance rules to discard sub-optimal labels early and thus reduce the number of explored paths. The first rule extends standard pairwise Pareto-dominance of scalar-valued labels to our function-based representation. The second rule is unique to function-based label representation and defines a \emph{set-based} dominance relationship to capture cases where the union of a set of labels dominates a single label. We note that a similar logic has been applied successfully to time-dependent \glspl{abk:vrp} and \glspl{abk:vrp} with weight-related costs \citep[cf.][]{DabiaRopkeEtAl2013, LuoQinEtAl2017}.
\begin{definition}{\textit{Pairwise dominance.}\hfill}
	\label{def:dominance}
	\break
	Let there be labels $\Label_a, \Label_b \in \SetOfLabels_{\Vertex}$ for some $\Vertex \in \SetOfVertices$.
	Then label $\Label_a \in \SetOfLabels$ dominates label $\Label_b $, denoted ${\Label_a \LabelDominatesLabel \Label_b}$, if and only if
	\begin{equation*}
		\CostProfile[{\Label_{a}}](c) \geq \CostProfile[{\Label_{b}}](c)\;\forall c \in \mathbb{R}\text{ and }\LabelVisitedTours[\Label_a] \supseteq \LabelVisitedTours[\Label_b].
	\end{equation*}
\end{definition}
\noindent Definition~\ref{def:dominance} states that a label $\Label_a$ dominates another label $\Label_b$ if $\Label_a$ achieves i) a higher or equally high \gls{abk:soc} at any given cost $c$ and ii) schedules a superset of operations. Accordingly, we do not define a dominance relationship between labels of paths ending at different vertices.
Note that as $\CostProfile$ are piecewise linear, it suffices to check $\CostProfile[{\Label_{a}}](c) \geq \CostProfile[{\Label_{b}}](c)$ at breakpoints $c \in \SetOfBreakpoints[{\CostProfile[{\Label_{a}}]}] \cup \SetOfBreakpoints[{\CostProfile[{\Label_{b}}]}]$ such that the dominance check's complexity remains linear. Unfortunately, pairwise dominance is relatively weak and often fails to detect superfluous labels. Specifically, this occurs if the images of $\CostProfile[{\Label_a}]$ and $\CostProfile[{\Label_b}]$ do not align, i.e., $\lbrack\CostProfileMinSoC[\Label_a], \CostProfileMaxSoC[\Label_a]\rbrack \neq \lbrack\CostProfileMinSoC[\Label_b], \CostProfileMaxSoC[\Label_b]\rbrack$. To mitigate this issue, we extend pairwise dominance to sets of labels as follows.
\begin{definition}{\textit{Set-based dominance.}\hfill}
	\label{def:set-dominance}
	\break
	A set of labels $\SetOfLabels' \subseteq \SetOfLabels_{\Vertex}$ at some vertex $\Vertex$ dominates label ${\Label \in \SetOfLabels_{\Vertex} \setminus \SetOfLabels'}$, denoted ${\SetOfLabels' \succeq \Label}$, if and only if
	\begin{equation*}
		\max_{\Label' \in \SetOfLabels'}\{\CostProfile[{\Label'}](c)\} \geq \CostProfile[{\Label}](c)\;\forall c \in \mathbb{R}\text{ and } \displaystyle\bigcap_{\Label' \in \SetOfLabels'}\LabelVisitedTours[\Label'] \supseteq \LabelVisitedTours[\Label].
	\end{equation*}
\end{definition}
\noindent Essentially, set-based dominance occurs if, for any given cost, at least one label in $\SetOfLabels'$ achieves a higher \gls{abk:soc} than label $\Label$, and each $\Label' \in \SetOfLabels'$ schedules a superset of operations.
\subsection{Label propagation}
\label{sec:label-propagation}
We distinguish propagating a label $\Label \in \SetOfLabels_{\OriginVertex}$ along a service or idling arc from propagating along a charging arc as the latter constitutes a station visit and thus allows to replenish additional energy. We detail our propagation methodology for both arc types separately in Sections~\ref{subsubsec:propagating-service-and-idling-arcs}~and~\ref{subsubsec:propagating-charging-arcs}.
\subsubsection{Service and idling arcs}\label{subsubsec:propagating-service-and-idling-arcs}
Propagating along service and idling arcs, denoted as ${\NewLabel \coloneqq \PropagateRegularArc}$, generates a new label at the target vertex $\TargetVertex$. This requires updating $\CostProfile$ according to arc consumption $\ArcConsumption$ and cost $\ArcCost$, which we obtain from shifting $\PrevCostProfile$ on the cost and \gls{abk:soc} axes, respectively: ${\NewCostProfile(c) \coloneqq \PrevCostProfile(c - \ArcCost) - \ArcConsumption}$.
Additional modifications are necessary when some charging decisions become infeasible after propagating $\Label$ along $(\Arc)$. Specifically, we need to maintain $\NewCostProfile(c) \in [\MinSoC, \MaxSoC] \cup \{-\infty\}$ for all $c \in \mathbb{R}$ such that we derive $\NewCostProfile$ as follows:
\begin{equation}
	\NewCostProfile(c) \coloneqq 
	\begin{cases}
		-\infty, & \text{if } \PrevCostProfile(c - \ArcCost) - \ArcConsumption < \MinSoC\\
		\MaxSoC & \text{if } \PrevCostProfile(c - \ArcCost) - \ArcConsumption \geq \MaxSoC\\
		\PrevCostProfile(c - \ArcCost) - \ArcConsumption & \text{otherwise.}
	\end{cases}
\end{equation}
Recall that $\ArcOperation = \{\Tour\}$ if $(\Arc)$ serves operation $\Tour$ and $\ArcOperation = \emptyset$ otherwise, such that $\NewLabel \coloneqq (\NewCostProfile, \LabelVisitedTours \cup \ArcOperation)$ yields the propagated label.
\subsubsection{Charging arcs}\label{subsubsec:propagating-charging-arcs}
Charging arc $(\Arc)$ represents a charging opportunity at station $f(\OriginVertex)$ where it is possible to charge for some duration $0 < \tau \leq \PeriodDuration$ at a cost resulting from the energy price $\PeriodEnergyPrice[{\PeriodOfVertex[\OriginVertex]}]$, battery degradation $\WearDensityFunction$, and fixed cost $\ArcCost$. For the sake of conciseness, we will interchangeably refer to a charging decision at $\OriginVertex$ using the charging time $\tau$ and the total cost according to the corresponding cost profile.

Propagating label $\Label \in \SetOfLabels_{\OriginVertex}$ along $(\Arc)$ realizes this charging opportunity such that the newly created label $\Label'$ should capture the arrival \gls{abk:soc} at vertex $\TargetVertex$. Recall that $\Label$ tracks charging decisions at the last visited station $\Vertex$ such that we are unaware of the actual arrival \gls{abk:soc} at $\OriginVertex$. The resulting label would thus be ill-defined: the arrival \gls{abk:soc} $q$ at vertex $\TargetVertex$ with cost $c$ depends on how we distribute our charging budget between stations $\Vertex$ and $\OriginVertex$.

Resolving this ambiguity requires a decision at either station. Specifically, we either need to fix a decision at $\Vertex$ and thus the arrival \gls{abk:soc} at $\OriginVertex$, or commit to a decision at $\OriginVertex$ and continue to track decisions at $\Vertex$. Intuitively, the first decision \textit{replaces} the currently tracked station $\Vertex$ with the new charging opportunity at $\OriginVertex$ such that we will refer to this operation as \textit{station replacement}, while the latter decision entails charging \textit{intermediately} at $\OriginVertex$ and will be referred to as such. Our propagation function covers both cases, such that charging arcs generate several labels. In what follows, we describe our methodology for both cases separately.

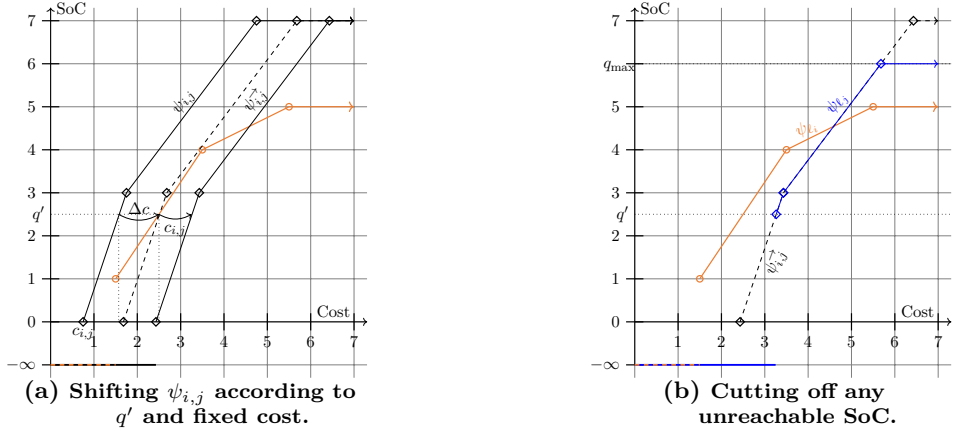
\begin{figure}
	\captionsetup[subfigure]{width=0.9\textwidth}
	\begin{subfigure}[T]{0.3\textwidth}
		\resizebox{\textwidth}{!}{
			\begin{tikzpicture} 

\tikzset{
	bp/.style={
		draw, thick, solid, inner sep=1.5pt
	},
	fbp/.style={
			circle, bp
		},
	phibp/.style={
		diamond, bp
	},
	f/.style={
		orange, thick
	},
	u/.style={
		blue
	},
	phi/.style={
		black
	}
}

\clip(-1, -1.3) rectangle (7.3, 7.5);

\draw[step=1cm, gray, very thin] (0, -1.3) grid (7.3, 7.3);
\draw[->] (0,0) -- (7.3, 0) node[above=0.25cm, left=0.25cm] {Cost};
\draw[->] (0,-1) -- (0, 7.3) node[right] {SoC};

\draw[thick] (2mm, -1) -- (-2mm, -1) node[left] {$-\infty$};
\foreach \y in {0, ..., 7} {
	\draw[thick] (2mm, \y) -- (-2mm, \y) node[left] {$\y$};
}

\foreach \x in {1, ..., 7} {
	\draw[thick] (\x, 2mm) -- (\x, -2mm) node[below] {$\x$};
}

\node[below] at (0.75, 0) {$\ArcCost$};
\draw[->,phi] (0.75, 0) node[phibp]{} -- ++(1, 3) node[phibp] {} -- node[rotate=55, above=-0.1cm]{$\StationProfile$} ++(3, 4) node[phibp]{} -- (7, 7);

\draw[phi, very thick] (0, -1) -- (2.43, -1);

\draw[f, dashed, very thick] (0, -1) -- (1.5, -1);
\draw[->,f] (1.5, 1) node[fbp] {} -- ++(2, 3) node[fbp] {} -- ++(2, 1) node[fbp]{} -- (7, 5);

\draw[dotted] (2.5, 0) -- (2.5, 2.5) -- (0, 2.5) node[left] {$q'$};

\draw[dotted] (1.57, 2.5) -- ++(0, -2.5);

\path[->, bend right] (1.57,2.5) edge node[above] {$\Delta c$} (2.5, 2.5);
\draw[->,phi,dashed] (1.68, 0) node[phibp]{} -- ++(1, 3) node[phibp] {} -- ++(3, 4) node[phibp]{} -- (7, 7);

\path[->, bend right] (2.5,2.5) edge node[below] {$\ArcCost$} (3.25, 2.5);
\draw[->,phi] (2.43, 0) node[phibp]{} -- ++(1, 3) node[phibp] {} -- node[rotate=55, above=-0.1cm]{$\ShiftedStationProfile$} ++(3, 4) node[phibp]{} -- (7, 7);

\end{tikzpicture}
		}
		\subcaption{\centering Shifting $\StationProfile$ according to $q'$ and fixed cost.}
	\end{subfigure}
	\hspace{0.15\textwidth}
	\begin{subfigure}[T]{0.3\textwidth}
		\resizebox{\textwidth}{!}{
			\begin{tikzpicture} 

\tikzset{
	bp/.style={
		draw, thick, solid, inner sep=1.5pt
	},
	f/.style={
		orange, thick
	},
	fbp/.style={
		circle, bp
	},
	ubp/.style={
		diamond, bp
	},
	phibp/.style = {
		diamond, bp
	},
	u/.style={
		blue
	},
	phi/.style={
		black
	}
}

\clip(-1, -1.3) rectangle (7.3, 7.5);

\draw[step=1cm, gray, very thin] (0, -1.3) grid (7.3, 7.3);
\draw[->] (0,0) -- (7.3, 0) node[above=0.25cm, left=0.25cm] {Cost};
\draw[->] (0,-1) -- (0, 7.3) node[right] {SoC};

\draw[thick] (2mm, -1) -- (-2mm, -1) node[left] {$-\infty$};
\foreach \y in {0, ..., 5} {
	\draw[thick] (2mm, \y) -- (-2mm, \y) node[left] {$\y$};
}

\draw[thick] (2mm, 6) -- (-2mm, 6) node[left=0.15cm,below=-0.2cm] {$q_{\max}$};
\draw[thick] (2mm, 7) -- (-2mm, 7) node[left] {$7$};

\foreach \x in {1, ..., 7} {
	\draw[thick] (\x, 2mm) -- (\x, -2mm) node[below] {$\x$};
}

\draw[->,f] (1.5, 1) node[fbp] {} -- ++(2, 3) node[fbp] {} -- node[above=2.5mm,rotate=25, left=2mm]{$\CostProfile[\Label_i]$} ++(2, 1) node[fbp]{} -- (7, 5);

\draw[dotted, name path=q] (0, 2.5) node[left] {$q'$} -- (7.3, 2.5);
\draw[dotted] (0, 6.0) -- (4.45, 6.0) -- (7.3, 6.0);

\draw[->,phi,dashed,name path=phi] (2.43, 0) node[phibp]{} -- node[rotate=70, below=0cm]{$\ShiftedStationProfile$} ++(1, 3) node[phibp] (test) {} -- ++(3, 4) node[phibp]{} -- (7, 7);

\path[name path=maxsoc] (0,6) -- (7.3,6);
\coordinate[name intersections={of=maxsoc and phi, by=stationcutoffUB}];
\coordinate[name intersections={of=q and phi, by=stationcutoffLB}];

\draw[u, very thick] (0, -1) -- (3.25, -1);
\draw[->, u] (stationcutoffLB) node[ubp]{} -- (test) node[ubp] {} -- node[rotate=55,right=5mm, above=-0.1cm]{$\CostProfile[\Label_j]$} (stationcutoffUB) node[ubp]{} -- (7, 6);

\draw[f, dashed, very thick] (0, -1) -- (1.5, -1);

\end{tikzpicture}
		}
		\subcaption{\centering Cutting off any unreachable \gls{abk:soc}.}
	\end{subfigure}
	\caption{Geometric interpretation of station replacement.\label{fig:station-replacement}}
	\begin{minipage}{\textwidth}
		\smaller\textit{Note.} The illustrated station replacement operation charges up to a cost of $c' = 2.5$, i.e., spends $c' - \CostProfileMinCost = 1.0$ on charging at the previous station. The orange (circles) and blue (diamonds) cost profiles correspond to the cost profile at the origin and target vertices ($\OriginVertex$, $\TargetVertex$), respectively. The precomputed (shifted) station cost profile $\StationProfile$ ($\ShiftedStationProfile$) is illustrated in black (diamonds). $q'$ denotes the arrival \gls{abk:soc} at station $i$. $\Delta c = c' - \InvStationProfile(q')$ offsets the station cost profile by the path's fixed cost. $q_{\max} = \ChargingFunction(q', \PeriodDuration)$ corresponds to the maximum reachable \gls{abk:soc} at station $i$.
	\end{minipage}
\end{figure}

\noindent Case \textit{i)}: \textit{station replacements}.
We denote this operation with $\Label' = \PropagateAndReplace[\Label][c']$. Here, cost value $c'$ with $c' \in [\CostProfileMinCost, \CostProfileMaxCost]$ corresponds to a decision on the sunk cost up to vertex $\OriginVertex$ and thus the arrival \gls{abk:soc} $q' = \CostProfile(c')$ at $\OriginVertex$. Recall that the generated label $\Label'$ should give the arrival \gls{abk:soc} at $\TargetVertex$ depending on the charging budget at $\OriginVertex$, i.e., $c - c'$. We compute the according cost profile $\CostProfile[\Label']$ based on the charging cost at vertex $\OriginVertex$, given as:
\begin{equation}
	\label{eq:charging-cost}
	\DeltaChargingCost[q'][\Delta q] \coloneqq \PeriodEnergyPrice[\PeriodOfVertex] \cdot \Delta q + \WearDensityFunction(q', q' + \Delta q).
\end{equation}
Here, $q'$ and $\Delta q$ correspond to the arrival \gls{abk:soc} and the \gls{abk:soc} recharged at $\OriginVertex$, respectively. We note that $\DeltaChargingCost[q'][\Delta q]$ is well defined for $0 \leq q' \leq \MaxSoC$ and $0 \leq \Delta q \leq \MaxSoC - q'$, such that we can formally state $\CostProfile[\Label']$ using the inverse of Equation~\eqref{eq:charging-cost}:
\begin{equation}
	\label{eq:cost-profile-replacement}
	\NewCostProfile(c) \coloneqq \begin{cases}
		-\infty&\text{if }c < c' + \ArcCost + \Delta c_{\PeriodOfVertex, \ChargerOfVertex}\langle q' \rangle(0)\\
		\NewPhi(q',\PeriodDuration[\PeriodOfVertex])&\text{if }c \geq c' + \ArcCost + \Delta c_{\PeriodOfVertex, \ChargerOfVertex}\langle q' \rangle (\NewPhi(q', \PeriodDuration[\PeriodOfVertex]) - q')\\
		q' + \InverseDeltaChargingCost[q'][c - \ArcCost - c']&\text{otherwise.}
	\end{cases}
\end{equation}
Here, we utilize precomputed \textit{station cost profiles} to compute Equation~\eqref{eq:cost-profile-replacement} efficiently. These give the arrival \gls{abk:soc} at vertex $\TargetVertex$ when spending $c$ on charging at $\OriginVertex$ with an empty battery:
\begin{equation}
	\label{eq:cost-profile-replacement-pre-computed}
	\NewStationProfile(c) \coloneqq \begin{cases}
		-\infty&\text{if } c < \ArcCost\\
		\MaxSoC&\text{if } c \geq \ArcCost + \Delta c_{\PeriodOfVertex, \ChargerOfVertex}\langle\MinSoC\rangle(\MaxSoC - \MinSoC)\\
		\InverseDeltaChargingCost[\MinSoC][c - \ArcCost]&\text{otherwise.}
	\end{cases}
\end{equation}
Station cost profiles allow to compute the replacement profile $\NewCostProfile$ in two steps: First, we shift $\NewStationProfile$ by $c' + \ArcCost - \InvNewStationProfile(q')$ on the cost axis to obtain $\ShiftedStationProfile$. Second, we establish upper and lower bounds on the \gls{abk:soc} and cut-off any unreachable \gls{abk:soc} levels. Finally, we get 
\begin{equation}
	\NewCostProfile(c) \coloneqq 
	\begin{cases}
		-\infty, & \text{if } c < \InvShiftedStationProfile(q')\\
		\MaxSoC & \text{if } c \geq \InvShiftedStationProfile(\ChargingFunction(q', \PeriodDuration))\\
		\ShiftedStationProfile(c) & \text{otherwise.}
	\end{cases}
\end{equation}
See Figure~\ref{fig:station-replacement} for an illustration of this procedure. We note that cost profiles derived according to this methodology remain, as compositions of concave piecewise-linear functions, also piecewise-linear and concave on $\lbrack\CostProfileMinCost[\NewLabel], \infty)$.

\noindent Case \textit{ii)}: \textit{intermediate charging}.
\newcommand{\InverseNewCostProfile}{\InvCostProfile[\NewLabel]}
Deciding to commit to charging for a fixed amount of time ${0 < \tau \leq \PeriodDuration}$ at vertex $\OriginVertex$ while continuing to track decisions at previous station $\Vertex$, denoted $\NewLabel \coloneqq \PropagateAndCharge$, transforms the cost profile non-linearly: decisions on charging at $\Vertex$ influence the arrival \gls{abk:soc} at $\OriginVertex$ and thus (possibly) the total energy recharged within timespan $\tau$. This, in turn, impacts the cost incurred from energy price and battery degradation. 
Formally, we can state the cost of charging at $\OriginVertex$ such that we arrive with \gls{abk:soc} $q$ at $\TargetVertex$ using Equation~\eqref{eq:cost-required-intermediate-station}:
\begin{equation}
	\begin{aligned}
		\label{eq:cost-required-intermediate-station}
		\InvNewCostProfile(q) \coloneqq&\;{\InvPrevCostProfile}(\overbrace{\NewPhi(\NewPhi^{-1}(q) - \tau)}^{\text{arrival SoC at }\OriginVertex})
		+ \underbrace{\overbrace{\ArcCost + \InvNewStationProfile(q)}^{\text{sunk cost after }\OriginVertex} - \overbrace{\InvNewStationProfile(\NewPhi(\NewPhi^{-1}(q) - \tau)}^{\text{sunk cost before } \OriginVertex})}_{\text{total cost of charging at }\OriginVertex}.
	\end{aligned}
\end{equation}
We note that the breakpoints of $\NewCostProfile$ correspond to the breakpoints of $\InverseNewCostProfile$ with reversed domain and co-domain such that we can derive an explicit representation of $\NewCostProfile$ by evaluating $\InverseNewCostProfile(q)$ at each breakpoint $q$ of $\InverseNewCostProfile$, computing segment slopes accordingly. We utilize the derivative $\frac{\partial \InverseNewCostProfile}{\partial q}(q)$ to obtain these breakpoints:
\begin{equation}
	\begin{aligned}
		\label{eq:inverse-intermediate-cost-profile-derivative}
		\PWLSlope[\InverseNewCostProfile][q](q) \coloneqq&
		\overbrace{\PWLSlope[\InvNewStationProfile][q](q) + \ArcCost}^{\mathclap{\text{cost of charging at }i}} 
		+ \underbrace{\PWLSlope[\NewPhi][q](\NewPhi^{-1}(q) - \tau) \cdot \PWLSlope[\NewPhi^{-1}][q](q)}_{\mathclap{\text{flow of charge from $i$ to $v$ (Term~\ref{eq:inverse-intermediate-cost-profile-derivative}.1)}}}
		\\&\cdot \lbrack
		\underbrace{\PWLSlope[{\InvPrevCostProfile}][q](\NewPhi(\NewPhi^{-1}(q) - \tau))}_{\mathclap{\text{additional charging cost at }v}}
		- \underbrace{\PWLSlope[\InvNewStationProfile][q](\NewPhi(\NewPhi^{-1}(q) - \tau))}_{\mathclap{\text{cost saving by charging less at }i}}
		\rbrack.
	\end{aligned}
\end{equation}
Every value $q$ at which Equation~\eqref{eq:inverse-intermediate-cost-profile-derivative} changes is a breakpoint of $\InverseNewCostProfile$. Hence, we can compute the breakpoints of $\InverseNewCostProfile$ based on the union of the breakpoints of functions $\PrevCostProfile$ and $\NewStationProfile$. We construct $\NewCostProfile$ from all breakpoints that lie within the \gls{abk:soc} interval 
\begin{equation*}
	\lbrack \NewPhi(\CostProfileMinSoC, \tau), \min\{\NewPhi(\CostProfileMaxSoC, \tau), \sum_{\Tour \in \SetOfTours \setminus \LabelVisitedTours} \TourConsumption,\MaxSoC\}\rbrack
\end{equation*}
and set $\CostProfileMinSoC, \CostProfileMinCost, \CostProfileMaxSoC$, and $\CostProfileMaxCost$ accordingly. Note that the derived cost profile $\NewCostProfile$ is not necessarily concave or even increasing. However, as we argue in the following, decreasing segments of such cost profiles are always dominated such that they can be discarded.

\begin{restatable}{proposition}{propone}
	\label{prop:more-expensive-when-non-increasing}
	The slope of a cost profile $\NewCostProfile$ obtained from $\NewLabel \coloneqq \PropagateAndCharge$ can only be negative at ${c \in [\CostProfileMinCost[\NewLabel], \CostProfileMaxCost[\NewLabel]]}$ if for $q' \coloneqq \NewPhi(\NewPhi^{-1}(\InvNewCostProfile(c)) - \tau)$, $\PWLSlope[{\InvPrevCostProfile}][q](q') < \PWLSlope[\InvNewStationProfile][q](q')$ holds.
\end{restatable}

\proof see Appendix~\ref{app:proofs}.\endproof

\noindent Intuitively, Proposition~\ref{prop:more-expensive-when-non-increasing} concerns cases where charging at $\Vertex$ is more expensive than charging at $\OriginVertex$, such that it pays off to \emph{shift} energy recharged at vertex $\OriginVertex$ to the previously tracked vertex $\Vertex$, i.e., use the charging opportunity captured by $\Label$. 

\begin{restatable}{proposition}{proptwo}
	\label{prop:decreasing-segments-are-dominated}
	For any $c$ on a decreasing segment $[c', c' + \epsilon]$ with $\epsilon > 0$ of some cost profile $\NewCostProfile$ obtained from $\NewLabel \coloneqq \PropagateAndCharge$, there exists some $\Label'' \in \SetOfLabels_j\setminus\{\NewLabel\}$ such that $\CostProfile[\Label''](c) \geq \NewCostProfile(c)$.
\end{restatable}

\proof see Appendix~\ref{app:proofs}.\endproof

\noindent With Proposition~\ref{prop:decreasing-segments-are-dominated} in mind, we ensure that cost profiles are non-decreasing by replacing non-concave cost profiles with their upper concave envelope. We argue that this preserves the correctness of our algorithm. Specifically, Proposition~\ref{prop:decreasing-segments-are-dominated} implies that for any $c$, there exists some label $\Label'' \in \SetOfLabels_{\TargetVertex}$ that reaches a higher \gls{abk:soc} at the same cost $c$. Hence, modifying $\Label'$ accordingly does not lead to the domination of any optimal labels.

\subsection{Non-dominated charging decisions}
\label{sec:committing-charging-operations}
\newcommand{\LHSProfile}{f}
\newcommand{\RHSProfile}{g}
\newcommand{\ShiftedOldLabel}{\overrightarrow{\Label}}
\newcommand{\ShiftedStationLabel}{\overrightarrow{\Label_{0}}}
\newcommand{\ShiftedOldProfile}{\CostProfile[\overrightarrow{\Label}]}
Propagating a label $\Label$ along a charging arc $(\Arc)$ creates new labels according to our propagation functions $\Label' \coloneqq \PropagateAndReplace[\Label][c']$ (station replacement) and $\Label' \coloneqq \PropagateAndCharge$ (intermediate charge).
Station replacements require a decision on the cost $c'$, i.e., the charging budget at the tracked station, and thus the respective arrival \gls{abk:soc} $q'$ at $\OriginVertex$. When charging intermediately, we need to decide on the time $\tau$ spent charging at $\OriginVertex$.
In both cases, time-continuous charging operations allow any decision that respects time and \gls{abk:soc} bounds, such that the set of choices for $c'$ and $\tau$ is unbounded in the general case. In what follows, we argue that in our problem setting only a finite set of values for $c'$ and $\tau$ generates non-dominated labels, such that it suffices to consider these during propagation.
Specifically, we show that labels $\SetOfLabels' \coloneqq \{\NoChargeLabel, \PropagateAndCharge[\OriginalLabel][\PeriodDuration]\} \cup \{\PropagateAndReplace[\OriginalLabel] \mid \forall c \in \SetOfBreakpoints[{\CostProfile[\OriginalLabel]}]\}$, with $\Label_j \coloneqq \PropagateRegularArc$, at $\TargetVertex$ capture all possibly optimal charging decisions.
\begin{samepage}
	\begin{restatable}{theorem}{theoremone}
		\label{theorem:one}
		For any charging arc $(\Arc)$ and label $\OriginalLabel \in \SetOfLabels_{\OriginVertex}$, 
		the set of labels $$\SetOfLabels' \coloneqq \{\NoChargeLabel, \PropagateAndCharge[\OriginalLabel][\PeriodDuration]\} \cup \{\PropagateAndReplace[\OriginalLabel] \mid \forall c \in \SetOfBreakpoints[{\CostProfile[\OriginalLabel]}]\},$$ with $\NoChargeLabel \coloneqq \PropagateRegularArc[\OriginalLabel]$ dominates any label $$\ToBeDominatedLabel \in \{\PropagateAndReplace[\OriginalLabel] \mid c \in \lbrack \CostProfileMinCost[\OriginalLabel], \CostProfileMaxCost[\OriginalLabel] \rbrack \} \cup \{\PropagateAndCharge[\OriginalLabel] \mid 0 \leq \tau \leq \PeriodDuration\} \setminus \SetOfLabels'.$$
	\end{restatable}
\end{samepage}
\noindent To prove Theorem~\ref{theorem:one}, we assume the contrary, i.e., that there exists some $\ToBeDominatedLabel$ not dominated by~$\SetOfLabels'$. We then show that for each $c \in \mathbb{R}$ there exists some $\Label \in \SetOfLabels'$ with $\CostProfile(c) \geq \ToBeDominatedProfile(\AssumedCost)$, hence contradicting the assumption. Our proof bases on the trade-off between charging at the station captured by $\Label$ versus charging at $\OriginVertex$: for any $\Delta c$ spent to charge using the charging opportunity captured by $\Label$, it will either be cheaper to spend additional money to continue charging at the tracked charger, or to utilize the new charging opportunity for this purpose. In the first case, the label spending ${\Delta c + \epsilon}$ for some $\epsilon > 0$ will dominate $\Label$, while the label spending ${\Delta c - \epsilon}$ will prevail in the second case. We can then show our claim using the piecewise linearity of our cost profiles. We refer to Appendix~\ref{app:proofs} for a full proof.
\newcommand{\IntermediateStation}{g}
\subsection{Example}
\label{subsec:propagation-example}
\newcommand{\ExampleNetwork}{\mathcal{G}}
\newcommand{\FirstChargerVertex}{\Vertex_{f}}
\newcommand{\SecondChargerVertex}{\Vertex_{g}}
The following example illustrates the behavior of our labeling algorithm on the example network illustrated in Figure~\ref{fig:propagation-example:configuration}. Figure~\ref{fig:propagation-example:propagation} depicts the labels created at each step of the algorithm. We provide the data basis for Figures~\ref{fig:propagation-example:configuration}~and~\ref{fig:propagation-example:propagation} in Appendix~\ref{app:example-details}.
\begin{paragraphs}
	\item[\emph{Figure~\ref{fig:propagation-example:root-propagation}:}] The algorithm initially extracts the empty path labeled $\SourceLabel \in \SetOfLabels_{\SourceNode}$ at the source $\SourceNode$. It then extends this path along the adjacent source arc $(\SourceNode, \Vertex_\Charger)$ at a cost of $\ArcCost[\SourceNode, \FirstChargerVertex] = 2$. Accordingly, we create a new label ${\Label_{v_f} \coloneqq \PropagateRegularArc[\SourceLabel][(\SourceNode, \FirstChargerVertex)]}$ with a minimum \gls{abk:soc} of one at a total cost of two at vertex $\FirstChargerVertex$. Geometrically, we obtain the new cost profile $\CostProfile[{\Label_f}]$ from \emph{shifting} $\CostProfile[\SourceLabel]$ on the \gls{abk:soc} axis. 
	\item[\emph{Figure~\ref{fig:propagation-example:first-station-propagation}:}] Our algorithm then continues and extracts $\Label_{v_f}$ at $\FirstChargerVertex$ in the next iteration, propagating it along charging arc $(\Vertex_f, v_2)$. Here, we can charge up to a maximum \gls{abk:soc} of $\ChargingFunction[{\Vertex_f}](q=0, \PeriodDuration = 4) = 3$ at a cost according to energy price $\PeriodEnergyPrice[\Vertex_f] = 2.5$, battery degradation $\WearDensityFunction$, and fixed cost $\ArcCost[(\Vertex_f, v_2)] = 0$. The algorithm creates a new label ${\Label_{v_2} \coloneqq \Label_f \PropagateAndReplaceAlone[f][(\Vertex_f, v_2)] 2}$ that captures this trade-off as detailed in Sections~\ref{sec:label-propagation} and \ref{sec:committing-charging-operations}. Note that it suffices to create a single label in this special case, as $\Vertex_f$ is the first charging opportunity considered such that the arrival \gls{abk:soc} at $\Vertex_f$ is unique.
	We obtain the updated cost profile $\StationProfile[\Label_{v_2}]$ from shifting $\StationProfile[{(v_f, v_2)}]$ on the cost axis such that it intersects $\CostProfile[\Label_{v_f}]$ at ${\CostProfileMinSoC[\Label_{v_f}] = 0}$, and subsequently cutting off any \gls{abk:soc} outside ${\lbrack 0, \ChargingFunction[{\ChargerOfVertex}](0, \PeriodDuration) \rbrack}$.
	\item[\emph{Figure~\ref{fig:propagation-example:first-tour-propagation}:}] The next iteration extracts $\Label_{v_2}$ at vertex $v_2$ and propagates it along service arc $(v_2, v_3)$ such that ${\Label_{v_3} \coloneqq \PropagateRegularArc[\Label_{v_2}][(v_2, v_3)]}$. Providing service to the respective operation consumes $1.5$ units of energy, which renders some charging decisions captured in $\CostProfile[{\Label_{v_2}}]$ infeasible: spending less than~$c = 6.5$ does not replenish sufficient energy to provide service. 
	Accordingly, we obtain the new cost profile $\CostProfile[{\Label_{v_{3}}}]$ from shifting $\CostProfile[{\Label_{v_2}}]$ by~$q = 1.5$ on the \gls{abk:soc} axis and subsequently cutting of any value below $\MinSoC = 0$. Essentially, we (implicitly) commit to charging at least~$1.5$ units of energy at $v_f$ to maintain feasibility.
	\item[\emph{Figure~\ref{fig:propagation-example:second-station-propagation}:}] The figure shows two iterations of our labeling algorithm. First, the algorithm propagates label $\Label_{v_3}$ along idle arc $(v_3, \SecondChargerVertex)$. This neither consumes \gls{abk:soc} nor incurs cost, such that $\CostProfile[{\Label_{\SecondChargerVertex}}] = \CostProfile[{\Label_{v_3}}]$ with ${\Label_{\SecondChargerVertex} \coloneqq \PropagateRegularArc[\Label_{v_3}][(v_3, \SecondChargerVertex)]}$.
	The next iteration propagates label $\Label_{\SecondChargerVertex}$ along charging arc $(\SecondChargerVertex, v_5)$, which again provides a charging opportunity. Here, in contrast to the previous charging opportunity $(\FirstChargerVertex, v_2)$, the arrival \gls{abk:soc}~$q$ at $\SecondChargerVertex$ is not unique.
	Our algorithm creates new labels according to Section~\ref{sec:committing-charging-operations} at $v_5$.
	Specifically, we create labels $\Label^{1}_{v_5} \coloneqq \Label_{\SecondChargerVertex} \PropagateAndReplaceAlone[g][(\SecondChargerVertex, v_5)] 6.5$, $\Label^{2}_{v_5} \coloneqq \PropagateAndChargeNoTau[\Label_{\SecondChargerVertex}][(\SecondChargerVertex, v_5)] \PeriodDuration\;(4)$, $\Label^{3}_{v_5} \coloneqq \Label_{\SecondChargerVertex} \PropagateAndReplaceAlone[g][(\SecondChargerVertex, v_5)] 8$, and 
	$\Label^{4}_{v_5} \coloneqq \Label_{\SecondChargerVertex} \PropagateAndReplaceAlone[g][(\SecondChargerVertex, v_5)] 11.7$.
	
	We observe two effects: first, some of the created labels are dominated according to Definitions~\ref{def:dominance} \& \ref{def:set-dominance}. Second, the cost profiles span segments of different lengths on the cost and \gls{abk:soc} axes. Here, non-linear charging functions $\ChargingFunction[]$ and the charging limit imposed by period length~$\PeriodDuration$ cause mismatches on the \gls{abk:soc} domain, while non-linear battery degradation is responsible for mismatching cost domains.
	\item[\emph{Figure~\ref{fig:propagation-example:second-tour-propagation-lower}:}] The algorithm proceeds to extract the labels in $\SetOfLabels_{v_5}$ in order of least cost. Hence, it first propagates label $\Label^{1}_{v_5}$ along service arc $(v_5, \SinkNode)$, which creates a new label ${\Label^{1}_{\SinkNode} \coloneqq \PropagateRegularArc[\Label^{1}_{v_5}][(v_5, \SinkNode)]}$ at $\SinkNode$ accordingly. 
	The algorithm then extracts label $\Label^{1}_{\SinkNode}$ at the sink as it has lower costs than label $\Label^{2}_{v_5}$. This terminates the algorithm and yields a path that reaches the sink. The cost profile of $\Label^{2}_{v_5}$ then gives the total cost of this shortest path as ${\CostProfileMinCost[\Label^{2}_{v_5}] = 11.875}$ with final \gls{abk:soc} ${\CostProfileMinSoC[\Label^{2}_{v_5}] = 0}$. 
	\item[\emph{Figure~\ref{fig:propagation-example:second-tour-propagation-higher}:}] Assuming a consumption of $\ArcConsumption[v_5, \SinkNode] \coloneqq 4.25$ for service arc $(v_5, \SinkNode)$, we observe a slightly different behavior. In this case, it does not suffice to charge as little as possible at $v_f$ such that label $\Label^{1}_{\SinkNode}$ is infeasible. Instead, label $\Label^{2}_{\SinkNode}$ now gives the optimal path. This constitutes an example where considering only station replacements does not yield an optimal solution.
\end{paragraphs}
\begin{figure}[t!]
	\captionsetup[subfigure]{width=0.9\textwidth}
	\centering
	\begin{subfigure}[t]{0.30\textwidth}
		\resizebox{\textwidth}{!}{
			\raisebox{2.5cm}{
				\begin{tikzpicture} 

\tikzset{
	every node/.style={
		inner sep=0pt, minimum size=0.7cm
	},
	every edge/.append style={
		thick,->
	},
	legend/.style = {
		minimum size=3mm
	},
legend_arc/.style = {
		thin, ->
	},
	dummy/.style={
		diamond, draw
	},
	station/.style={
		regular polygon, regular polygon sides = 3, draw
	},
	idle/.style={
		rectangle, draw, minimum size=0.5cm
	},
	charging_arc/.style={
		black, solid
	},
	idle_arc/.style={
		green!50!black, dashed
	},
	service_arc/.style={
		blue, dotted, thick
	},
	arc_cost/.style={
		above
	},
	arc_consumption/.style={
		below
	},
}

\node[draw, dummy] (source) at (0, 0) {$s^-$};
\node[draw, station] (first_station) at (2, 0) {$v_f$};
\node[draw, idle] (v_2) at (4, 0) {$v_2$};
\node[draw, idle] (v_3) at (-0.25, -1.25) {$v_3$};
\node[draw, station] (second_station) at (1.75, -1.25) {$v_g$};
\node[draw, idle] (v_4) at (3.75, -1.25) {$v_5$};
\node[draw, dummy] (sink) at (5.75, -1.25) {$s^+$};

\path[charging_arc] (source) edge node[arc_cost, above=3mm, left=-3.5mm]{$c = 2$} (first_station);
\path[charging_arc] (first_station) edge node[arc_cost]{} node[arc_consumption] {} (v_2);

\draw[service_arc](v_2) -- node[arc_consumption,below=3.5mm,right=-6mm]{$q = 1.5$} ++(1.5, 0);
\draw[service_arc,<-](v_3) -- ++ (-1, 0);

\path[idle_arc] (v_3) edge node[arc_cost]{} node[arc_consumption] {} (second_station);
\path[charging_arc] (second_station) edge node[arc_cost]{} node[arc_consumption] {} (v_4);
\path[service_arc] (v_4) edge node[arc_cost]{} node[arc_consumption] {$q = 3.5$} (sink);

\path (-0.5, -2.25) node[dummy, legend] {} -- node[legend,anchor=west] {\tiny Source/Sink} +(6mm, 0) -- ++(0, -0.5) node[idle, legend] {} -- node[legend,anchor=west] {\tiny Garage} +(6mm, 0)
 -- ++(2, 0.5) node[station, legend] {} -- node[legend,anchor=west] {\tiny Station} +(6mm, 0) --  ++(-2mm, -0.5) edge[legend_arc, charging_arc] +(4mm, 0) -- node[legend,anchor=west] {\tiny Charging} +(10mm, 0)
 -- ++(2, 0.5) edge[legend_arc, idle_arc] +(4mm,0) -- node[legend,anchor=west] {\tiny Idling} +(10mm, 0) -- ++(0, -0.5) edge[legend_arc, service_arc] +(4mm,0) -- node[legend,anchor=west] {\tiny Service} +(10mm, 0);

\end{tikzpicture}
			}
		}
		\subcaption{The example network.\label{fig:propagation-example:network}}
	\end{subfigure}
	\begin{subfigure}[t]{0.30\textwidth}
			\resizebox{\textwidth}{!}{
				\begin{tikzpicture} 

\tikzmath{
		\maxX = 6; \maxXBoundary = \maxX + 0.3;
		\minX = 0; \minXBoundary = \minX - 0.3;
		\maxY = 7; \maxYBoundary = \maxY + 0.3;
		\minY = 0; \minYBoundary = \minY - 0.3;
}

\tikzset{
	bp/.style={
		circle, draw, thick, solid, inner sep=1.5pt
	},
	profile/.style={
		thick
	},
	replaced/.style={
		orange, profile
	},
	intermediate/.style={
		green, profile
	},
	prev/.style={
		blue, profile, dashed
	},
	g/.style={
		blue, profile
	},
gbp/.style={
	bp, diamond
},
f/.style={
		orange, profile
	},
fbp/.style={
	bp, circle
},
	transformation/.style={
		black, dashed, bend left
	},
}
\clip(-1, -0.6) rectangle (6.3, 7.5);
\draw[step=1cm, gray, very thin] (0, -0.3) grid (\maxXBoundary, \maxYBoundary);
\draw[->] (0,0) -- (\maxXBoundary, 0) node[above=0.25cm, left=0.25cm] {Cost};
\draw[->] (0,\minYBoundary) -- (0, \maxYBoundary) node[right] {SoC};

\foreach \y in {0, ..., \maxY} {
	\draw[thick] (2mm, \y) -- (-2mm, \y) node[left] {$\y$};
}

\foreach \x in {0, 5, ..., 30} {
		\tikzmath{\xscaled = (\x) / 5;}
	\draw[thick] (\xscaled, 2mm) -- ++(0, -4mm) node[below] {$\x$};
}

\coordinate (f_bp1) at (0.0, 0.0);
\coordinate (f_bp2) at (1.5, 2.0);
\coordinate (f_bp3) at (6.125, 7.0);
\coordinate (f_bp4) at (\maxXBoundary, 7.0);

\coordinate (g_bp1) at (0, 0.0);
\coordinate (g_bp2) at (0.625, 2.0);
\coordinate (g_bp3) at (3.5, 7.0);
\coordinate (g_bp4) at (\maxXBoundary, 7.0);

\draw[g, ->] (g_bp1) node[gbp]{} -- (g_bp2) node[gbp]{} -- node[rotate=60, above=-0.5mm]{$\CostProfile[(v_g, v_5)]$} (g_bp3) node[gbp]{} -- (g_bp4);
\draw[f, ->] (f_bp1) node[fbp]{} -- (f_bp2) node[fbp]{} -- node[rotate=50, below=-1mm]{$\CostProfile[(v_f, v_2)]$} (f_bp3) node[fbp]{} -- (f_bp4);

\end{tikzpicture}
			}
		\subcaption{Station cost profiles.\label{fig:propagation-example:station-cost-profiles}}
	\end{subfigure}
	\begin{subfigure}[t]{0.30\textwidth}
		\resizebox{\textwidth}{!}{
			\begin{tikzpicture} 

\tikzmath{
		\maxX = 6; \maxXBoundary = \maxX + 0.3;
		\minX = 0; \minXBoundary = \minX - 0.3;
		\maxY = 7; \maxYBoundary = \maxY + 0.3;
		\minY = 0; \minYBoundary = \minY - 0.3;
}

\tikzset{
	bp/.style={
		circle, draw, thick, solid, inner sep=1.5pt
	},
	profile/.style={
		thick
	},
	replaced/.style={
		orange, profile
	},
	intermediate/.style={
		green, profile
	},
	prev/.style={
		blue, profile, dashed
	},
	g/.style={
		blue, profile
	},
gbp/.style={
	bp, diamond
},
f/.style={
	orange, profile
},
fbp/.style={
	bp, circle
},
	transformation/.style={
		black, dashed, bend left
	},
}
\clip(-1, -0.6) rectangle (6.3, 7.5);
\draw[step=1cm, gray, very thin] (0, -0.3) grid (\maxXBoundary, \maxYBoundary);
\draw[->] (0,0) -- (\maxXBoundary, 0) node[above=0.25cm, left=0.25cm] {Time};
\draw[->] (0,\minYBoundary) -- (0, \maxYBoundary) node[right] {Cost};

\foreach \y in {0, ..., \maxY} {
	\draw[thick] (2mm, \y) -- (-2mm, \y) node[left] {$\y$};
}

\foreach \x in {0, 2, ..., 12} {
		\tikzmath{\xscaled = (\x) / 2;}
	\draw[thick] (\xscaled, 2mm) -- ++(0, -4mm) node[below] {$\x$};
}

\coordinate (f_bp1) at (0.0, 0.0);
\coordinate (f_bp2) at (4.0, 6.0);
\coordinate (f_bp3) at (6.0, 7.0);
\coordinate (f_bp4) at (\maxXBoundary, 7.0);

\coordinate (g_bp1) at (0, 0.0);
\coordinate (g_bp2) at (3.0, 6.0);
\coordinate (g_bp3) at (4.5, 7.0);
\coordinate (g_bp4) at (\maxXBoundary, 7.0);

\draw[g, ->] (g_bp1) node[gbp]{} -- (g_bp2) node[gbp]{} -- node[rotate=35, above=-0.7mm]{$\Phi_g$} (g_bp3) node[gbp]{} -- (g_bp4);
\draw[f, ->] (f_bp1) node[fbp]{} -- (f_bp2) node[fbp]{} -- node[rotate=30, below=-1mm]{$\Phi_f$} (f_bp3) node[fbp]{} -- (f_bp4);

\end{tikzpicture}
		}
		\subcaption{Charging functions.\label{fig:propagation-example:infrastructure}}
	\end{subfigure}
	\caption{Example network (\ref{fig:propagation-example:network}), station cost profiles (\ref{fig:propagation-example:station-cost-profiles}), and charging functions (\ref{fig:propagation-example:infrastructure}).\label{fig:propagation-example:configuration}}
	\begin{minipage}{\textwidth}
		\smaller\textit{Note.} The example network corresponds to a simplified time-expanded network as introduced in Section~\ref{sec:network-design}. We assume a period length of $\PeriodDuration = 4$, $\MinSoC = 0$, and $\MaxSoC = 7$. Orange profiles with circle markers indicate station $f$, blue profiles with diamond markers station $g$.
	\end{minipage}
\end{figure}
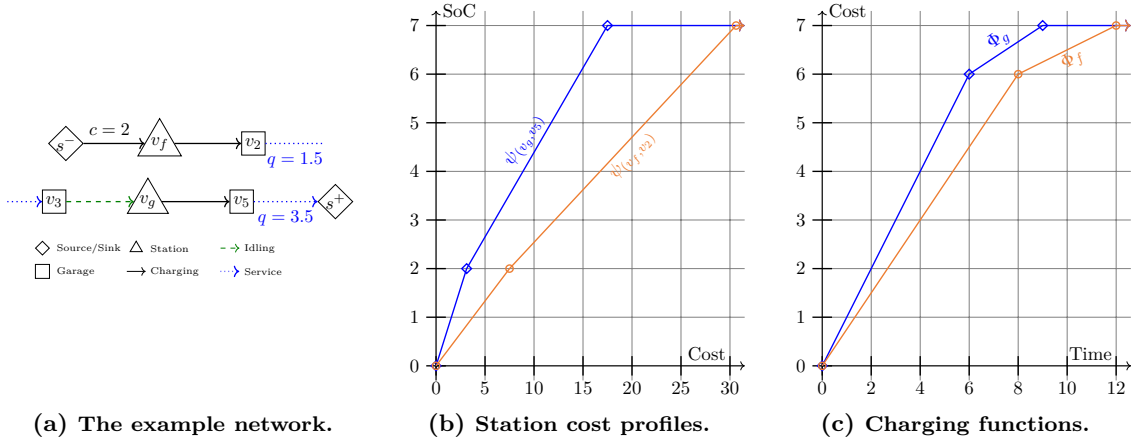
\begin{figure}[p]
	\def\ExampleFigureSize{13.25cm}%
	\captionsetup[subcaption]{width=0.9\textwidth}
	\caption{Running our labeling algorithm on the example network illustrated in Figure~\ref{fig:propagation-example:configuration}. \label{fig:propagation-example:propagation}}
		\noindent
		\begin{minipage}[t][9.25cm][t]{0.31\textwidth}
			\vspace{0pt}
			\begin{subfigure}{\textwidth}
				\resizebox{1\textwidth}{!}{
					\begin{tikzpicture} 

\tikzmath{
		\maxX = 6; \maxXBoundary = \maxX + 0.3;
		\minX = 0; \minXBoundary = \minX - 0.3;
		\maxY = 6; \maxYBoundary = \maxY + 0.3;
		\minY = -1; \minYBoundary = \minY - 0.3;
}

\tikzset{
	bp/.style={
		circle, draw, thick, solid, inner sep=1.5pt
	},
	profile/.style={
		thick
	},
	propagated/.style={
		orange, profile
	},
	prev/.style={
		blue, profile
	},
	prevbp/.style={
		bp, diamond
	},
	phi/.style={
		black
	},
	transformation/.style={
		black, dashed, bend left
	},
}

\draw[step=1cm, gray, very thin] (0, -1.3) grid (\maxXBoundary, \maxYBoundary);
\draw[->] (0,0) -- (\maxXBoundary, 0) node[above=0.25cm, left=0.25cm] {Cost};
\draw[->] (0,-1.3) -- (0, \maxYBoundary) node[right] {SoC};

\draw[thick] (2mm, -1) -- (-2mm, -1) node[left] {$-\infty$};
\foreach \y in {0, ..., 5} {
	\draw[thick] (2mm, \y) -- (-2mm, \y) node[left] {$\y$};
}
\draw[thick] (2mm, \maxY) -- (-2mm, \maxY) node[left] {$6$};

\foreach \x in {1, ..., \maxX} {
	\draw[thick] (\x, 2mm) -- (\x, -2mm) node[below] {$\x$};
}

\draw[propagated, very thick] (0,-1) -- ++(2, 0);
\draw[propagated, ->] (2, 0) node[bp]{} -- node[above]{$\CostProfile[\Label_{v_f}]$} (\maxXBoundary, 0);

\draw[prev, dashed, ->] (0, 0) node[prevbp]{}  -- node[above=3.5mm,left=-5mm]{$\CostProfile[\Label_{\SourceNode}]$} (\maxXBoundary, 0);

\draw[->] (0,0) edge[transformation] node[above=-0.25mm]{$\ArcCost=2$} (2, 0);

\end{tikzpicture}
				}
				\subcaption{\centering Propagating along source arc $(\SourceNode, v_f)$.\label{fig:propagation-example:root-propagation}}
				\begin{minipage}{\textwidth}
					\smaller \textit{Note.} Source arc $(\SourceNode, v_f)$ has a fixed cost of $2$, such that we shift $\CostProfile[\Label_{\SourceNode}]$ (blue, dashed, diamonds) by $2$ on the cost axis to obtain $\CostProfile[\Label_{v_f}]$ (orange, solid, circles).
				\end{minipage}
			\end{subfigure}
		\end{minipage}%
		\hspace{.02\textwidth}%
		\begin{minipage}[t][9.25cm][t]{0.31\textwidth}
			\vspace{0pt}
			\begin{subfigure}{\textwidth}
				\resizebox{1\textwidth}{!}{
					\begin{tikzpicture} 

\tikzmath{
		\maxX = 6; \maxXBoundary = \maxX + 0.3;
		\minX = 0; \minXBoundary = \minX - 0.3;
		\maxY = 6; \maxYBoundary = \maxY + 0.3;
		\minY = -1; \minYBoundary = \minY - 0.3;
}

\tikzset{
	bp/.style={
		circle, draw, thick, solid, inner sep=1.5pt
	},
	profile/.style={
		thick
	},
	propagated/.style={
		orange, profile
	},
	prev/.style={
		blue, profile
	},
	prevbp/.style={
		bp, diamond
	},
	stationbp/.style={
		bp, regular polygon, regular polygon sides=3
},
	station_profile/.style={
		black, dashed
	},
	transformation/.style={
		black, dashed, bend left
	},
}

\clip(-1, -1.3) rectangle (6.3, 6.5);
\draw[step=1cm, gray, very thin] (0, \minYBoundary) grid (\maxXBoundary, \maxYBoundary);
\draw[->] (0,0) -- (\maxXBoundary, 0) node[above=0.25cm, left=0.25cm] {Cost};
\draw[->] (0,-1.3) -- (0, \maxYBoundary) node[right] {SoC};

\draw[thick] (2mm, -1) -- (-2mm, -1) node[left] {$-\infty$};
\foreach \y in {0, ..., \maxX} {
	\draw[thick] (2mm, \y) -- (-2mm, \y) node[left] {$\y$};
}

\foreach \x in {2, 4, ..., 12} {
		\tikzmath{\xscaled = \x / 2;}
	\draw[thick] (\xscaled, 2mm) -- ++(0, -4mm) node[below] {$\x$};
}

\path[name path=station] (0, 0)-- (3.0, 2.0) -- (14.0, 8.0);
\path[name path=xmaxbound] (6.3, 0) -- (6.3, 6);

\coordinate[name intersections={of=station and xmaxbound, by=stationcutoff}];

\draw[station_profile, ->, name path=station] (0, 0) node[stationbp]{} -- node[above=4mm, left=0mm, rotate=36]{$\StationProfile[(\Vertex_f, \Vertex_2)]$} (3.0, 2.0) node[stationbp]{} -- (stationcutoff);

\draw[prev, dashed, very thick] (0,-1) -- ++(1.0, 0);
\draw[prev, dashed, ->] (1.0, 0.0) node[prevbp]{} -- node[above]{$\CostProfile[\Label_{v_f}]$} (\maxXBoundary, 0.0);
\draw[propagated, very thick] (0, -1) -- ++(1.0, 0);
\draw[propagated, ->] (1.0, 0.0) node[bp]{} -- (4.0, 2.0) node[bp]{} -- node[below, rotate=30]{$\CostProfile[\Label_{v_2}]$} (5.833333, 3.0) node[bp]{} -- (\maxXBoundary, 3.0);

\end{tikzpicture}
				}
				\subcaption{\centering Propagating along charging arc $(v_f, v_2)$.\label{fig:propagation-example:first-station-propagation}}
				\begin{minipage}{\textwidth}
					\smaller \textit{Note.} Propagating profile $\CostProfile[\Label_{v_{\Vertex_f}}]$ (blue, dashed, diamonds) captures the charging trade-off at $v_f$ in $\CostProfile[\Label_{v_2}]$ (orange, solid, circles). The dashed black profile with triangle markers corresponds to the station cost profile of arc $(v_f, v_2)$.
				\end{minipage}
			\end{subfigure}
		\end{minipage}%
		\hspace{.02\textwidth}%
		\begin{minipage}[t][9.25cm][t]{0.31\textwidth}
			\vspace{0pt}
			\begin{subfigure}{\textwidth}
				\resizebox{1\textwidth}{!}{
					\begin{tikzpicture} 

\tikzmath{
		\maxX = 6; \maxXBoundary = \maxX + 0.3;
		\minX = 0; \minXBoundary = \minX - 0.3;
		\maxY = 6; \maxYBoundary = \maxY + 0.3;
		\minY = -1; \minYBoundary = \minY - 0.3;
}

\tikzset{
	bp/.style={
		circle, draw, thick, solid, inner sep=1.5pt
	},
	profile/.style={
		thick
	},
	propagated/.style={
		orange, profile
	},
	prev/.style={
		blue, profile
	},
	prevbp/.style={
		bp, diamond
	},
	station_profile/.style={
		black, dashed
	},
	transformation/.style={
		black, dashed, bend left
	},
}
\clip(-1, -1.3) rectangle (6.3, 6.5);
\draw[step=1cm, gray, very thin] (0, -1.3) grid (\maxXBoundary, \maxYBoundary);
\draw[->] (0,0) -- (\maxXBoundary, 0) node[above=0.25cm, left=0.25cm] {Cost};
\draw[->] (0,-1.3) -- (0, \maxYBoundary) node[right] {SoC};

\draw[thick] (2mm, -1) -- (-2mm, -1) node[left] {$-\infty$};
\foreach \y in {0, ..., \maxY} {
	\draw[thick] (2mm, \y) -- (-2mm, \y) node[left] {$\y$};
}

\foreach \x in {2, 4, ..., 12} {
		\tikzmath{\xscaled = \x / 2;}
	\draw[thick] (\xscaled, 2mm) -- ++(0, -4mm) node[below] {$\x$};
}

\draw[propagated] (0, -1) -- ++(3.25, 0);

\draw[prev, dashed, very thick] (0, -1) -- (1, -1);
\draw[prev, dashed, ->] (1, 0) node[prevbp]{} -- (4, 2) node[prevbp]{} -- node[above, rotate=40, font=\large]{$\CostProfile[\Label_{v_2}]$} (5.833333, 3.0) node[prevbp]{} -- (\maxXBoundary, 3.0);

\path[name path=propagated_label] (1, -1.5) -- (4, 0.5) -- (5.833333, 1.5) -- (\maxXBoundary, 1.5);
\path[name path=infty] (0,-1) -- (\maxXBoundary, -1);

\coordinate[name intersections={of=infty and propagated_label, by=propagated_cutoff}];

\draw[propagated, dashed, very thick] (propagated_cutoff) -- (3.25, 0);
\draw[propagated, ->] (3.25, 0) node[bp]{} -- (4, 0.5) node[bp]{} -- node[above, rotate=40, font=\large]{$\CostProfile[\Label_{v_3}]$} (5.8333333, 1.5) node[bp]{} -- (\maxXBoundary, 1.5);

\draw[transformation, <-] (3.25, 0) -- ++(0, 1.5) node[bp, dotted] (test) {};
\draw[transformation, ->] (4, 2) -- ++(0, -1.5);
\draw[transformation, ->] (5.833333, 3.0) -- ++(0, -1.5);

\draw[transformation, dashed] (0, 1.5) node[left] {$q$} -- (test);

\end{tikzpicture}
				}
				\subcaption{\centering Propagating along service arc $(v_2, v_3)$.\label{fig:propagation-example:first-tour-propagation}}
				\begin{minipage}{\textwidth}
					\smaller \textit{Note.} We shift profile $\CostProfile[\Label_{v_2}]$ (blue, dashed, diamonds) according to the consumption of service arc $(v_2, v_3)$, specifically by $q=1.5$, on the \gls{abk:soc} axis. The dashed part of profile $\CostProfile[\Label_{v_3}]$ (orange, solid, circles) is cut off.
				\end{minipage}
			\end{subfigure}
		\end{minipage}%
		
		\begin{minipage}[t][\ExampleFigureSize][t]{0.31\textwidth}
			\vspace{0pt}
			\begin{subfigure}{\textwidth}
				\resizebox{1\textwidth}{!}{
					\begin{tikzpicture} 

\tikzmath{
		\maxX = 6; \maxXBoundary = \maxX + 0.3;
		\minX = 0; \minXBoundary = \minX - 0.3;
		\maxY = 6; \maxYBoundary = \maxY + 0.3;
		\minY = -1; \minYBoundary = \minY - 0.3;
}

\tikzset{
	bp/.style={
		circle, draw, thick, solid, inner sep=1.5pt
	},
	profile/.style={
		thick
	},
	replaced/.style={
		red, profile
	},
	replbp/.style={
		bp, regular polygon, regular polygon sides = 4
},
	intermediate/.style={
		green, profile
	},
intermediatebp/.style={
		bp, regular polygon, regular polygon sides = 3
	},
	prev/.style={
		blue, profile
	},
prevbp/.style={
		bp, diamond
	},
	after_idle/.style={
		orange, profile
	},
	station_profile/.style={
		black, dashed
	},
	transformation/.style={
		black, dashed, bend left
	},
}
\clip(-1, -1.3) rectangle (6.3, 6.5);
\draw[step=1cm, gray, very thin] (0, -1.3) grid (\maxXBoundary, \maxYBoundary);
\draw[->] (0,0) -- (\maxXBoundary, 0) node[above=0.25cm, left=0.25cm] {Cost};
\draw[->] (0,-1.3) -- (0, \maxYBoundary) node[right] {SoC};

\draw[thick] (2mm, -1) -- (-2mm, -1) node[left] {$-\infty$};
\foreach \y in {0, ..., \maxY} {
	\draw[thick] (2mm, \y) -- (-2mm, \y) node[left] {$\y$};
}

\foreach \x in {6, 9, ..., 24} {
		\tikzmath{\xscaled = (\x - 6) / 3;}
	\draw[thick] (\xscaled, 2mm) -- ++(0, -4mm) node[below] {$\x$};
}

\coordinate (v3_bpinf) at (0.1666666, -1.0);
\coordinate (v3_bp1) at (0.1666666, 0.0);
\coordinate (v3_bp2) at (0.6666666, 0.5);
\coordinate (v3_bp3) at (1.8888888, 1.5);
\coordinate (v3_bp4) at (\maxXBoundary, 1.5);

\coordinate (vg_bpinf) at (0.1666666, -1.0);
\coordinate (vg_bp1) at (0.1666666, 0.0);
\coordinate (vg_bp2) at (0.6666666, 0.5);
\coordinate (vg_bp3) at (1.8888888, 1.5);
\coordinate (vg_bp4) at (\maxXBoundary, 1.5);

\coordinate (v5_1_bpinf) at (0.1666666, -1.0);
\coordinate (v5_1_bp1) at (0.1666666, 0.0);
\coordinate (v5_1_bp2) at (1.0, 2.0);
\coordinate (v5_1_bp3) at (2.2777777, 4.0);
\coordinate (v5_1_bp4) at (\maxXBoundary, 4.0);

\coordinate (v5_2_bpinf) at (2.2777777, -1.0);
\coordinate (v5_2_bp1) at (2.2777777, 4.0);
\coordinate (v5_2_bp2) at (2.8888888, 4.5);
\coordinate (v5_2_bp3) at (4.3333333, 5.5);
\coordinate (v5_2_bp4) at (\maxXBoundary, 5.5);

\coordinate (v5_3_bpinf) at (0.6666666, -1.0);
\coordinate (v5_3_bp1) at (0.6666666, 0.5);
\coordinate (v5_3_bp2) at (1.2916666, 2.0);
\coordinate (v5_3_bp3) at (2.8888888, 4.5);
\coordinate (v5_3_bp4) at (\maxXBoundary, 4.5);

\coordinate (v5_4_bpinf) at (1.8888888, -1.0);
\coordinate (v5_4_bp1) at (1.8888888, 1.5);
\coordinate (v5_4_bp2) at (2.0972222, 2.0);
\coordinate (v5_4_bp3) at (4.3333333, 5.5);
\coordinate (v5_4_bp4) at (\maxXBoundary, 5.5);

\draw[intermediate, very thick] (0, -1) -- (v5_2_bpinf);
\draw[replaced, dashed, very thick] (0, -1) -- (v5_4_bpinf);
\draw[prev, very thick] (0, -1) -- (v3_bpinf);
\draw[after_idle, very thick] (0, -1) -- (vg_bpinf);

\draw[prev, ->] (v3_bp1) -- (v3_bp2) -- (v3_bp3) node[prevbp]{} -- node[below]{$\CostProfile[\Label_{v_3}]$} (v3_bp4);
\draw[after_idle, dashed, ->] (vg_bp1) node[prevbp]{} -- (vg_bp2) node[prevbp]{} -- (vg_bp3) node[prevbp]{} -- node[above]{$\CostProfile[\Label_{v_g}]$} (vg_bp4);

\draw[replaced, ->] (v5_1_bp1) node[replbp]{} -- (v5_1_bp2) node[replbp]{} -- node[above=-2mm, rotate=70]{$\CostProfile[{\Label^{1}_{v_5}}]$} (v5_1_bp3) node[replbp]{} -- (v5_1_bp4);

\draw[replaced, dashed, ->] (v5_3_bp1) node[replbp]{} -- (v5_3_bp2) node[replbp]{} -- node[below=-2mm, rotate=70]{$\CostProfile[{\Label^{3}_{v_5}}]$}  (v5_3_bp3) node[replbp]{} -- (v5_3_bp4);

\draw[replaced, dashed] (v5_4_bp1) node[replbp]{} -- (v5_4_bp2) node[replbp]{} -- node[below=8mm, right=0mm, rotate=70]{$\CostProfile[{\Label^{4}_{v_5}}]$} (v5_4_bp3);
\draw[replaced, ->] (v5_4_bp3) node[replbp]{} -- (v5_4_bp4);

\draw[intermediate] (v5_2_bp1) node[intermediatebp]{} -- (v5_2_bp2) node[intermediatebp]{} -- node[above, rotate=40]{$\CostProfile[{\Label^{2}_{v_5}}]$} (v5_2_bp3) node[intermediatebp]{};
\draw[intermediate, dashed, ->] (v5_2_bp3) -- (v5_2_bp4);

\end{tikzpicture}
				}
				\subcaption{\centering Propagating along idle and charging arcs $(v_3, v_g)$, $(v_g, v_5)$.\label{fig:propagation-example:second-station-propagation}}
				\begin{minipage}{\textwidth}
					\smaller \textit{Note.} Labels $\Label^{1}_{v_5}, \Label^{3}_{v_5}$, and $\Label^{5}_{v_5}$ (red, squares) settle on charging $1.5$, $2$, and $3$ units of energy at $v_f$, such that the arrival \gls{abk:soc} at $v_g$ equals $0$, $0.5$, and $1.5$ respectively. Label $\Label^2_{v_5}$ (green, triangles) commits to charge for $\tau = \PeriodDuration$ at $v_g$. Together, the solid red and green profiles dominate the dashed red profiles as they provide higher \gls{abk:soc} at a lower cost. The profile of label $\Label^3_{v_5}$ would not be dominated by the profile of label $\Label^1_{v_5}$ or $\Label^2_{v_5}$ alone.
				\end{minipage}
			\end{subfigure}
		\end{minipage}%
		\hspace{.02\textwidth}%
		\begin{minipage}[t][\ExampleFigureSize][t]{0.31\textwidth}
			\vspace{0pt}
			\begin{subfigure}{\textwidth}
				\resizebox{1\textwidth}{!}{
					\begin{tikzpicture} 

\tikzmath{
		\maxX = 6; \maxXBoundary = \maxX + 0.3;
		\minX = 0; \minXBoundary = \minX - 0.3;
		\maxY = 6; \maxYBoundary = \maxY + 0.3;
		\minY = -1; \minYBoundary = \minY - 0.3;
}

\tikzset{
	bp/.style={
		regular polygon, regular polygon sides = 4, draw, thick, solid, inner sep=1.5pt
	},
	profile/.style={
		thick
	},
	replaced/.style={
		orange, profile
	},
	intermediate/.style={
		green, profile
	},
intermediatebp/.style={
		bp, regular polygon, regular polygon sides = 3
	},
	prev/.style={
		blue, profile, dashed
	},
	prevbp/.style={
		bp, regular polygon, regular polygon sides =4
	},
	station_profile/.style={
		black, dashed
	},
	transformation/.style={
		black, dashed, bend left
	},
}
\clip(-1, -1.3) rectangle (6.3, 6.5);
\draw[step=1cm, gray, very thin] (0, -1.3) grid (\maxXBoundary, \maxYBoundary);
\draw[->] (0,0) -- (\maxXBoundary, 0) node[above=0.25cm, left=0.25cm] {Cost};
\draw[->] (0,-1.3) -- (0, \maxYBoundary) node[right] {SoC};

\draw[thick] (2mm, -1) -- (-2mm, -1) node[left] {$-\infty$};
\foreach \y in {0, ..., \maxY} {
	\draw[thick] (2mm, \y) -- (-2mm, \y) node[left] {$\y$};
}

\foreach \x in {6, 9, ..., 24} {
		\tikzmath{\xscaled = (\x - 6) / 3;}
	\draw[thick] (\xscaled, 2mm) -- ++(0, -4mm) node[below] {$\x$};
}

\coordinate (v5_1_bp1) at (0.1666666, 0.0);
\coordinate (v5_1_bp2) at (1.0, 2.0);
\coordinate (v5_1_bp3) at (2.2777777, 4.0);
\coordinate (v5_1_bp4) at (\maxXBoundary, 4.0);

\coordinate (v5_2_bp1) at (2.2777777, 4.0);
\coordinate (v5_2_bp2) at (2.8888888, 4.5);
\coordinate (v5_2_bp3) at (4.3333333, 5.5);
\coordinate (v5_2_bp4) at (\maxXBoundary, 5.5);

\coordinate (vs_1_bpinf) at (1.9583333, -1);
\coordinate (vs_1_bp1) at (1.9583333, 0.0);
\coordinate (vs_1_bp2) at (2.2777777, 0.5);
\coordinate (vs_1_bp3) at (\maxXBoundary, 0.5);

\coordinate (vs_2_bpinf) at (2.2777777, -1);
\coordinate (vs_2_bp1) at (2.2777777, 0.5);
\coordinate (vs_2_bp2) at (2.8888888, 1.0);
\coordinate (vs_2_bp3) at (4.3333333, 2.0);
\coordinate (vs_2_bp4) at (\maxXBoundary, 2.0);

\draw[intermediate, very thick] (0, -1) -- (vs_2_bpinf);
\draw[replaced, dashed, very thick] (0, -1) -- (vs_1_bpinf);

\draw[prev, dashed, ->, name path=v5_1] (v5_1_bp1) node[prevbp]{} -- (v5_1_bp2) node[prevbp]{} -- node[above, rotate=70]{$\CostProfile[{\Label^{1}_{v_5}}]$} (v5_1_bp3) node[prevbp]{} -- (v5_1_bp4);
\draw[prev, dashed, ->] (v5_2_bp1) node[intermediatebp]{} -- (v5_2_bp2) node[intermediatebp]{} -- node[above, rotate=40]{$\CostProfile[{\Label^{2}_{v_5}}]$} (v5_2_bp3) node[intermediatebp]{} -- (v5_2_bp4);

\draw[intermediate,->] (vs_2_bp1) node[intermediatebp]{} -- (vs_2_bp2) node[intermediatebp]{} -- node[above, right=-7.5mm, rotate=35]{$\CostProfile[{\Label^{2}_{\SinkNode}}]$} (vs_2_bp3) node[intermediatebp]{} -- (vs_2_bp4);

\draw[replaced,->] (vs_1_bp1) node[bp]{} -- (vs_1_bp2) node[bp]{} -- node[above]{$\CostProfile[{\Label^{1}_{\SinkNode}}]$} (vs_1_bp3);

\path[name path=vs_1_bp1_y] (vs_1_bp1) -- ++(0, 5);
\coordinate[name intersections={of=vs_1_bp1_y and v5_1, by=vs_1_cutoff}];

\draw[transformation, ->] (vs_1_cutoff) node[bp, dashed]{} -- (vs_1_bp1);
\draw[transformation, ->] (v5_2_bp1) -- (vs_2_bp1);
\draw[transformation, ->] (v5_2_bp2) -- (vs_2_bp2);
\draw[transformation, ->] (v5_2_bp3) -- (vs_2_bp3);

\end{tikzpicture}
				}
				\subcaption{\centering Propagating along service arc $(v_5, \SinkNode)$ with $\ArcConsumption[v_5, \SinkNode] = 3.5$. \label{fig:propagation-example:second-tour-propagation-lower}}
				\begin{minipage}{\textwidth}
				\smaller \textit{Note.} The optimal path reaches the sink at a cost of $\CostProfileMinCost = 11.875$, charging $1.5$ and $3.5$ units of energy at $v_f$ and $v_g$, respectively. Note that the algorithm terminates after propagating $\Label^1_{v_{5}}$ (blue, dashed, squares) as label $\Label^1_{v_{\SinkNode}}$ (orange, squares) is extracted before $\Label^2_{v_{5}}$ (blue, dashed, triangles). Label $\Label^2_{v_{\SinkNode}}$ (green, triangles) is thus never created.
				\end{minipage}
			\end{subfigure}
		\end{minipage}%
		\hspace{.02\textwidth}%
		\begin{minipage}[t][\ExampleFigureSize][t]{0.31\textwidth}
			\vspace{0pt}
			\begin{subfigure}{\textwidth}
				\resizebox{1\textwidth}{!}{
					\begin{tikzpicture} 

\tikzmath{
		\maxX = 6; \maxXBoundary = \maxX + 0.3;
		\minX = 0; \minXBoundary = \minX - 0.3;
		\maxY = 6; \maxYBoundary = \maxY + 0.3;
		\minY = -1; \minYBoundary = \minY - 0.3;
}

\tikzset{
	bp/.style={
		circle, draw, thick, solid, inner sep=1.5pt
	},
	profile/.style={
		thick
	},
	replaced/.style={
		orange, profile
	},
	replbp/.style={
		bp, regular polygon, regular polygon sides = 4, dashed
	},
	intermediate/.style={
		green, profile
	},
intermediatebp/.style={
		bp, regular polygon, regular polygon sides = 3
	},
	prev/.style={
		blue, profile, dashed
	},
	prevbp/.style={
		bp, regular polygon, regular polygon sides = 4, dashed
	},
	station_profile/.style={
		black, dashed
	},
	transformation/.style={
		black, dashed, bend left
	},
}
\clip(-1, -1.3) rectangle (6.3, 6.5);
\draw[step=1cm, gray, very thin] (0, -1.3) grid (\maxXBoundary, \maxYBoundary);
\draw[->] (0,0) -- (\maxXBoundary, 0) node[above=0.25cm, left=0.25cm] {};
\draw[->] (0,-1.3) -- (0, \maxYBoundary) node[right] {SoC};

\draw[thick] (2mm, -1) -- (-2mm, -1) node[left] {$-\infty$};
\foreach \y in {0, ..., \maxY} {
	\draw[thick] (2mm, \y) -- (-2mm, \y) node[left] {$\y$};
}

\foreach \x in {6, 9, ..., 24} {
		\tikzmath{\xscaled = (\x - 6) / 3;}
	\draw[thick] (\xscaled, 2mm) -- ++(0, -4mm) node[below] {$\x$};
}

\coordinate (v5_1_bpinf) at (0.1666666, -1.0);
\coordinate (v5_1_bp1) at (0.1666666, 0.0);
\coordinate (v5_1_bp2) at (1.0, 2.0);
\coordinate (v5_1_bp3) at (2.2777777, 4.0);
\coordinate (v5_1_bp4) at (\maxXBoundary, 4.0);

\coordinate (v5_2_bpinf) at (2.2777777, -1.0);
\coordinate (v5_2_bp1) at (2.2777777, 4.0);
\coordinate (v5_2_bp2) at (2.8888888, 4.5);
\coordinate (v5_2_bp3) at (4.3333333, 5.5);
\coordinate (v5_2_bp4) at (\maxXBoundary, 5.5);

\coordinate (v5_3_bpinf) at (0.6666666, -1.0);
\coordinate (v5_3_bp1) at (0.6666666, 0.5);
\coordinate (v5_3_bp2) at (1.2916666, 2.0);
\coordinate (v5_3_bp3) at (2.8888888, 4.5);
\coordinate (v5_3_bp4) at (\maxXBoundary, 4.5);

\coordinate (vs_1_bpinf) at (0.166666, -1);
\coordinate (vs_1_bp1) at (0.166666, -4.25);
\coordinate (vs_1_bp2) at (1.0, -2.25);
\coordinate (vs_1_bp3) at (2.27777, -0.25);
\coordinate (vs_1_bp4) at (\maxXBoundary, -0.25);
\coordinate (vs_2_bpinf) at (2.277777, -1);
\coordinate (vs_2_bp1) at (2.277777, -0.25);
\coordinate (vs_2_bp2) at (2.888888, 0.25);
\coordinate (vs_2_bp3) at (4.333333, 1.25);
\coordinate (vs_2_bp4) at (\maxXBoundary, 1.25);
\coordinate (vs_3_bpinf) at (0.6666666, -1.0);
\coordinate (vs_3_bp1) at (0.6666666, -3.75);
\coordinate (vs_3_bp2) at (1.2916666, -2.25);
\coordinate (vs_3_bp3) at (2.8888888, 0.25);
\coordinate (vs_3_bp4) at (\maxXBoundary, 0.25);

\draw[intermediate, very thick] (0, -1) -- (v5_2_bpinf);

\draw[prev, dashed, ->] (v5_1_bp1) node[prevbp]{} -- (v5_1_bp2) node[prevbp]{} -- node[above, rotate=70]{$\CostProfile[{\Label^{1}_{v_5}}]$} (v5_1_bp3) node[prevbp]{} -- (v5_1_bp4);
\draw[prev, dashed, ->] (v5_2_bp1) node[intermediatebp]{} -- (v5_2_bp2) node[intermediatebp]{} -- node[above, rotate=40]{$\CostProfile[{\Label^{2}_{v_5}}]$} (v5_2_bp3) node[intermediatebp]{} -- (v5_2_bp4);
\draw[replaced, red, dashed, ->] (v5_3_bp1) node[replbp]{} -- (v5_3_bp2) node[replbp]{} -- node[below=-4mm, rotate=70]{$\CostProfile[{\Label^{3}_{v_5}}]$}  (v5_3_bp3) node[replbp]{} -- (v5_3_bp4);

\begin{scope}
\clip(-1, -1) rectangle (6.3, 6.5);
\draw[replaced, red, dashed, ->] (vs_3_bp1) -- (vs_3_bp2) -- (vs_3_bp3)-- node[above]{$\CostProfile[{\Label^{3}_{\SinkNode}}]$} (vs_3_bp4);

\draw[replaced, dashed, ->] (vs_1_bp2) -- (vs_1_bp3) node[replbp]{} -- node[below=4mm, right=-2mm]{$\CostProfile[{\Label^{1}_{\SinkNode}}]$} (vs_1_bp4);

\path[name path=xaxis] (0, 0) -- (\maxXBoundary, 0);
\path[name path=vs_2] (vs_2_bp1) -- (vs_2_bp2) -- (vs_2_bp3) -- (vs_2_bp4);
\coordinate[name intersections={of=vs_2 and xaxis, by=A}];
\draw[intermediate, dashed] (vs_2_bp1) node[intermediatebp, dashed]{} -- (A);
\draw[intermediate, ->] (A) node[intermediatebp]{} -- (vs_2_bp2) node[intermediatebp]{} -- node[above, right=-7.5mm, rotate=35]{$\CostProfile[{\Label^{2}_{\SinkNode}}]$} (vs_2_bp3) node[intermediatebp]{} -- (vs_2_bp4);

\end{scope}

\draw[transformation, ->] (v5_2_bp1) -- (vs_2_bp1);
\draw[transformation, ->] (v5_2_bp2) -- (vs_2_bp2);
\draw[transformation, ->] (v5_2_bp3) -- (vs_2_bp3);

\end{tikzpicture}
				}
				\subcaption{\centering Propagating along service arc $(v_5, \SinkNode)$ with $\ArcConsumption[v_5, \SinkNode] = 4.25$.\label{fig:propagation-example:second-tour-propagation-higher}}
				\begin{minipage}{\textwidth}
					\smaller \textit{Note.} In this case, propagating label $\Label^1_{v_{5}}$ (blue, dashed, squares) yields an infeasible label $\Label^1_{\SinkNode}$ (orange, dashed, squares). Propagating next label $\Label^2_{v_{5}}$ (blue, dashed, triangles) yields optimal label $\Label^2_{v_{5}}$ (green, solid, triangles), which reaches the sink at a cost of $\CostProfileMinCost = 14.1875$, charging $1.75$ and $4$ at $v_g$ and $v_f$ respectively. This illustrates a case where none of the labels obtained from station replacements at $v_g$ yields the optimal solution.
				\end{minipage}
			\end{subfigure}
		\end{minipage}
\end{figure}
\subsection{Speedup techniques}
\label{subsec:speedup-techniques}
A straightforward application of our label-setting algorithm suffers from several phenomena which lead to the generation of many superfluous labels: 
first, it detects infeasible charge and service scheduling decisions only when extending a service arc or reaching the sink. 
Second, our algorithm is biased towards schedules that charge late and little, which results in the generation of many infeasible paths before reaching the sink and further yields schedules with low diversity. 
Third, redundancy in the network's structure creates many (cost-)equivalent labels.\\
We mitigate the first issue by establishing lower bounds on the \gls{abk:soc} and propagating service operation time windows to station and garage vertices such that we can discard infeasible paths early. We further derive a \textit{potential function} for each $\Vertex \in \SetOfVertices$ which estimates the remaining cost required to reach the sink. This potential function then offsets costs of partial paths accordingly to eliminate the bias towards labels with low \gls{abk:soc}. Finally, we reduce symmetry issues caused by network redundancy through additional checks in our feasibility condition.

\paragraph{\gls{abk:soc} bounds:}

\newcommand{\MaxChargingRate}{\underset{{\Charger \in \SetOfChargers}}{\max} \PWLSlope[\ChargingFunction][\tau]}
We seek to establish a lower bound on the \gls{abk:soc} at vertex $\Vertex \in \SetOfVertices$ for each label $\Label \in \SetOfLabels_\Vertex$, such that we can discard $\Label$ if $\CostProfileMaxSoC$ falls below this bound.
We express this bound using an auxiliary function $\texttt{ub}(\Period, \Period', \SetOfTours[]')$, which gives an upper bound on the maximum \gls{abk:soc} rechargeable in the interval $\lbrack\SetOfPeriods_i,\SetOfPeriods_j\rbrack$, $1 \leq i < j \leq |\SetOfPeriods|$, when servicing operations $\SetOfTours[]' \subseteq \SetOfTours$:
\begin{equation*}
	\texttt{ub}(\SetOfPeriods_i, \SetOfPeriods_j, \SetOfTours[]') \coloneqq \MaxChargingRate \cdot (j - i - \sum_{\Tour \in \SetOfTours[]'} (\TourDuration + 1)) \cdot \PeriodDuration - \sum_{\Tour \in \SetOfTours[]'} \TourConsumption.
\end{equation*}
In other words, $\texttt{ub}(\SetOfPeriods_i, \SetOfPeriods_j, \SetOfTours[]')$ corresponds to the \gls{abk:soc} reached when charging with the maximum charging rate $\MaxChargingRate$ in all non-service periods, accounting for service-related consumption. With this in mind, the following set of conditions needs to hold for a path labeled with label $\Label$ to be feasibly extensible to the sink:
\begin{equation*}
	\forall\Tour \in \SetOfTours \setminus \LabelVisitedTours: \overbrace{\texttt{ub}(\PeriodOfVertex[\Vertex], \LatestTourDeparture, \underbrace{\{\Tour' \in \SetOfTours \setminus \LabelVisitedTours \mid \Tour' \TourPrecedence \Tour\}}_{\mathclap{\text{not yet serviced operations that precede }\Tour}})}^{\mathclap{\text{maximum SoC reachargeable before }\LatestTourDeparture}} + \CostProfileMaxSoC \geq \TourConsumption.
\end{equation*}
Here, $\TourPrecedence$ is a precedence relation between operations $\Tour_a, \Tour_b \in \SetOfTours$, i.e., it indicates whether $\Tour_a$ has to be served before $\Tour_b$, formally, $\Tour_a \TourPrecedence \Tour_b \Leftrightarrow \EarliestTourDeparture[\Tour_b] + \TourDuration[\Tour_b] > \LatestTourDeparture[\Tour_a]$.
Essentially, for each $\Tour \in \SetOfTours$ not yet serviced, we check whether the residual \gls{abk:soc} and the maximum \gls{abk:soc} reachable before having to leave for $\Tour$ jointly suffice to service operation $\Tour$.

\paragraph{Potential functions:}
We recall that our label-setting search extracts the label with the least cost in each iteration. Accordingly, the algorithm is biased towards schedules that charge close to service operations, which negatively impacts the schedule diversity and may lead to increased computational times. We attenuate the impact of this effect by computing a lower bound on the cost required to feasibly reach the sink in a potential function $h(\Label, \Vertex)$ for a label $\Label \in \SetOfLabels_{\Vertex}$ at vertex $\Vertex$:
\begin{equation}
	\label{eq:potential-function}
	\begin{aligned}
		h(\Label, \Vertex) \coloneqq \CostProfileMinCost + \max(0, (\overbrace{\sum_{\Tour \in \SetOfTours \setminus \LabelVisitedTours} \TourConsumption}^{\mathclap{\text{remaining consumption}}}) - \CostProfileMinSoC) \\
		\cdot \bigg(\underbrace{\min_{q'\in [\MinSoC, \MaxSoC]} \frac{\partial \WearDensityFunction}{\partial q}(q')}_{\mathclap{\text{minimum battery degradation cost}}} + \overbrace{\min_{p' \in \SetOfPeriods, p' \geq \PeriodOfVertex[\Vertex]} \PeriodEnergyPrice[{p'}]}^{\mathclap{\text{minimum energy price}}}\bigg) + c(\Vertex).
	\end{aligned}
\end{equation}
Equation~\eqref{eq:potential-function} comprises three cost components. First, the cost required to feasibly reach $\Vertex$. Second, a lower bound on the charging-related costs required to reach the sink. Third, the network-related cost $c(\Vertex)$ of the shortest path to the sink vertex. 
\paragraph{Extended feasibility check:}

We perform the following additional checks when testing label feasibility to detect energy infeasible or service-incomplete schedules early, reduce label redundancy, and avoid superfluous iterations:
\begin{enumerate}
	\setlength{\itemsep}{0pt}
	\setlength{\parskip}{0pt}
	\item We terminate the search when we extract a label with positive cost at any vertex. This remains correct as our algorithm extracts labels in order of lowest cost, such that path cost monotonically increases due to non-positive station capacity duals. Hence, the corresponding vehicle schedule cannot have negative reduced cost.
	\item We propagate service arcs only if the associated operation has not been covered yet.
	\item We discard labels $\Label$ extracted at vertex $\OriginVertex$ if $\PeriodOfVertex[\OriginVertex] > \LatestTourDeparture$ for any $\Tour \notin \LabelVisitedTours, \Tour \in \SetOfTours$. We relax this strict inequality for non-garage vertices, i.e., $\PeriodOfVertex[\OriginVertex] \geq \LatestTourDeparture$.
	\item We do not propagate charging arcs if a label's minimum reachable \gls{abk:soc} suffices to cover all operations, such that additional charging is superfluous.
\end{enumerate}

\section{Branch-and-Price}
\label{sec:branch-and-price}

	We embed our column generation procedure into a \gls{abk:bnb} algorithm to address cases where the \gls{abk:rmp}'s final solution is fractional. The resulting \gls{abk:bnp} algorithm relies on a problem-specific branching rule (Section~\ref{sec:branching}), partial pricing (Section~\ref{sec:partial-pricing}), a primal heuristic (Section~\ref{sec:primal-heuristics}), and uses a two-stage vertex selection strategy (Section~\ref{sec:node-selection}).
\subsection{Branching rule}
\label{sec:branching}

Our branching rule bases on the observation that a basic feasible solution of the \gls{abk:rmp} is only fractional if charger capacity constraints are binding. In other words, when two or more schedules compete for a charger that is already at capacity.

\begin{restatable}{proposition}{branching}
	\label{prop:fractional-only-if-charger-capacity}
	Let $\sigma$ be a fractional basic feasible solution to the \gls{abk:rmp}. Then it holds that $\exists (\Period, \Charger) \in \SetOfPeriods \times \SetOfChargers$ such that $\sum_{\ScheduleUsed \in \sigma} \ScheduleUsed \ScheduleMatrix_{\Period, \Charger} = \ChargerCapacity$, and there exist at least two $\ScheduleUsed \in \sigma$ with $0 < \ScheduleUsed < 1$.
\end{restatable}

\proof see Appendix~\ref{app:proofs}.

We refer to pairs $(\Period, \Charger) \in \SetOfPeriods \times \SetOfChargers$ where this occurs as \emph{conflicts} and resolve these by creating a new branch for each vehicle $\Vehicle \in \SetOfVehicles$ that participates in the conflict, formally where $\sum_{\Schedule \in \ReducedSetOfVehicleSchedules} \ScheduleUsed \ScheduleMatrix_{\Period, \Charger} > 0$. Each of these branches cuts off solutions where vehicle $k$ uses charger $\Charger$ in period $\Period$. We enforce these constraints in our subproblems to avoid additional dual variables in our master problem. For this purpose, we purge all violating columns from $\ReducedSetOfVehicleSchedules$ and remove the station vertex representing the charging opportunity from $\TimeExpandedNetwork$.

Our algorithm relies on a hierarchial selection strategy to resolve cases with multiple conflicts. 
Specifically, we prioritize conflicts $(\Period, \Charger) \in \SetOfPeriods \times \SetOfChargers$ by i) the most fractional participating schedule, ii) the number of non-integral columns, iii) the amount of energy recharged, iv) charging speed of $\Charger$, and v) the lowest vehicle index.

\subsection{Partial pricing}
\label{sec:partial-pricing}
Schedules generated in the same iteration often show similar charging patterns due to shared dual variables, overlapping time windows, and equal energy prices, such that charger capacity constraints limit the number of simultaneously generated schedules that can be part of the same basic solution. Hence, many of the schedules generated do not contribute to the convergence of the dual variables and the lower bound, especially when considering large fleets and a high number of low-capacity chargers such that limiting the number of vehicles priced in each iteration often reduces total runtime, even if this requires additional iterations of the column generation procedure.

We implement this so-called \textit{partial pricing} approach as follows:
After solving the \gls{abk:rmp}, we solve the arising pricing subproblems in a round-robin fashion until a total number of $\PartialPricingScheduleCount$ schedules with negative reduced costs have been generated or all subproblems have been considered. Our termination criterion remains unchanged, that is, we still terminate the column generation procedure when no subproblem produces a schedule with negative reduced cost.
The main factor that drives the computational effectiveness of this approach is the number of columns to generate, i.e., $\PartialPricingScheduleCount$. Our experiments show that aligning $\PartialPricingScheduleCount$ with the minimum charger capacity performs best.

One of the major drawbacks of partial pricing is that the lower bound generated at each iteration is relatively weak \citep[cf.][]{DesrosiersLuebbecke2005}. We address this issue by periodically performing a full iteration, i.e., temporarily set $\PartialPricingScheduleCount = |\SetOfVehicles|$.
\subsection{Primal heuristic}
\label{sec:primal-heuristics}
We utilize a primal heuristic to quickly find upper bounds, thus speeding up the solution procedure by allowing to prune nodes of the \gls{abk:bnb} tree early. To this end, we use a diving heuristic that explores an auxiliary branch-and-bound tree in a depth-first fashion, branching on the variables $\ScheduleUsed$ of the extensive formulation (\ipref{mips:MP-mip}).
At each node of this auxiliary \gls{abk:bnb} tree, the algorithm forces some fractional $\ScheduleUsed$ to one until an integral or infeasible solution is found. 
We rely on strong branching to boost the success rate and solution quality of this diving algorithm: at each node explored in the diving phase, we bound the impact of fixing a candidate schedule by solving the accordingly modified \gls{abk:rmp}. We then fix the schedule promising the most improvement based on the derived lower bounds.

As this procedure is expensive for large numbers of schedules, our algorithm considers only a small subset of schedules at each node. We select these from the column set at the current node ($\ReducedSetOfSchedules^N$) according to a roulette wheel criterion based on \textit{dissimilarity} and \textit{quality}. For this purpose, we define the distance $d(\Schedule_1, \Schedule_2)$ between two columns $\Schedule_1, \Schedule_2 \in \ReducedSetOfSchedules^N_\Vehicle$ as the hamming distance between charger allocation matrices $\{\ScheduleMatrix[\Schedule_1]_{\Period, \Charger} \mid \Period \in \SetOfPeriods, \Charger \in \SetOfChargers\}$ and $\{\ScheduleMatrix[\Schedule_2]_{\Period, \Charger} \mid \Period \in \SetOfPeriods, \Charger \in \SetOfChargers\}$. Accordingly, the distance between two schedules is inversely proportional to the number of shared charging operations between the two schedules. For each vehicle without an assigned schedule, we then build a pool of columns $\Theta_\Vehicle$ as follows:
Let $r_Q(\Schedule)$ be the \textit{rank} of $\Schedule$ in $\ReducedSetOfSchedules^N_\Vehicle$ according to its schedule's cost. Furthermore, let $D(\Schedule, \Theta_\Vehicle)$ be the average distance to all $\Schedule' \in \Theta_\Vehicle$ or $0$ if $\Theta_\Vehicle = \emptyset$. Again, $r_D(\Schedule, \Theta_\Vehicle)$ gives the rank of $\Schedule$ according to $D(\cdot)$. We then build $\Theta_\Vehicle$ greedily such that Equation~\eqref{eq:adaptive-column-weight} is maximized:
\begin{equation}
	\label{eq:adaptive-column-weight}
	F(\Schedule, \Theta_\Vehicle) \coloneqq \alpha r_Q(\Schedule) + (1 - \alpha) r_D(\Schedule, \Theta_\Vehicle).
\end{equation}
We stop once the column pool reaches the desired size ($\bar{\Theta}_{\Vehicle}$).

We execute this primal heuristic on every node of the \gls{abk:bnb} tree until the algorithm finds an integral solution. We switch to a periodic evaluation strategy at this point and run our heuristic only on every $|\SetOfVehicles|\textsuperscript{th}$ node.

\subsection{Node selection strategy}
\label{sec:node-selection}
We place the nodes generated by our branching procedure into a node queue. The order of this queue dictates the order in which open nodes are solved and hence influences the performance of the \gls{abk:bnp} algorithm. Our node selection strategy combines the benefits of the two most common branching strategies, namely \textit{depth-first}, which prioritizes nodes with the highest distance from the root node, and \textit{best-bound-first}, which selects nodes according to their lower bounds \citep[cf.][]{Wolsey1998}, in a two-stage approach. We first select nodes according to the depth-first strategy, breaking ties by vehicle index, and switch to best-bound-first when we find a feasible integer solution during branching or using our primal heuristic. Here, we order nodes with equal lower bounds by node depth, prioritizing nodes deeper in the tree. We resolve any remaining ties by picking the leftmost node.
\section{Design of experiments}
\label{sec:instances}

The aim of our computational study is twofold:
first, we validate the correctness, investigate the performance, and analyze the scalability of our \gls{abk:bnp} algorithm in a set of numerical experiments.
These benchmark our algorithm against a \gls{abk:mip}-based formulation (cf. Appendix~\ref{sec:compact-formulation}), on a set of small, randomly generated instances.
We further test our algorithm on a set of larger instances, where we assess the impact of various instance parameters (cf. Table~\ref{table:param-exp-1-scale}), on our algorithm's runtime.
Second, we assess the impact of integrated charge and service operation scheduling in a potential real-world scenario. Here, we analyze to which extent flexible service operations improve overall vehicle utilization, investigate how service flexibility affects the amount of charging infrastructure required, and assess the impact of energy prices on the cost savings obtainable through integrated charge and service scheduling.
To keep this paper concise, we refer to Appendix~\ref{app:benchmark-instance-generation} for details on the instance derivation of our numerical experiments and focus on the design of our managerial study in the following:
\begin{table}[b]
	\centering
	\caption{Instance generation parameter values used in the numerical experiments. \label{table:param-exp-1-scale}}
	\begin{threeparttable}
		{\footnotesize{
				\begin{tabular}{cccccccc}
					\toprule
					& \makecell{Fleet size} & \makecell{Planning horizon\\length (days)} & \makecell{Time window\\length (periods)} & \makecell{Charger\\count} & \makecell{Total charger \\capacity} & \makecell{Segment\\ count $(\WearDensityFunction)$} & \makecell{Segment\\ count $(\Phi_f)$}\\
					\midrule
					Small & 3 & 1 & 6 & 1 & 1 & 3 & 3 \\
					\midrule
					Base & 12 & 2 & 4 & 2 & 6 & 4 & 3\\
					Min & 12 & 1 & 0 & 1 & 6 & 2 & 2\\
					Max & 68 & 5 & 8 & 6 & 12 & 8 & 8\\
					Step & 8 & 1 & 1 & 1 & 1 & 1 & 1\\
					\bottomrule
				\end{tabular}	
		}}
		{}
	\end{threeparttable}
\end{table}
\label{subsec:base-case-generation}
\begin{table}[t]
	\begin{subtable}[T]{0.32\textwidth}
		\centering
		\subcaption{\centering Considered \gls{abk:wdf}.\label{table:wdf-param}}
		\begin{threeparttable}
			{\footnotesize{
					\begin{tabular}{ccc}
						\toprule
						SoC & Cost & Unit cost \\
						\midrule
						$25\%$ & $1.59$\euro&$0.14$\euro\\
						$50\%$ & $3.30$\euro&$0.15$\euro\\
						$75\%$ & $5.20$\euro&$0.17$\euro\\
						$100\%$ & $7.79$\euro&$0.23$\euro\\
						\bottomrule
						Avg. & - & $0.17$\euro\\
					\end{tabular}	
			}}
			{}
		\end{threeparttable}
	\end{subtable}
	\begin{subtable}[T]{0.66\textwidth}
		\centering
		\subcaption{Considered charging functions.\label{table:station-parameters}}
		\begin{threeparttable}
			{\footnotesize{
					\begin{tabular}{cc|cc}
						\toprule
						\multicolumn{2}{c|}{Fast} & \multicolumn{2}{c}{Slow}\\
						\midrule
						Time (minutes) & SoC & Time (minutes) & SoC \\
						\midrule
						$72.32$&$34.90$&$435$&$45.0$\\
						$92.6$&$42.49$&&\\
						$120$&$45.00$&&\\
						\bottomrule
						Avg. rate&$22.5$ kW/h&Avg. Rate&$6.21$ kW/h\\
						Full charge&$2$ h&Full charge&$7.25$ h\\
					\end{tabular}	
			}}
			{}
		\end{threeparttable}
	\end{subtable}
	\caption{Wear-cost density function and charging infrastructure considered in the case study.}
\end{table}

Our managerial study captures the planning problem of a \gls{abk:lsp} supplying retail stores in an urban area over a planning horizon of two days. We assume a fleet of sixteen vehicles, each operating three shifts a day, i.e., one night, one morning, and one afternoon shift. Each shift comprises a delivery tour that takes six hours and consumes 15 kWh of energy, such that the battery is depleted once a day. We randomly distribute operation departure times such that vehicles spend a minimum of one hour before each operation at the depot. We extend this slack to two hours for the first operation assigned to each vehicle and distribute earliest and latest departure times symmetrically around the departure time of the case without any service flexibility.
Each vehicle is equipped with a 45 kWh battery, purchased at a price of \euro5.406. This corresponds to an average price per kWh of \$120/kWh \citep{BloombergNEF} at an exchange rate of 0.877\euro/\$. We use the battery wear data provided in \cite{HanHanEtAl2014} calibrated according to our battery price to derive the \gls{abk:wdf} (cf. Table~\ref{table:wdf-param}). Energy prices are based on the hourly day-ahead spot market price for the 16.01.2021 and 17.01.2021 \citep{entsoe2021}, linearly interpolated to fit 30-minute time steps. We scale the reported prices according to the average European energy price of \euro$0.2127$ in 2020 \citep{eurostatTEN00117}.
As we aim to capture potential charging trade-offs during off-service periods, our planning horizon starts with the previous day's end-of-operations (22:00), assuming an initially empty battery.
Charging may be conducted at the depot using one of two different charger types: a CCS 50 kW DC fast charger, capable of delivering a full 45 kWh charge in around two hours~(cf. Table~\ref{table:station-parameters}), and a slow charger with an average charging rate of around six kWh that recharges the battery in around seven hours. These slow chargers match the standard level 2 single-phase AC on-board chargers of most \glspl{abk:ecv} \citep[cf. Table~\ref{table:station-parameters},][]{IEC61851-1}. We hence assume that one of the latter is available for each vehicle. For the DC fast charger, our basecase assumes a capacity of six. 

\section{Results and discussion}
\label{sec:results}
We conducted all of our experiments on a standard desktop computer equipped with an \texttt{Intel(R) Core(TM) i9-9900, 3.1 GHz CPU} and \texttt{16 GB} of \texttt{RAM}, running \texttt{Ubuntu 20.04}. We have implemented the \gls{abk:bnp} algorithm in \texttt{Python (3.8.11)} using IBM CPLEX (Version 20.01) to solve the \gls{abk:rmp}. The pricing problem is implemented in \texttt{C++} (GCC~11.1.0). 
We considered solutions with a relative gap of less than 0.0001 optimal and ran all our experiments in a single thread. We further limited the computational time to 3600 seconds. Setup time is negligible and thus not reported. We refer to Appendix~\ref{app:online} for instances and detailed results.
\subsection{Computational performance}
We benchmark our algorithm against the \gls{abk:mip} proposed in Appendix~\ref{sec:compact-formulation} on a set of $50$ small instances, generated according to Appendix~\ref{app:benchmark-instance-generation} and the parameter values listed as \textit{small} in Table~\ref{table:param-exp-1-scale}. Table~\ref{table:results-mip-vs-bp-summary} summarizes our results and compares runtime, bounds, gap, size of the \gls{abk:bnb} tree, the total number of instances, and the number of instances solved to optimality. 
\begin{table}[t]
	\centering
	\caption{Aggregated results for the small benchmark instances. \label{table:results-mip-vs-bp-summary}}
	\begin{threeparttable}
		\small
		\begin{tabular}{lrrrrrrr}
			\toprule
			{} &  Avg. t[s] &  Avg. obj. &  Avg. LB &  Avg. \#nodes & \#optimal & \#unsolved & \#total \\
			Algorithm        &                   &                 &             &                       &          &&        \\
			\midrule
			Branch \& Price &              0.85 &          396.56 &      396.56 &                  4.98 &       50 & 0 &    50 \\
			MIP              &           3600.00 &          397.46 &      114.74 &            2183022.26 &        1 & 2 &    50 \\
			\bottomrule
		\end{tabular}
		\begin{tablenotes}[flushleft]
			\item\smaller Abbreviations hold as follows: t[s] - runtime in seconds, obj. - objective value, LB - lower bound, \#nodes - size of the \gls{abk:bnb} tree, \#optimal - number of optimally solved solutions, \#unsolved - number of instances where no incumbent was identified, \#total - total number of instances. Averages do not include unsolved instances.
		\end{tablenotes}
	\end{threeparttable}
\end{table}
As can be seen, the proposed \gls{abk:bnp} algorithm outperforms the \gls{abk:mip} on all instances, proving optimality in less than a second on average.
 
\begin{figure*}[p]
	\centering
	\begin{minipage}[t][5.5cm][t]{0.48\textwidth}
		\vspace{0pt}
		\begin{subfigure}{\textwidth}
			\includegraphics[width=1\textwidth]{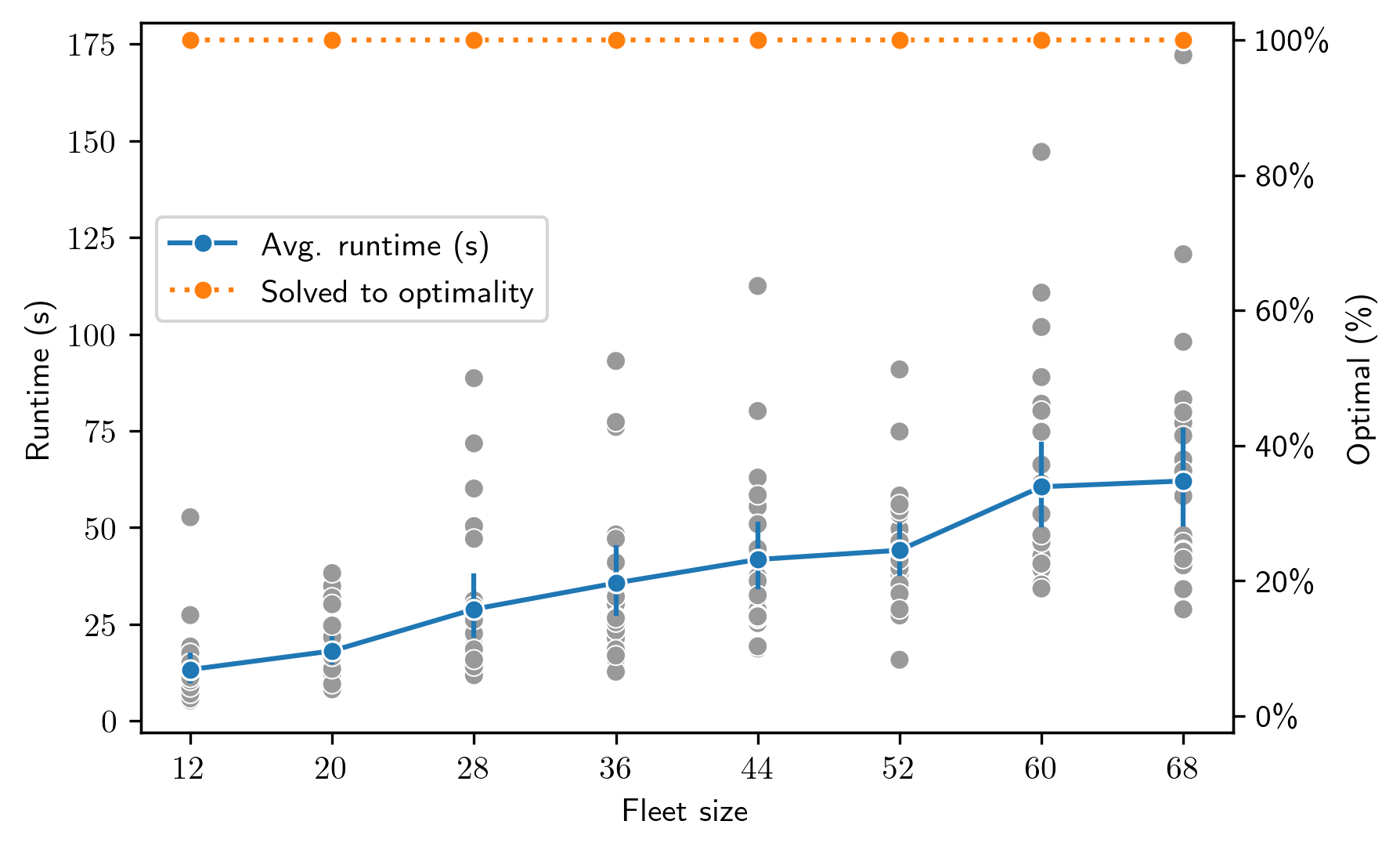}
			\subcaption{\centering Runtime for varying fleet sizes. \label{fig:results-scalability-fleet-size}}
		\end{subfigure}
	\end{minipage}%
	\hspace{.03\textwidth}%
	\begin{minipage}[t][5.5cm][t]{0.48\textwidth}
		\vspace{0pt}
		\begin{subfigure}{\textwidth}
			\includegraphics[width=1\textwidth]{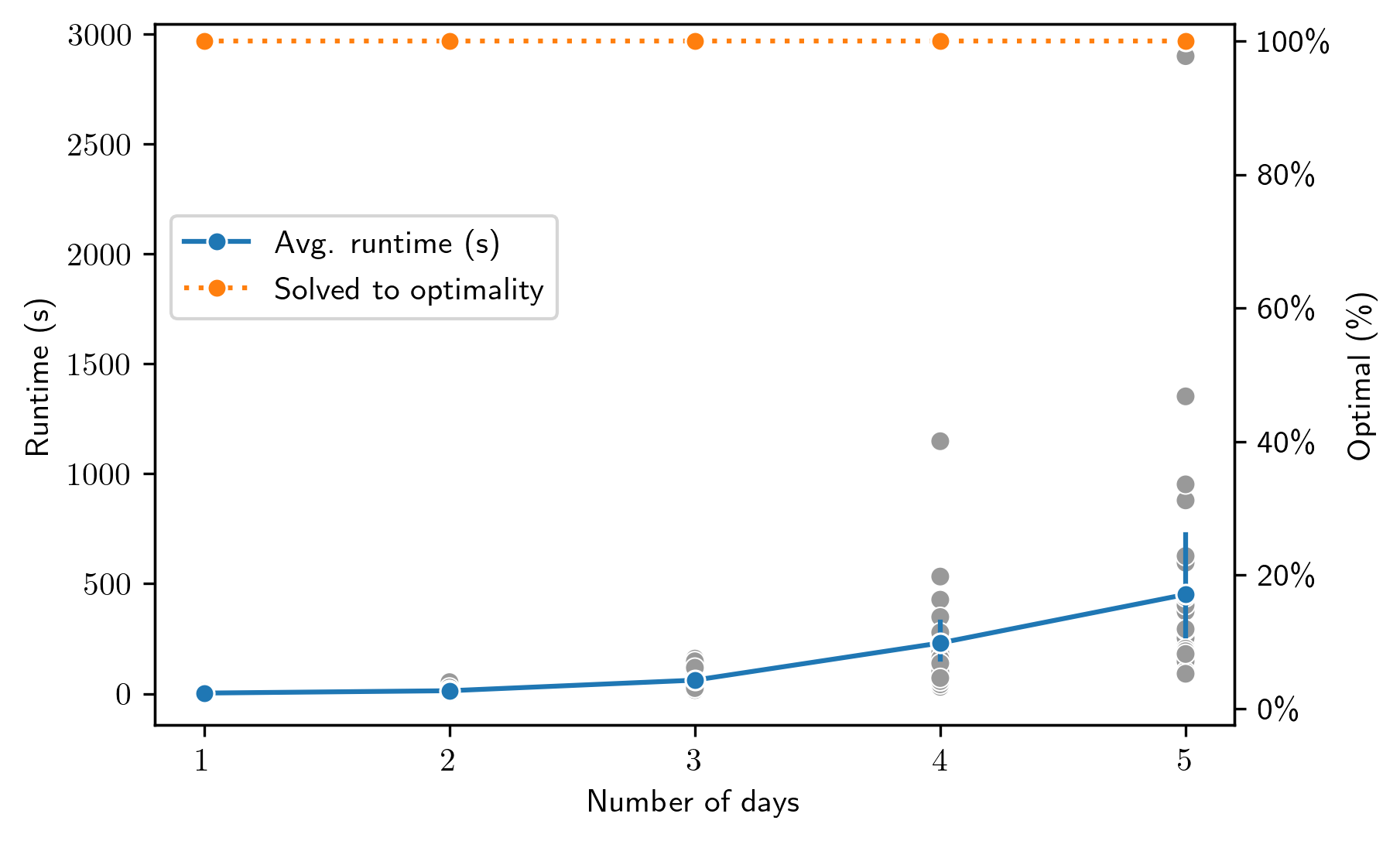}
			\subcaption{\centering Runtime for varying planning horizon length.\label{fig:results-scalability-number-of-days}}
		\end{subfigure}
	\end{minipage}%
	
	\begin{minipage}[t][6cm][t]{0.48\textwidth}
		\vspace{0pt}
		\begin{subfigure}{\textwidth}
			\includegraphics[width=1\textwidth]{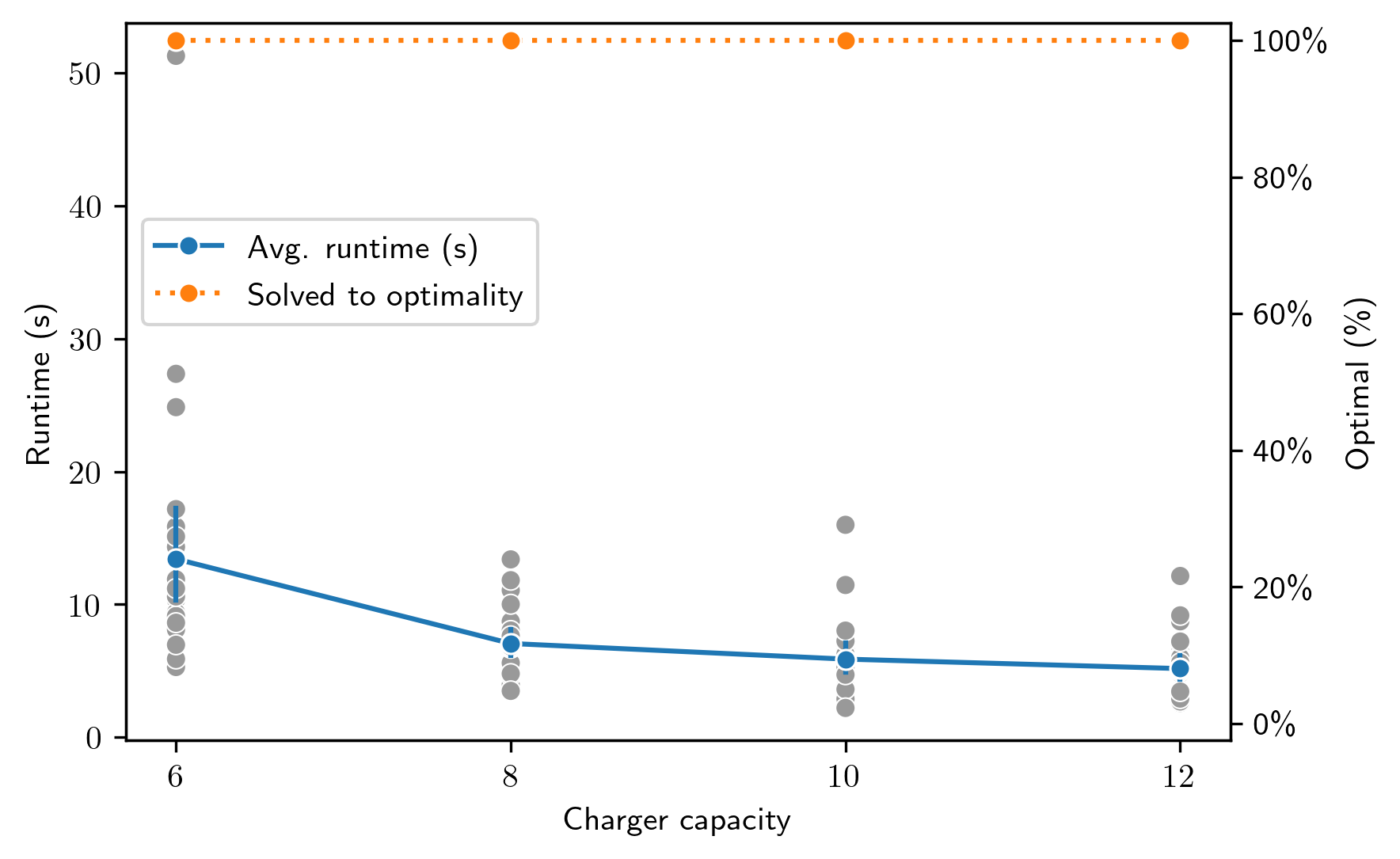}
			\subcaption{\centering Runtime for varying charger capacities.\newline \label{fig:results-scalability-base-charger-capacity}}
		\end{subfigure}
	\end{minipage}%
	\hspace{.03\textwidth}%
	\begin{minipage}[t][6cm][t]{0.48\textwidth}
		\vspace{0pt}
		\begin{subfigure}{\textwidth}
			\includegraphics[width=1\textwidth]{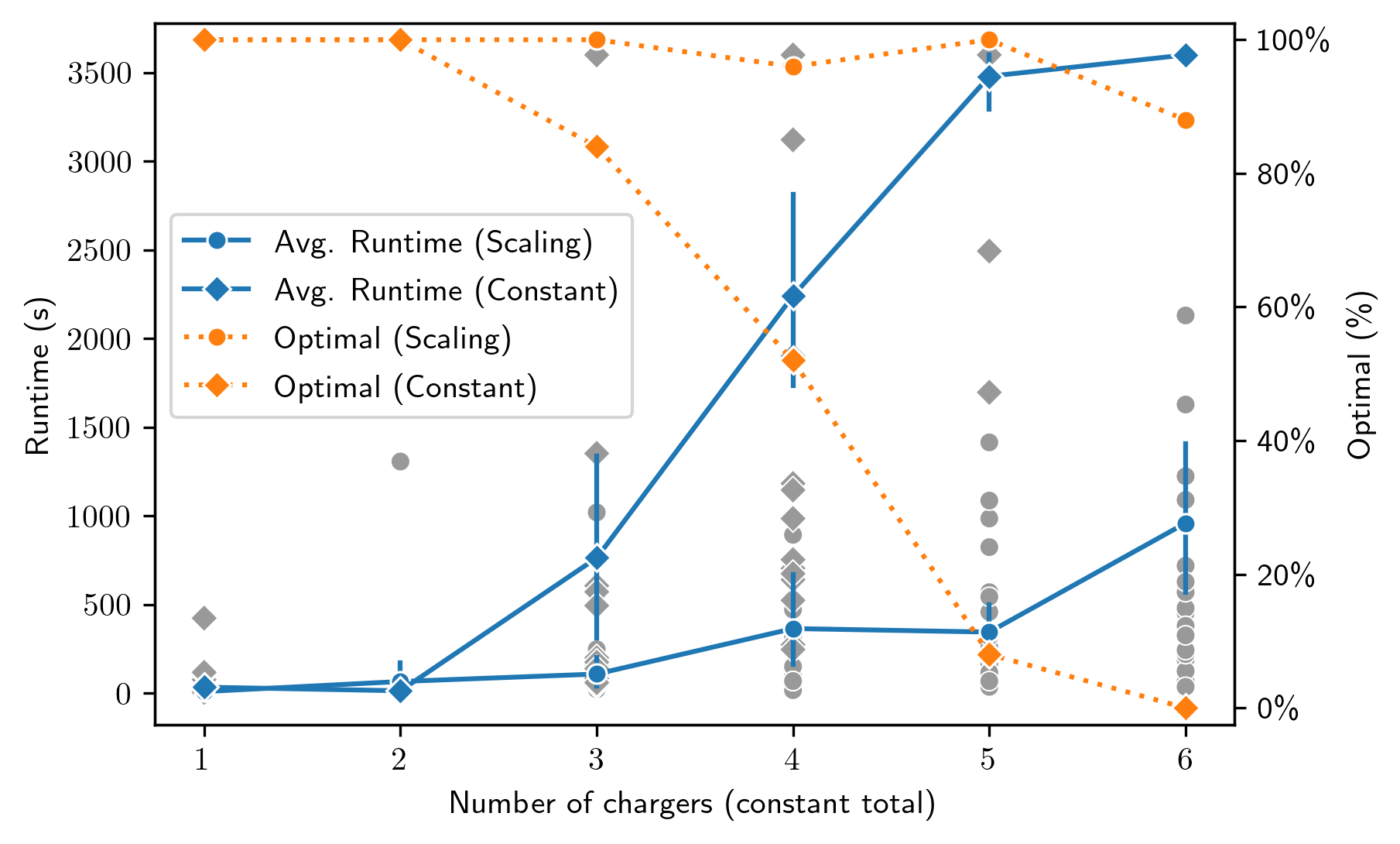}
			\subcaption{\centering Runtime for varying numbers of distinct charger types.\label{fig:results-scalability-number-of-charger-types}}
		\end{subfigure}
	\end{minipage}
	
	\begin{minipage}[t][6.5cm][t]{0.48\textwidth}
		\vspace{0pt}
		\begin{subfigure}{\textwidth}
			\includegraphics[width=1\textwidth]{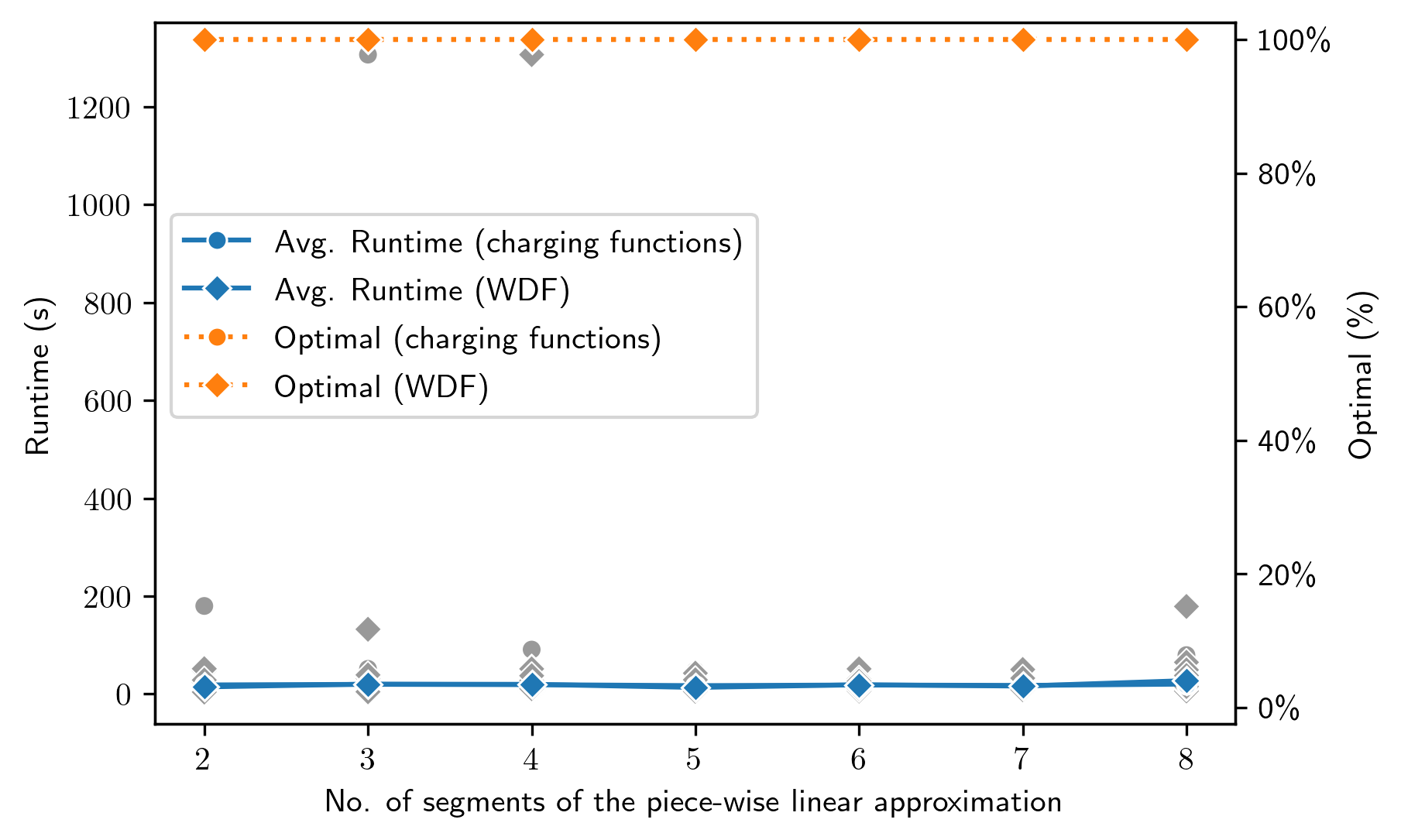}
			\subcaption{\centering Runtime for varying degree of approximation\newline quality.\label{fig:results-scalability-approximation-complexity}}
		\end{subfigure}
	\end{minipage}
	\hspace{.03\textwidth}%
	\begin{minipage}[t][6.5cm][t]{0.48\textwidth}
		\vspace{0pt}
		\begin{subfigure}{\textwidth}
			\includegraphics[width=1\textwidth]{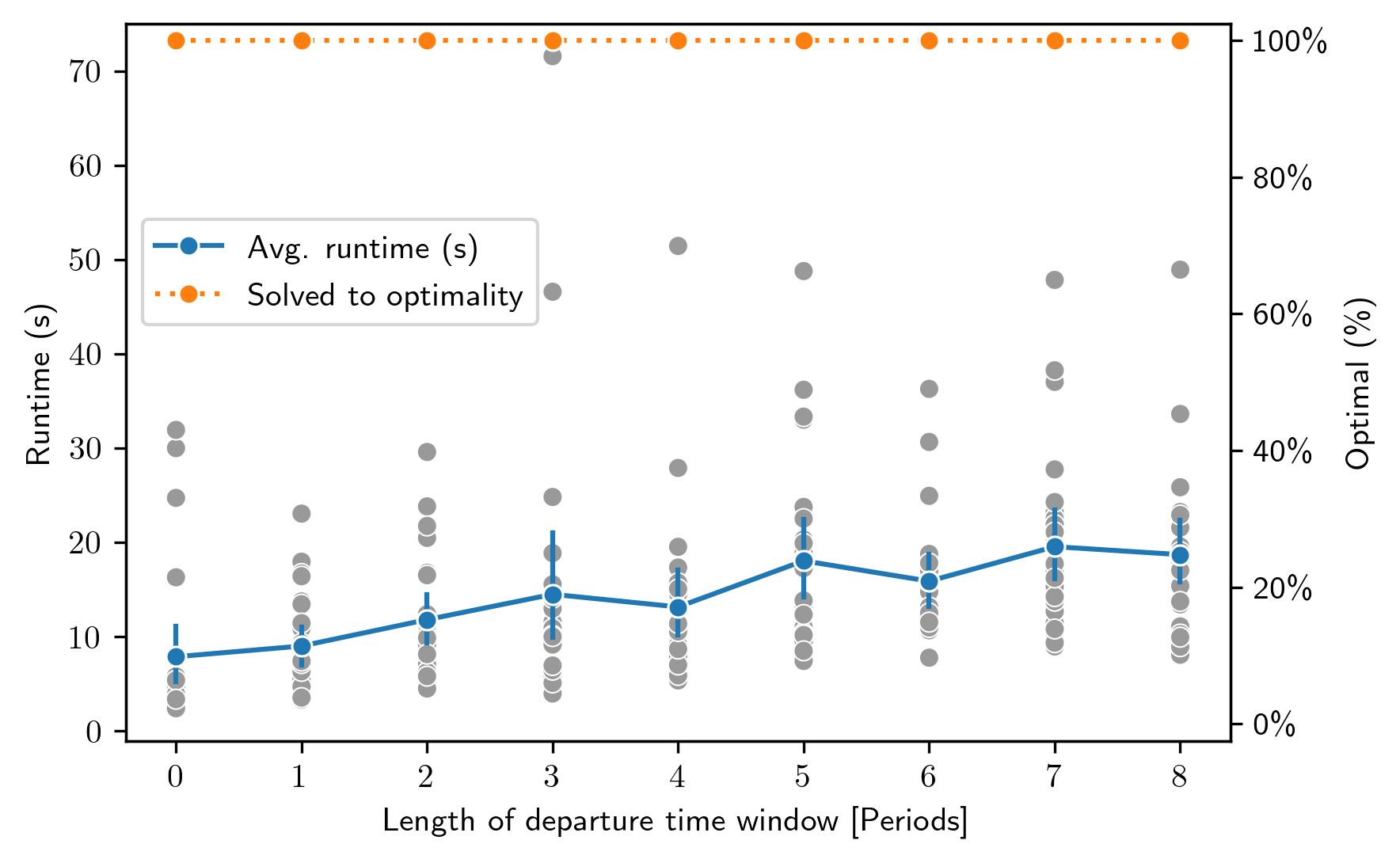}
			\subcaption{\centering Runtime for varying departure time\newline window size.\label{fig:results-scalability-time-window-size}}
		\end{subfigure}
	\end{minipage}
	\begin{minipage}{\textwidth}
		\small \emph{Note.} Each gray dot represents a specific instance. The solid blue line corresponds to the average runtime in seconds. The dashed orange line shows the percentage of instances where the algorithm did not prove optimality within 3600 seconds.
	\end{minipage}
	\caption{Results for the large benchmark instances. \label{fig:results-scalability}}
\end{figure*}

In the remainder of this section, we focus on the performance of our \gls{abk:bnp} algorithm on larger instances. Here, we generate a total of $25$ instances according to Appendix~\ref{app:benchmark-instance-generation}, varying each parameter in Table~\ref{table:param-exp-1-scale} separately, such that our study comprises a total of $1300$ runs.
Figure~\ref{fig:results-scalability} summarizes our results on these.
\begin{paragraphs}
\item[\emph{Impact of fleet size (Figure~\ref{fig:results-scalability-fleet-size}):}]  Our results indicate a linear increase in runtime with increasing fleet size, which results from the number of subproblems scaling linearly with the number of vehicles. Note that we keep the ratio of vehicles to charger capacity constant in this experiment to isolate the effect of increasing fleet size.

\item[\emph{Impact of planning horizon length (Figure~\ref{fig:results-scalability-number-of-days}):}]  Our results indicate a strong correlation between planning horizon length and the average runtime. We can attribute this to the computational complexity of the pricing subproblem: longer planning horizons increase the network size linearly, such that the number of generated labels increases exponentially in the worst case. Additionally, longer planning horizons increase the probability of charger conflicts such that the number of branches increases and more iterations are necessary to solve the individual \gls{abk:bnb} nodes.

\item[\emph{Impact of charger capacity (Figure~\ref{fig:results-scalability-base-charger-capacity}):}] Here, the runtime decreases with increasing charger capacity. This can be attributed to lower charger contention, which accelerates the convergence of both the branch and bound and column generation procedures.
	
	\item[\emph{Impact of the number of charger types (Figure~\ref{fig:results-scalability-number-of-charger-types})}:]
	We consider a \textit{scaling} and a \textit{constant} setting to isolate the effect of the number of charger types on runtime.
	In the scaling setting, we increase the total charger capacity proportional to the number of chargers added, such that adding a charger always increases the total number of available charging spots. In the constant setting, we instead distribute the base case charger capacity across available chargers such that the total capacity remains constant.
	In both experiments, the average runtime increases with the number of chargers available as the size of the pricing network increases. When keeping the total capacity constant, increasing the number of chargers also increases charger contention as vehicles compete for more chargers with lower individual capacities. Together, these effects cause an exponential runtime increase. 
	Increasing the charger capacity proportional to the number of chargers weakens both effects. Here, runtime scales linearly with the number of chargers.
	
	\item[\emph{Impact of approximation quality (Figure~\ref{fig:results-scalability-approximation-complexity})}:] %
	This experiment asses the effect of increasing the number of breakpoints, and thus the approximation quality, of our piecewise linear functions. Recall that the number of created labels increases with the number of breakpoints of both the \gls{abk:wdf} and the charging functions (cf. Sections~\ref{sec:label-propagation}~\&~\ref{sec:committing-charging-operations}). Nevertheless, the runtime of our algorithm varies only slightly with increasing approximation quality. This underlines the strength of our dominance criteria.
	
	\item[\emph{Impact of departure time window size (Figure~\ref{fig:results-scalability-time-window-size})}:]
	By design of the time-expanded network, one would expect a substantial increase in computation time with increasing departure time window length: longer time windows add additional arcs and thus increase the number of feasible paths through the network. Nevertheless, our algorithm scales well to longer time windows, and we observe only a slight increase in runtime.
\end{paragraphs}
	
In conclusion, our algorithm manages to reliably solve instances with 68 vehicles or a planning horizon of 5 days within an hour, allowing for day-ahead planning in practice.
\subsection{Managerial study}
\begin{figure}[t]
	\captionsetup[subcaption]{width=0.9\textwidth}
	\centering
	\begin{subfigure}[t]{0.48\textwidth}
		\includegraphics[width=0.87\textwidth]{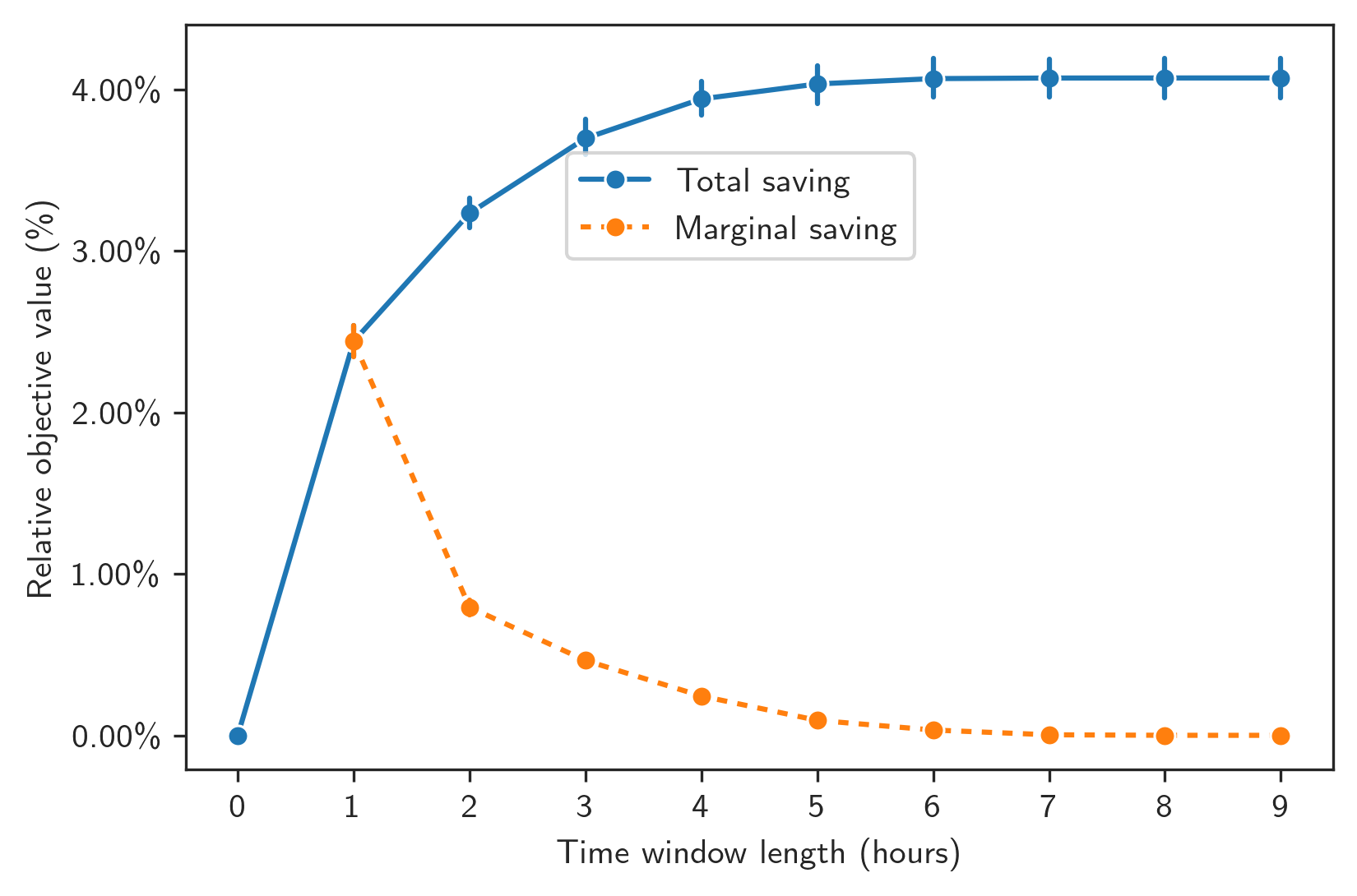}
		\subcaption{Relative objective value savings.\label{fig:basecase-results-objective}}
		\begin{minipage}{0.9\textwidth}
			\smaller\textit{Note.} The solid blue line shows the total objective value saving relative to the static scenario. The dashed orange line shows the marginal saving.\vspace{3ex}
		\end{minipage}
	\end{subfigure}
	\hspace{.02\textwidth}%
	\begin{subfigure}[t]{0.48\textwidth}
		\includegraphics[width=\textwidth]{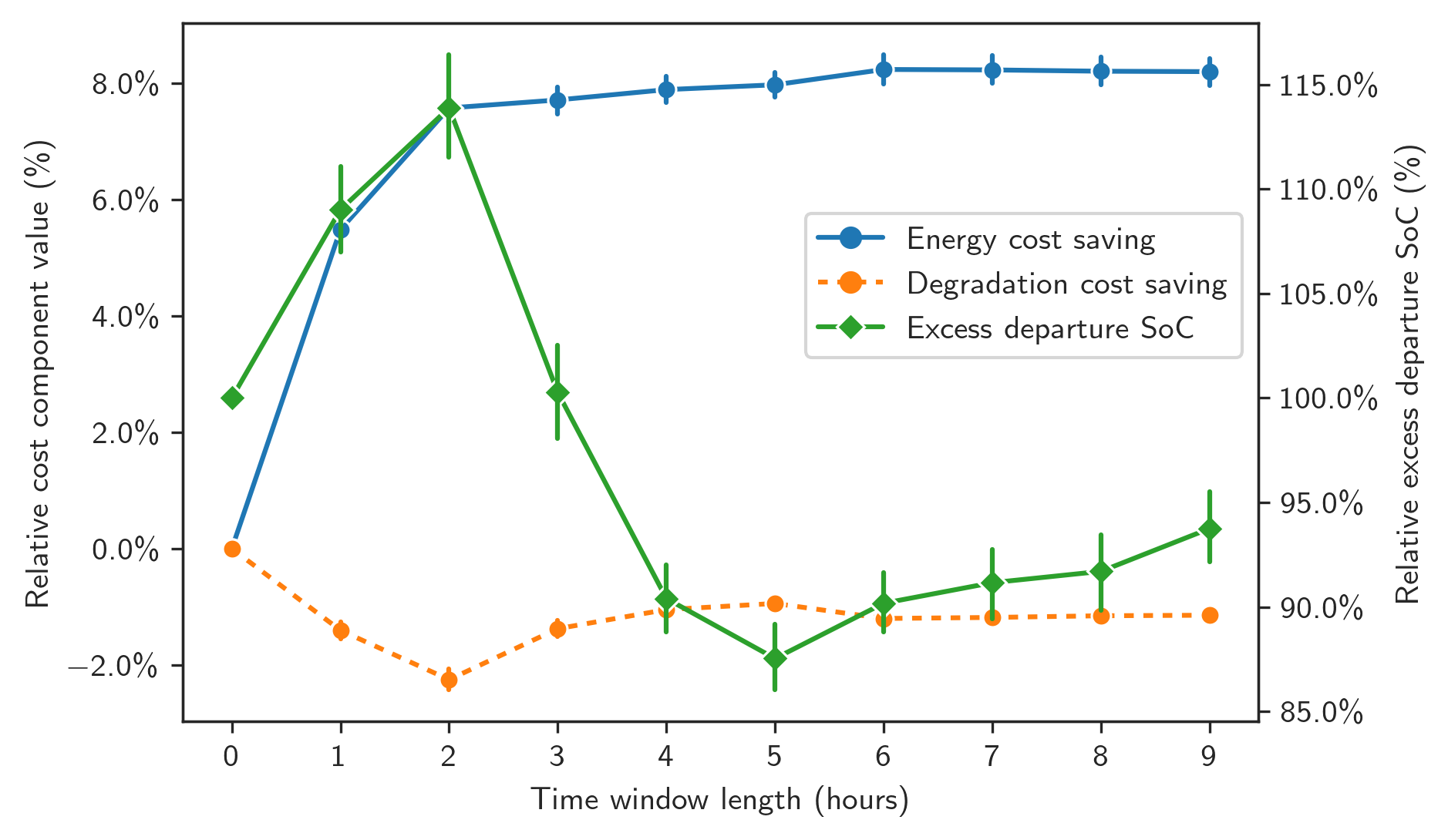}
		\subcaption{Cost components of relative savings.\label{fig:basecase-results-energy-and-degradation}}
		\begin{minipage}{0.9\textwidth}
			\smaller\textit{Note.} The solid blue and dashed orange lines show the average energy and degradation cost savings compared to the static scenario. The green line shows the average excess departure \gls{abk:soc}, i.e., the charge upon departure beyond $\TourConsumption$ averaged over all operations $\Tour \in \SetOfTours[]$, relative to the static scenario.
		\end{minipage}
	\end{subfigure}
	\caption{Results of the basecase experiment.\label{fig:basecase-results}}
\end{figure}

\paragraph{Impact on total cost savings}
This study aims to derive managerial insights on the impact of service operation flexibility on total cost savings. For this purpose, we have run the developed algorithm on a set of $50$ realistic instances derived according to Section~\ref{subsec:base-case-generation} with varying degrees of schedule flexibility, resulting in a total of $500$ runs. Figures~\ref{fig:basecase-results-objective} and~\ref{fig:basecase-results-energy-and-degradation} summarize our results.

Figure~\ref{fig:basecase-results-objective} shows the average total and marginal objective value savings of instances with varying time window lengths compared to the same instances with static departure times.
We observe that flexible service operations have an overall positive impact on the objective value. Specifically, allowing a service time window of one hour already yields a cost saving of $2.5\%$. The marginal saving decreases sharply with increasing flexibility, such that one hour of flexibility already exploits $50\%$ of the savings potential, while four hours of flexibility exploit $80\%$. The relative saving converges to $5\%$ at nine hours of flexibility.

\textsc{Result 1.} Flexible service operations have a positive impact on the total cost. The obtainable savings increase with increasing flexibility, converging to maximum cost savings of $5\%$. Marginal savings decrease with increasing time windows, such that one hour of flexibility already exploits $50\%$ of the savings potential, while four hours of flexibility yield $80\%$ of the obtainable savings.

Figure~\ref{fig:basecase-results-energy-and-degradation} shows the impact of increasingly flexible service operations on the individual cost components and excess departure \gls{abk:soc}. Here, we observe a trade-off between battery degradation and energy cost. Specifically, higher schedule flexibility allows utilizing periods with energy prices cheap enough to outweigh additional battery degradation costs caused by cycling the battery at higher \gls{abk:soc} levels, such that energy costs decrease while average excess departure \gls{abk:soc}, and thus battery degradation costs, increase.

\begin{figure}[b]
	\includegraphics[width=0.5\textwidth]{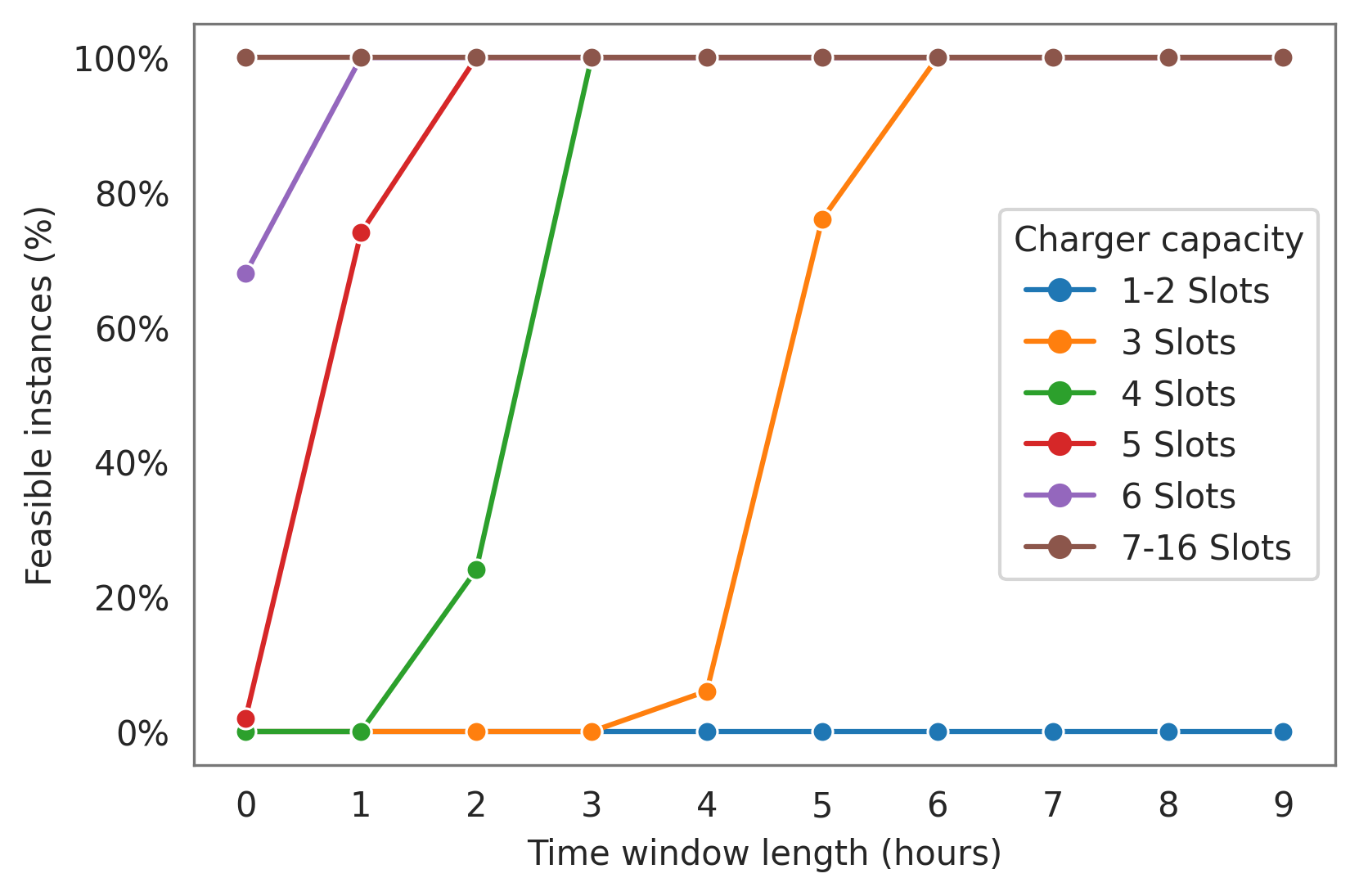}
	\caption{Percentage of feasible instances for varying charger capacity and time window length\label{fig:charger-capacity-results-feasiblity}.\vspace{-0.25cm}}
\end{figure}

\begin{figure}[t]
\includegraphics[width=\textwidth]{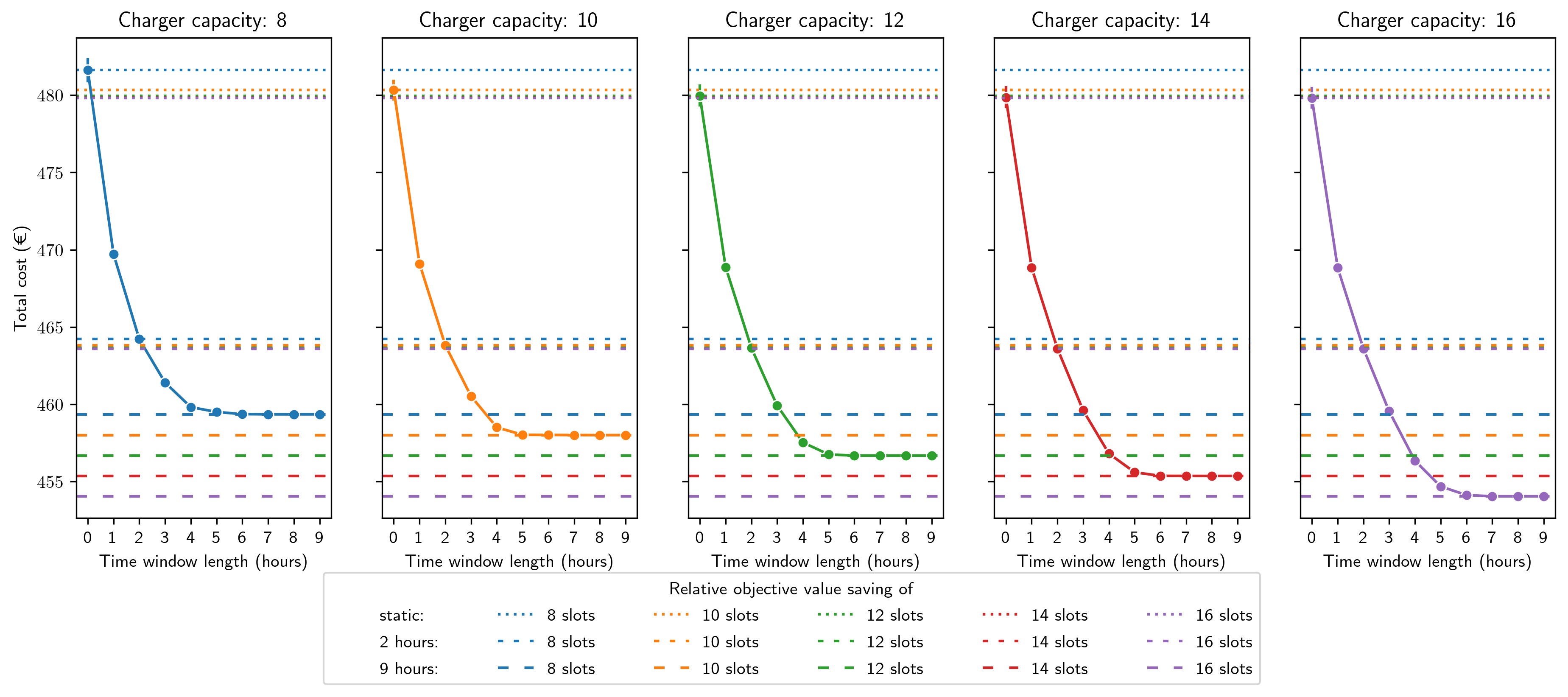}
\caption{Objective value for varying charger capacity and time window length\label{fig:charger-capacity-results-obj}.}
\begin{minipage}{0.9\textwidth}
	\smaller\textit{Note.} The dotted and dashed lines give the objective for different charger capacities assuming static time windows, a time window size of two hours, and nine hours, respectively.
\end{minipage}
\end{figure}

\textsc{Result 2.} Dynamic service operations allow utilizing the trade-off between battery degradation cost and energy prices.
\paragraph{Impact of charger capacity}
We slightly adapt our instance generation procedure and limit charging operations to the fast charger (Table~\ref{table:station-parameters}) to isolate the impact of integrated charge and service operation scheduling on charging infrastructure utilization. Again, we generate $50$ base instances on which we then vary departure time window length and charger capacity, resulting in a total of $8.000$ instances. Figures~\ref{fig:charger-capacity-results-feasiblity}~\&~\ref{fig:charger-capacity-results-obj} illustrate the results of our charger capacity analysis. 

Figure~\ref{fig:charger-capacity-results-feasiblity} shows the percentage of feasible instances for varying charger capacities and time window lengths. Our scenario requires a minimum charger capacity of seven to support fleet operations when planning service and charging operations separately. Integrated planning of charge and service operations lowers the required charger capacity from seven to three, such that only $86\%$ of the original charger capacity is necessary with one hour of flexibility. A departure time window of three hours, which amounts to half of the shift length, reduces the required capacity to $57\%$. Further doubling the departure time window length yields an additional reduction of $14\%$ to $43\%$ of the basecase's capacity.

\textsc{Result 3.} Integrated planning of charge and service operations reduces the amount of charging infrastructure required for fleet operation but shows decreasing marginal benefits: one hour of flexibility reduces the number of chargers required by $14\%$, three hours by $43\%$, and six hours by $57\%$.

Figure~\ref{fig:charger-capacity-results-obj} shows the total operational costs for varying charger capacities and time window lengths. As we can see, both adding additional chargers and increasing service flexibility reduces the total operational costs. Here, integrated planning has a stronger effect than increasing charger capacity. Specifically, a departure time window of one hour already outperforms the static scenario, even when doubling the number of available chargers such that a dedicated charger is available for each vehicle. Moreover, with a charger capacity of eight, the total cost of static operations, two, and five hours of planning flexibility amounts to \euro $481.63$, \euro $464.21$ ($3.62\%$ saving), and \euro $459.50$ ($4.59\%$ saving), respectively. Increasing the charger capacity by four lowers these costs to \euro $479.96$, \euro $463.65$ ($3.40\%$ saving), and \euro $456.75$ ($4.83\%$ saving), respectively. In comparison, the cost saving of adding four additional chargers to a scenario with eight chargers amounts to $0.34\%$, $0.12\%$, and $0.60\%$ for static operations, two, and five hours of planning flexibility , respectively.

\textsc{Result 4.} Increasing either service operation flexibility or charger capacity lowers operational costs. Here, increasing flexibility yields higher savings than additional investments into charging infrastructure.

Generally, increasing the number of chargers available offsets total savings, such that the number of chargers available limits the maximum cost savings obtainable through integrated planning: with eight chargers, average savings converge to $4.62\%$, with 12 chargers, the maximum saving obtainable is $4.84\%$, and reaches $5.37\%$ when each vehicle has a dedicated charging spot.

\textsc{Result 5.} Charger capacity offsets the cost savings obtainable and thus limits the maximum cost savings of integrated charge and service operation planning. With the minimum number of chargers required in the static scenario, savings peak at $4.62\%$; increasing the charger capacity to $150\%$ and $200\%$ further increases peak savings to $4.84\%$ and $5.37\%$, respectively.

\paragraph{Impact of energy price distribution}
We modify the energy price distribution considered in our basecase to investigate its effect on the cost savings obtainable through integrated charge and service operation scheduling. Specifically, we draw energy prices from a normal distribution $\mathcal{N}(\mu' \cdot \overline{\WearDensityFunction}, \sigma' \cdot \overline{\WearDensityFunction})$, where $\overline{\WearDensityFunction}$ gives the average unit wear cost of our \gls{abk:wdf} (cf. Table~\ref{table:wdf-param}). We consider scenarios with high ($\mu' \coloneqq 2.0$), medium ($\mu' \coloneqq 1.0$), and low ($\mu' \coloneqq 0.5$) energy rates. Analogously, we compare \gls{abk:tou} plans with high ($\sigma' \coloneqq 0.1250$), medium ($\sigma' \coloneqq 0.0625$), and low ($\sigma' \coloneqq 0.0375$) variance. We generate instances with departure time windows of up to five hours in a full factorial design, such that our study comprises a total of $4.950$ runs.

\begin{figure}[t]
	\captionsetup[subcaption]{width=0.9\textwidth}
	\centering
	\begin{subfigure}{0.48\textwidth}
		\includegraphics[width=\textwidth]{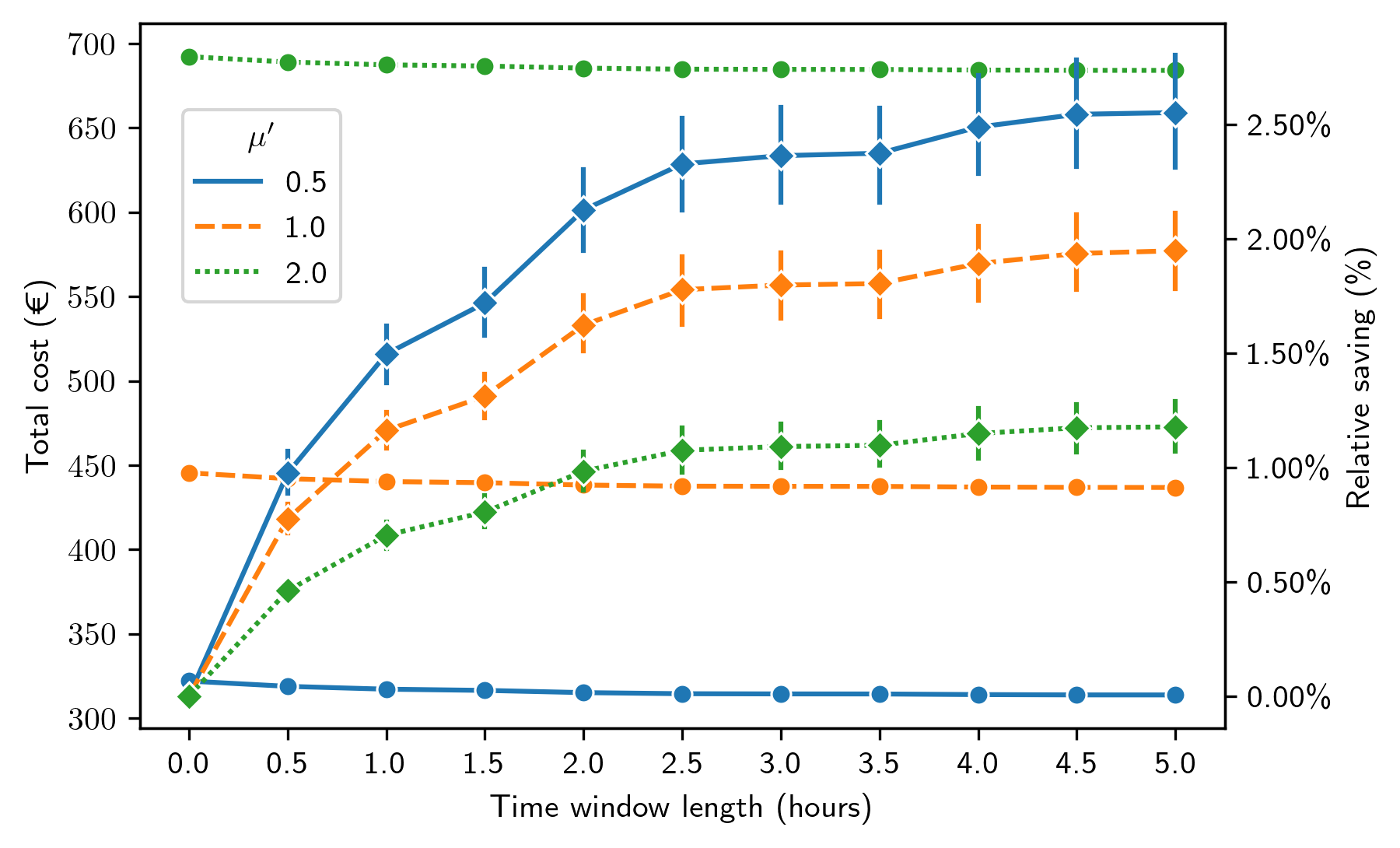}
		\subcaption{Impact of mean energy price $(\sigma' \coloneqq 0.0625)$.\label{fig:energy-price-results-mean}}
	\end{subfigure}
	\begin{subfigure}{0.48\textwidth}
		\includegraphics[width=\textwidth]{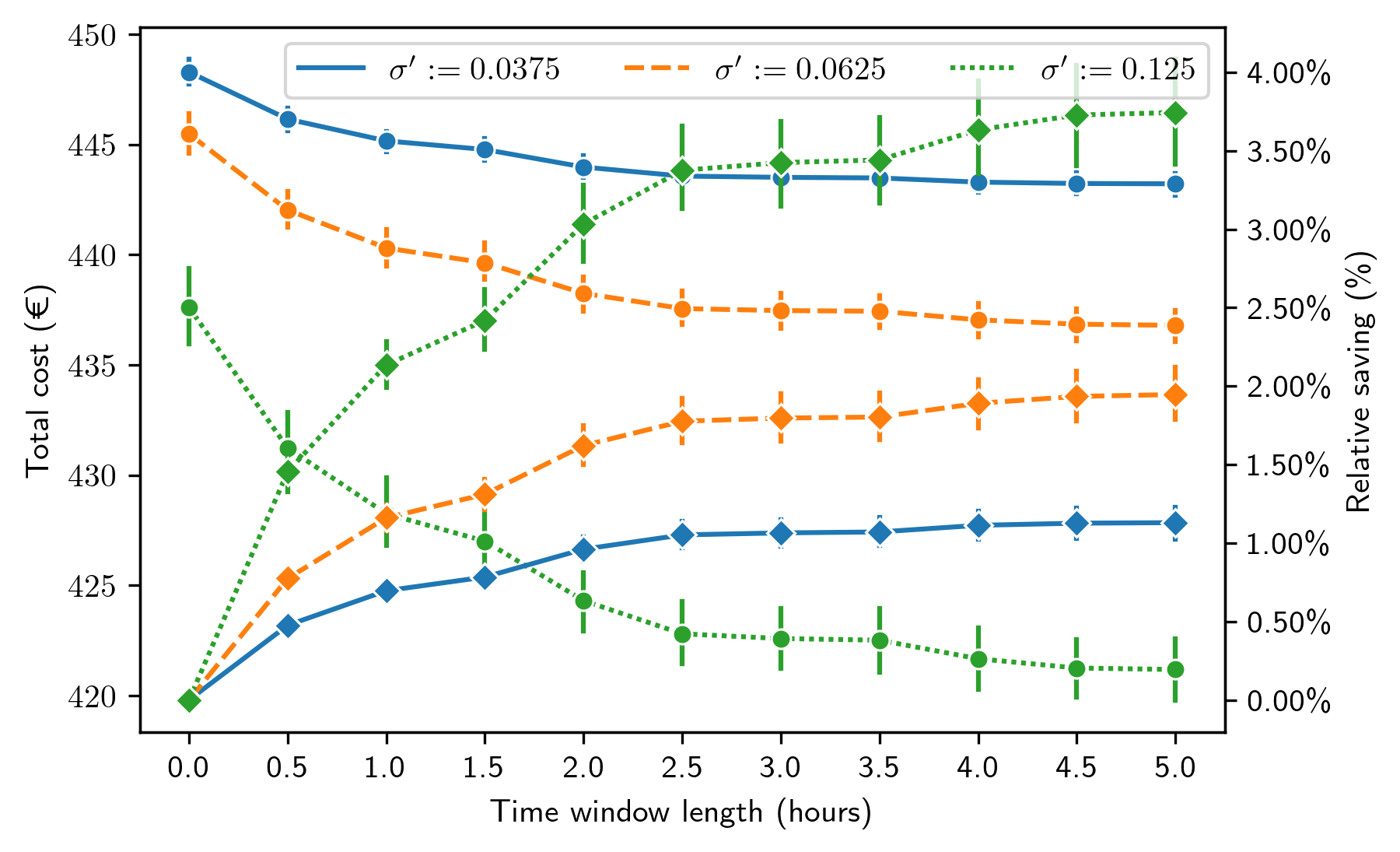}
		\subcaption{Impact of energy price variance $(\mu' \coloneqq 1.0)$.\label{fig:energy-price-results-sigma}}
	\end{subfigure}
	\caption{Results of the energy price experiment.\label{fig:energy-price-mean-sigma-comparison}}
	\begin{minipage}{0.9\textwidth}
		\smaller\textit{Note.} Lines with round markers show the total cost. Lines with diamond markers show relative savings.
	\end{minipage}
\end{figure}

Figure~\ref{fig:energy-price-mean-sigma-comparison} compares the cost savings relative to statically scheduled service operations across time windows of varying size and different values for $\mu'$ and $\sigma'$, respectively.
The mean energy price has an offsetting effect on the total cost. Specifically, we observe an average $56.59\%$ increase in total cost when doubling energy prices. Energy rates discounted by $50\%$ yield average savings of $28.23\%$. 
Regarding the relative cost saving of adding additional flexibility, we observe the opposite effect. Here, relative savings increase with lower mean energy price: with two hours of flexibility, the high, medium, and low rate plans provide an average relative saving of $0.98\%$, $1.62\%$, and $2.12\%$, respectively. At five hours, the average relative savings amounts to $1.18\%$, $1.95\%$, and $2.55\%$, respectively. The marginal saving is unaffected by mean energy price: high, medium, and low rates each reach $80\%$ of the maximum saving at a time window length of two hours. This is related to the higher impact of battery degradation in scenarios with a low mean energy price.

\textsc{Result 6.} The mean energy price offsets total cost. The impact of integrated charge and service operations planning is most pronounced when battery degradation costs are relatively high.

Concerning the energy price standard deviation, we observe two effects: first, \gls{abk:tou} plans with highly variable energy prices lead to greater cost savings in the considered scenario. Specifically, the \gls{abk:tou} plans with high, medium, and low variance converge to a saving of $3.74\%$, $1.95\%$, and $1.12\%$, respectively, such that the savings potential increases threefold under highly variable energy prices compared to a rate plan with low variance.
Second, the relative savings of additional service flexibility increase with increasing variance. In other words, the higher the energy price variance, the higher the impact of service flexibility.

\textsc{Result 7.} Integrated planning of charge and service operations provides the largest savings in scenarios with highly variable energy prices. Specifically, doubling and quadrupling the energy price variance roughly doubles and triples relative savings.
\section{Conclusion}
\label{sec:conclusion}
We presented a novel charge- and service operation scheduling problem where a fleet of electric vehicles fulfills a set of service operations under the assumption of limited charging station capacity, variable energy prices, battery degradation, and non-linear charging behavior. We developed an exact algorithm based on \gls{abk:bnp} to solve the proposed problem. A novel labeling algorithm with efficient dominance criteria, a primal heuristic, and a problem-specific branching rule establish the efficiency of our algorithm, which we demonstrated in numerical experiments. This numerical study asserts the competitiveness of our algorithm through a benchmark against an equivalent mixed-integer formulation, showing that our algorithm significantly outperforms commercial solvers. This study further shows the algorithm's scalability to instances of larger size, optimally solving instances with planning horizons of 5 days or 68 vehicles within the hour, allowing for day-ahead planning in practice. We further derived several managerial insights concerning the impact of service flexibility. Specifically, we find that integrated scheduling of charge and service operations allows to better utilize the trade-off between battery degradation costs and energy price, such that cost savings of up to $5\%$ can be realized. Moreover, service flexibility reduces charger contention, allowing to reduce the number of chargers installed by up to $57\%$. Finally, we analyze the impact of different \gls{abk:tou} plans on the benefit of flexible service operations. Here, we find that integrated charge and service operation scheduling performs best in scenarios with highly variable energy rates.

\section*{Acknowledgments}
This work was supported by the German Federal Ministry for Economic Affairs and Energy within the project MILAS (01MV21020B).

\singlespacing{
	\raggedright
	\footnotesize
	\bibliographystyle{apalike}
	\bibliography{./literature/main,./literature/oscm}
}
\newpage
\begin{appendices}
	\normalsize
	\section{Counterexample}
\label{app:counterexample-baum}
\Revision{Straightforwardly adopting the labeling algorithms proposed in \citet{BaumDibbeltEtAl2019} and \citet{FrogerMendozaEtAl2019} to our problem setting yields an algorithm with a slightly modified label representation. Specifically, cost profiles track only the \textit{last} visited station in these works. Accordingly, intermediate charging is not possible such that realizing a charging opportunity at $i$ when propagating $\Label$ along $(\Arc)$ creates new labels according to the station replacement operation (Section~\ref{sec:label-propagation}) only. Specifically, the algorithm creates a set of new labels ${\{\PropagateAndReplace \mid \forall c \in \SetOfBreakpoints[{\CostProfile[\Label]}]\}} \cup \{\PropagateRegularArc\}$ based on the breakpoints of cost profile $\CostProfile$ at target vertex $\TargetVertex$. 

The following example illustrates a case where this approach does not yield an optimal solution: consider a planning horizon of two periods $\SetOfPeriods_1, \SetOfPeriods_2$ of duration $\PeriodDuration = 5$ with energy prices $\PeriodEnergyCost[\SetOfPeriods_1] = 10c$ and $\PeriodEnergyCost[\SetOfPeriods_2] = c$, a single service operation with consumption $q$, and a single charger $\Charger$ with constant charging rate $\frac{q}{8}$. For the sake of simplicity, we ignore battery degradation in this example and assume zero fixed costs for each arc in the time-expanded network. The optimal path through this network spends~$3\cdot10c$ on charging at vertex $(\SetOfPeriods_1, \Charger)$ and~$5\cdot c$ at ${(\SetOfPeriods_2, \Charger)}$, such that the minimal cost is ${3\cdot10c + 5\cdot c}$.
According to \citep{BaumDibbeltEtAl2019, FrogerMendozaEtAl2019}, a visit to vertex $(\SetOfPeriods_1, \Charger)$ generates a single label with cost profile $\CostProfile[\Label_1](c) = \frac{c}{10} \forall c \in [0, 5]$. Realizing the second charging opportunity at vertex $(\SetOfPeriods_2, \Charger)$ then creates two labels, one for each breakpoint of $\CostProfile[\Label_1]$, i.e., $\Label_2 \coloneqq \PropagateAndReplace[\Label_1][0]$ and $\Label_3 \coloneqq \PropagateAndReplace[\Label_1][5\cdot10c]$. Here, $\Label_2$ is infeasible as $\CostProfileMaxSoC[\Label_2] = \frac{5}{8}q < q$ and $\Label_3$ is not optimal: $\InvCostProfile[\Label_3](q) = 5\cdot10c + 3c > 3\cdot10c + 5c$.}
	\section{Proofs}
\label{app:proofs}
\propone*
\proof We first observe that Term \ref{eq:inverse-intermediate-cost-profile-derivative}.1 is never negative as $\NewPhi$ is piecewise-linear and concave, which, together with $\NewPhi^{-1}(q) - \tau \leq \NewPhi^{-1}(q)$, implies $\PWLSlope[\NewPhi][q](\NewPhi^{-1}(q) - \tau) \geq \PWLSlope[\NewPhi][q](\NewPhi^{-1}(q))$:
\begin{equation*}
	\PWLSlope[\NewPhi][q](\NewPhi^{-1}(q) - \tau) \cdot \PWLSlope[\NewPhi^{-1}][q](q) = \frac{\PWLSlope[\NewPhi][q](\NewPhi^{-1}(q) - \tau)}{\PWLSlope[\NewPhi][q](\NewPhi^{-1}(q))} \geq 1.
\end{equation*}
As $\PWLSlope[\InvNewStationProfile][q]$ is non-negative, we have $\PWLSlope[{\InvNewCostProfile}][q](q') < 0$, which implies $\PWLSlope[{\InvPrevCostProfile}][q](q') < \PWLSlope[\InvNewStationProfile][q](q')$.
\endproof
\proptwo*
\proof Let there be some $c$ where $\CostProfile[\Label']$ is decreasing. Further let ${q \coloneqq \InvCostProfile[\Label'](c)}$. Since $\CostProfile[\Label']$ is decreasing on $[c', c'+\epsilon]$, we have ${\CostProfileMinCost[\Label'] < c < \CostProfileMaxCost[\Label']}$ such that there exists some ${\Label'' \coloneqq \PropagateAndReplace[\Label][c'']}$ with ${\CostProfileMaxSoC[\Label''] = q}$.
As we have ${\PWLSlope[{\InvPrevCostProfile}][q](q') < \PWLSlope[\InvNewStationProfile][q](q')}$ for ${q' \coloneqq \InvCostProfile[\Label''](c'')}$ (Proposition~\ref{prop:more-expensive-when-non-increasing}), we can apply the same reasoning as in the proof of Theorem~\ref{theorem:one} (Case 2.2) and argue that there exists some set of labels $\SetOfLabels' \subseteq \SetOfLabels_j \setminus \{\Label'\}$ such that $\SetOfLabels' \geq \CostProfile[\Label'](c)$.

\theoremone*
\proof Assume the contrary, i.e., that there exists some $\ToBeDominatedLabel$ not dominated by $\SetOfLabels'$. Then there exists some $\AssumedCost \in \mathbb{R}$ such that $\ToBeDominatedProfile(\AssumedCost) > \max_{\Label'' \in \SetOfLabels'}\{\CostProfile[{\Label''}](\AssumedCost)\}$. 

\newcommand{\AssumedSoC}{q}
\newcommand{\TauLabel}{\Label_{\PeriodDuration}}
\newcommand{\TauProfile}{\CostProfile[{\TauLabel}]}	
\newcommand{\InvTauProfile}{\InvCostProfile[{\TauLabel}]}
\newcommand{\DominatingReplacementLabel}{\Label'''}
\newcommand{\DominatingReplacementProfile}{\CostProfile[{\DominatingReplacementLabel}]}
\newcommand{\InverseDominatingReplacementProfile}{\InvCostProfile[{\DominatingReplacementLabel}]}
\newcommand{\EntryCostDominatingReplacement}{c_{j}'''}
\newcommand{\EntrySoCDominatingReplacement}{q'''}

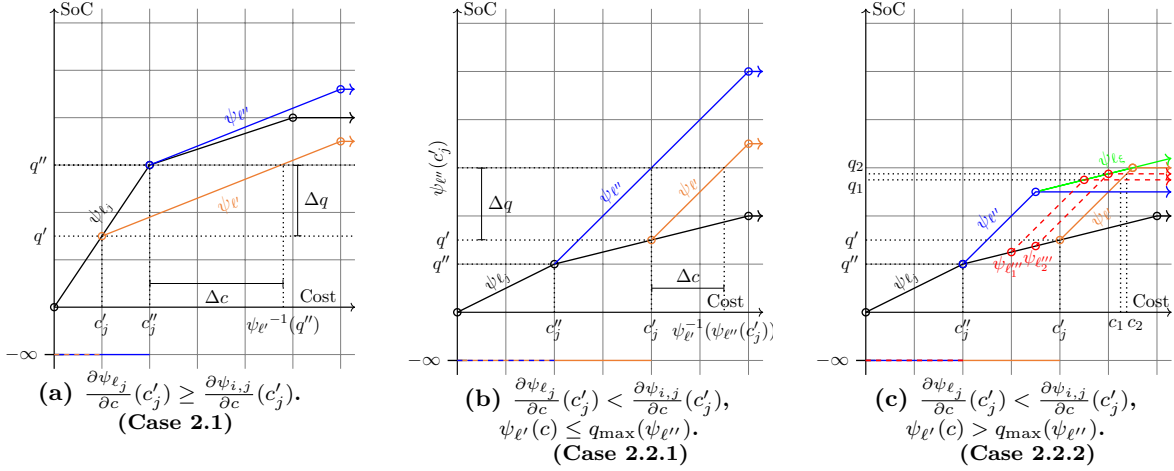
\begin{figure}
	\begin{subfigure}[T]{0.3\textwidth}
		\resizebox{\textwidth}{!}{%
			\usetikzlibrary{intersections}
\begin{tikzpicture}

\tikzmath{
		\maxX = 6; \maxXBoundary = \maxX + 0.3;
		\minX = 0; \minXBoundary = \minX - 0.3;
		\maxY = 6; \maxYBoundary = \maxY + 0.3;
		\minY = -1; \minYBoundary = \minY - 0.3;
}

\tikzset{
	bp/.style={
		circle, draw, thick, solid, inner sep=1.5pt
	},
	profile/.style={
		thick
	},
	dominated/.style={
		orange, profile
	},
	dominating/.style={
		blue, profile
	},
	nocharge/.style={
		black, profile
	},
	hint/.style={
		black, dotted, thick
	}
}

\clip(-1, -1.3) rectangle (6.5, 6.5);

\draw[step=1cm, gray, very thin] (0, -1.3) grid (\maxXBoundary, \maxYBoundary);
\draw[->] (0,0) -- (\maxXBoundary, 0) node[above=0.25cm, left=0.25cm] {Cost};
\draw[->] (0,-1.3) -- (0, \maxYBoundary) node[right] {SoC};

\draw[thick] (2mm, -1) -- (-2mm, -1) node[left] {$-\infty$};
\foreach \y in {0, ..., \maxX} {
}

\foreach \x in {2, 4, ..., 12} {
		\tikzmath{\xscaled = \x / 2;}
}

\draw[nocharge, ->] (0, 0) node[bp]{} -- node[above=1.5mm, right=-2.5mm, rotate=60, font=\large]{$\NoChargeProfile$}  (2.0, 3) node[bp]{} -- (5.0, 4.0) node[bp]{} -- (\maxXBoundary, 4.0);

\draw[dominating] (0, -1) -- ++(2.0, 0);
\draw[dominating, ->] (2.0, 3) node[bp]{} -- node[above, rotate=30, font=\large]{$\DominatingProfile$} ++(4.0, 1.6) node[bp]{} -- ++(0.3, 0);

\draw[dominated, dashed] (0, -1) -- ++(1.0, 0);
\draw[dominated, ->] (1.0, 1.5) node[bp]{} -- node[below, rotate=30, font=\large]{$\ToBeDominatedProfile$} ++(5.0, 2) node[bp]{} -- (\maxXBoundary, 3.5);

\draw[|-|] (2.0, 0.5) -- node[below]{$\Delta c$} (4.8, 0.5);
\draw[|-|] (5.1, 3.0) -- node[right]{$\Delta q$} (5.1, 1.5);

\draw[hint] (0, 3.0) node[left]{$\EntrySoCDominating$}  -- 
								(2.0, 3.0) -- (2.0, 0) node[below]{$\EntryCostDominating$};
\draw[hint] (0, 3.0) -- (4.8, 3.0) -- ++(0.0, -3.0) node[below]{$\ToBeDominatedProfile^{-1}(\EntrySoCDominating)$};
\draw[hint] (4.8, 3.0) -- (\maxXBoundary, 3.0);

\draw[hint] (0, 1.5) node[left]{$\EntrySoCDominatedStation$}  -- (1.0, 1.5) -- (1.0, 0) node[below]{$\EntryCostDominatedStation$};
\draw[hint] (1.0, 1.5) -- (\maxXBoundary, 1.5);

\end{tikzpicture}
		}
		\subcaption{\centering ${\PWLSlope[\NoChargeProfile][c](\EntryCostDominatedStation) \geq \PWLSlope[\StationProfile][c](\EntryCostDominatedStation)}$.\newline(Case~2.1)\newline\label{fig:proof-example:old-better}}
		\begin{minipage}{\textwidth}
			\smaller\textit{Note.} $\Delta c$ corresponds to the additional cost incurred by using vertex $\OriginVertex$ to recharge up to the \gls{abk:soc} at which $\Label''$ stops charging using the charging opportunity captured by the previous label. As the previous charging opportunity provides a lower price, $\Delta c$ is positive.
		\end{minipage}
	\end{subfigure}
	\hspace{0.01\textwidth}
	\begin{subfigure}[T]{0.3\textwidth}
		\resizebox{\textwidth}{!}{%
			\begin{tikzpicture}

\tikzmath{
		\maxX = 6; \maxXBoundary = \maxX + 0.3;
		\minX = 0; \minXBoundary = \minX - 0.3;
		\maxY = 6; \maxYBoundary = \maxY + 0.3;
		\minY = -1; \minYBoundary = \minY - 0.3;
}

\tikzset{
	bp/.style={
		circle, draw, thick, solid, inner sep=1.5pt
	},
	profile/.style={
		thick
	},
	dominated/.style={
		orange, profile
	},
	dominating/.style={
		blue, profile
	},
	nocharge/.style={
		black, profile
	},
	hint/.style={
		black, dotted, thick
	}
}

\clip(-1, -1.3) rectangle (6.5, 6.5);

\draw[step=1cm, gray, very thin] (0, -1.3) grid (\maxXBoundary, \maxYBoundary);
\draw[->, name path=xaxis] (0,0) -- (\maxXBoundary, 0) node[above=0.25cm, left=0.25cm] {Cost};
\draw[->, name path=yaxis] (0,-1.3) -- (0, \maxYBoundary) node[right] {SoC};

\draw[thick] (2mm, -1) -- (-2mm, -1) node[left] {$-\infty$};
\foreach \y in {0, ..., \maxX} {
}

\foreach \x in {2, 4, ..., 12} {
		\tikzmath{\xscaled = \x / 2;}
}

\draw[nocharge, ->] (0, 0) node[bp]{} -- node[rotate=45, above=-1mm]{$\NoChargeProfile$} (2.0, 1) node[bp](primeprime){} -- coordinate (prime) (6.0, 2) node[bp]{} -- (\maxXBoundary, 2);

\path let \p1 = (prime) in coordinate (cprime) at (0,\y1){};
\path let \p1 = (prime) in coordinate (qprime) at (\x1,0){};

\path let \p1 = (primeprime) in coordinate (cprimeprime) at (0,\y1){};
\path let \p1 = (primeprime) in coordinate (qprimeprime) at (\x1,0){};

\draw[dominated] (0, -1) -- (4.0, -1);
\draw[dominated,  name path=lprime, ->] (prime) node[bp]{} -- node[rotate=45,above=2.5mm, left=-3mm]{$\ToBeDominatedProfile$} ++(2.0, 2) node[bp]{} -- ++(0.3, 0.0);

\draw[dominating, dashed] (0, -1) -- ++(2.0, 0);
\draw[dominating, name path=lprimeprime, ->] (2.0, 1) -- node[rotate=45, above=2.5mm, left=5mm]{$\DominatingProfile$} ++(4.0, 4) node[bp]{} -- ++(0.3, 0);

\path[name path=deltac] (qprime) -- ++(0, 6);
\coordinate[name intersections={of=lprimeprime and deltac,by=deltacpoint}];
\path let \p1 = (deltacpoint) in coordinate (deltacq) at (0, \y1);
\path[name path=deltacx] (deltacq) -- ++(6, 0);
\coordinate[name intersections={of=deltacx and lprime,by=deltacprime}];
\path let \p1 = (deltacprime) in coordinate (deltacprimeq) at (\x1, 0);

\draw[hint] (cprime) node[left]{$\EntrySoCDominatedStation$} -- (prime) -- (qprime) node[below]{$\EntryCostDominatedStation$};
\draw[hint] (cprimeprime) node[left]{$\EntrySoCDominating$} -- (primeprime) -- (qprimeprime) node[below]{$\EntryCostDominating$};
\draw[hint] (deltacq) node[above, rotate=90]{$\DominatingProfile(\EntryCostDominatedStation)$} -- (deltacpoint) -- (prime);
\draw[hint] (deltacpoint) -- (deltacprime) -- (deltacprimeq) node[below]{$\InverseToBeDominatedProfile(\DominatingProfile(\EntryCostDominatedStation))$};

\coordinate (deltaqA) at ($(deltacq) + (0.5,0)$);
\coordinate (deltaqB) at ($(cprime) + (0.5,0)$);

\draw[|-|] (deltaqA) -- node[right]{$\Delta q$} (deltaqB);

\coordinate (deltacA) at ($(deltacprimeq) + (0, 0.5)$);
\coordinate (deltacB) at ($(qprime) + (0,0.5)$);
\draw[|-|] (deltacA) -- node[above]{$\Delta c$} (deltacB);

\end{tikzpicture}
		}
		\subcaption{\centering ${\PWLSlope[\NoChargeProfile][c](\EntryCostDominatedStation) < \PWLSlope[\StationProfile][c](\EntryCostDominatedStation)}$, $\ToBeDominatedProfile(\AssumedCost) \leq \CostProfileMaxSoC[\DominatingLabel]$.\newline(Case 2.2.1)\label{fig:proof-example:new-better}}
		\begin{minipage}{\textwidth}
			\smaller\textit{Note.} Here, $\Delta c$ corresponds to the cost saved by using $\OriginVertex$ to charge up to the arrival \gls{abk:soc} of $\Label'$ at $\TargetVertex$. $\Delta c$ is positive as station $\OriginVertex$ provides a lower charging price in this case.
		\end{minipage}
	\end{subfigure}
\hspace{0.01\textwidth}
	\begin{subfigure}[T]{0.3\textwidth}
		\resizebox{\textwidth}{!}{%
			\begin{tikzpicture}

\tikzmath{
		\maxX = 6; \maxXBoundary = \maxX + 0.3;
		\minX = 0; \minXBoundary = \minX - 0.3;
		\maxY = 6; \maxYBoundary = \maxY + 0.3;
		\minY = -1; \minYBoundary = \minY - 0.3;
}

\tikzset{
	bp/.style={
		circle, draw, thick, solid, inner sep=1.5pt
	},
	profile/.style={
		thick
	},
	dominated/.style={
		orange, profile
	},
	dominating/.style={
		blue, profile
	},
	nocharge/.style={
		black, profile
	},
	lppp/.style={
		red, profile, dashed
	},
	hint/.style={
		black, dotted, thick
	}
}

\clip(-1, -1.3) rectangle (6.5, 6.5);

\draw[step=1cm, gray, very thin] (0, -1.3) grid (\maxXBoundary, \maxYBoundary);
\draw[->, name path=xaxis] (0,0) -- (\maxXBoundary, 0) node[above=0.25cm, left=-0.1cm] {Cost};
\draw[->, name path=yaxis] (0,-1.3) -- (0, \maxYBoundary) node[right] {SoC};

\draw[thick] (2mm, -1) -- (-2mm, -1) node[left] {$-\infty$};
\foreach \y in {0, ..., \maxX} {
}

\foreach \x in {2, 4, ..., 12} {
		\tikzmath{\xscaled = \x / 2;}
}

\draw[nocharge, ->, name path=nocharge] (0, 0) node[bp]{} -- node[rotate=45, above=-1mm]{$\NoChargeProfile$} (2.0, 1) node[bp](primeprime){} -- coordinate (prime) (6.0, 2) node[bp]{} -- (\maxXBoundary, 2);

\path let \p1 = (prime) in coordinate (cprime) at (0,\y1){};
\path let \p1 = (prime) in coordinate (qprime) at (\x1,0){};

\path let \p1 = (primeprime) in coordinate (cprimeprime) at (0,\y1){};
\path let \p1 = (primeprime) in coordinate (qprimeprime) at (\x1,0){};

\coordinate (ppp) at ($(primeprime)!0.75!(prime)$);
\path let \p1 = (ppp) in coordinate (cppp) at (0,\y1){};
\path let \p1 = (ppp) in coordinate (qppp) at (\x1,0){};

\coordinate (ppp2) at ($(primeprime)!0.5!(prime)$);
\path let \p1 = (ppp2) in coordinate (cppp2) at (0,\y1){};
\path let \p1 = (ppp2) in coordinate (qppp2) at (\x1,0){};

\draw[dominated, dashed] (0, -1) -- (2, -1);
\draw[dominated] (2, -1) -- (4, -1);
\draw[dominated,  name path=lprime, ->] (prime) node[bp]{} -- node[rotate=45,below=2.5mm, left=-3mm]{$\ToBeDominatedProfile$} ++(1.5, 1.5) node[bp](qmaxlprime){} -- ++(0.8, 0.0);

\draw[dominating] (0, -1) -- ++(2.0, 0);
\draw[dominating, name path=lprimeprime, ->] (2.0, 1) node[bp]{} -- node[rotate=45, above]{$\DominatingProfile$} ++(1.5, 1.5) node[bp](qmaxlprimeprime){} -- ++(2.8, 0);

\draw[green, profile,->] (qmaxlprime) -- node[rotate=30,above=1mm,right=3mm]{$\TauProfile$} (qmaxlprimeprime)  -- ($(qmaxlprime)!-0.4!(qmaxlprimeprime)$);

\node[bp,dominated] at (qmaxlprime){};

\draw[lppp, dashed] (0, -1) -- ++(2.0, 0);
\draw[lppp, name path=lppp, ->] (ppp) node[below=3mm,right=-4mm]{$\CostProfile[{\DominatingReplacementLabel_2}]$} node[bp]{}  -- ++(1.5, 1.5) node[bp](assumed){} -- ++(1.3, 0);

\draw[lppp, dashed] (0, -1) -- ++(2.0, 0);
\draw[lppp, name path=lppp2, ->] (ppp2) node[below=3mm,left=-4mm]{$\CostProfile[{\DominatingReplacementLabel_1}]$} node[bp]{} --  ++(1.5, 1.5) node[bp](assumed2){} -- ++(1.8, 0);

\path let \p1 = (assumed) in coordinate (assumedq) at (0, \y1);
\path[name path=assumedyaxis] (assumed) -- ++(0, -6);
\path[name path=assumedxaxis] (assumed) -- ++(6, 0);
\coordinate[name intersections={of=assumedxaxis and lprime,by=lprimeassumed}];
\path let \p1 = (lprimeassumed) in coordinate (assumedc) at (\x1, 0);

\path let \p1 = (assumed2) in coordinate (assumedq2) at (0, \y1);
\path[name path=assumed2xaxis] (assumed2) -- ++(6, 0);
\coordinate[name intersections={of=assumed2xaxis and lprime,by=lprimeassumed2}];
\path let \p1 = (lprimeassumed2) in coordinate (assumedc2) at (\x1, 0);

\draw[hint] (assumedc) node[below=2.5mm,left=-1mm]{$c_1$} -- (lprimeassumed);
\draw[hint] (assumed) -- (assumedq) node[left=2mm,above=-1mm]{$\AssumedSoC_2$};

\draw[hint] (assumedc2) node[below=2.5mm,right=0mm]{$c_2$} -- (lprimeassumed2);
\draw[hint] (assumed2) -- (assumedq2) node[left=2mm,below=-1mm]{$\AssumedSoC_1$};

\draw[hint] (cprime) node[left]{$\EntrySoCDominatedStation$} -- (prime) -- (qprime) node[below]{$\EntryCostDominatedStation$};
\draw[hint] (cprimeprime) node[left]{$\EntrySoCDominating$} -- (primeprime) -- (qprimeprime) node[below]{$\EntryCostDominating$};

\end{tikzpicture}
		}
		\subcaption{\centering ${\PWLSlope[\NoChargeProfile][c](\EntryCostDominatedStation) < \PWLSlope[\StationProfile][c](\EntryCostDominatedStation)}$, $\ToBeDominatedProfile(\AssumedCost) > \CostProfileMaxSoC[\DominatingLabel]$.\newline(Case 2.2.2)\label{fig:proof-example:new-better-intermediate}}
		\begin{minipage}{\textwidth}
			\smaller\textit{Note.} $c_1$ and $c_2$ provide example values for $c$. The red dashed lines illustrate the corresponding cost profiles $\CostProfile[{\DominatingReplacementLabel_1}]$ and $\CostProfile[{\DominatingReplacementLabel_2}]$, respectively. The green profile illustrates charging intermediately. Note that, for the sake of simplicity, we show only the segment of $\TauProfile$ relevant to this proof.
		\end{minipage}
	\end{subfigure}
	\caption{Illustration of cases 2.1 and 2.2 in the proof of Theorem~\ref{theorem:one}.}
\end{figure}

\begin{paragraphs}
	\item[Case 1:] $c < \CostProfileMinCost[\ToBeDominatedLabel]$. \\
	Then $\ToBeDominatedProfile(c) = -\infty \leq \max_{\CostProfile[{\Label''}] \in \SetOfLabels'}\{\Label''(c)\}$ holds by definition.
	\item[Case 2:] $\CostProfileMinCost[\ToBeDominatedLabel] \leq c \leq \CostProfileMaxCost[\ToBeDominatedLabel]$. \\
	W.l.o.g., we assume $\ToBeDominatedLabel = \PropagateAndReplace[\OriginalLabel][c']$ for some $c'$ and let $\EntryCostDominatedStation \coloneqq c' + \ArcCost$ and $\EntrySoCDominatedStation \coloneqq \OriginalProfile(c') = \NoChargeProfile(\EntryCostDominatedStation)$. Recall from Section~\ref{sec:label-propagation} that 
	\begin{align*}
		\ToBeDominatedProfile(c) &= \CostProfile[\OriginalLabel](c') + \ArcCost + \InverseDeltaChargingCost[\EntrySoCDominatedStation][c - \ArcCost - c']\\
		&= \NoChargeProfile(\EntryCostDominatedStation) + \InverseDeltaChargingCost[\EntrySoCDominatedStation][c - \ArcCost - c']\\
		&= \NoChargeProfile(\EntryCostDominatedStation) + \InverseDeltaChargingCost[\EntrySoCDominatedStation][c - \EntryCostDominatedStation].
	\end{align*}
	\begin{paragraphs}
		\item[Case 2.1:] $\PWLSlope[\NoChargeProfile][c](\EntryCostDominatedStation) \geq \PWLSlope[\ShiftedStationProfile][c](\EntryCostDominatedStation)$.\\ %
		Let $\DominatingLabel \coloneqq \PropagateAndReplace[\OriginalLabel][c'']$, with $c'' = \min\{c \mid c \in \SetOfBreakpoints[{\CostProfile[\OriginalLabel]}] \wedge c \geq c'\}$, and let $\EntryCostDominating \coloneqq c'' + \ArcCost$ such that 
		\begin{align*}
			\DominatingProfile(c) &= \OriginalProfile(c') + \ArcCost + \InverseDeltaChargingCost[\EntrySoCDominating][c - \ArcCost - c'']\\
			&= \NoChargeProfile(\EntryCostDominating) + \InverseDeltaChargingCost[\EntrySoCDominating][c - \ArcCost - c'']\\
			&= \NoChargeProfile(\EntryCostDominating) + \InverseDeltaChargingCost[\EntrySoCDominating][c - \EntryCostDominating].
		\end{align*}
		We argue that $\DominatingLabel$ exists and $\DominatingLabel \in \SetOfLabels'$ holds since $c' \in \lbrack \CostProfileMinCost[\OriginalLabel], \CostProfileMaxCost[\OriginalLabel] \rbrack$.
		\begin{paragraphs}
			\item[Case 2.1.1:] $\CostProfileMinCost[\ToBeDominatedLabel] \leq \AssumedCost \leq \CostProfileMinCost[\DominatingLabel]$. \\
			Then we have $\CostProfileMinCost[\NoChargeLabel] \leq \AssumedCost < \CostProfileMinCost[\DominatingLabel] \leq \CostProfileMaxCost[\NoChargeLabel]$. From the definition of $\DominatingLabel$, it further follows that $\PWLSlope[\NoChargeProfile][c]$ is constant on $\lbrack \EntryCostDominatedStation, \AssumedCost \rbrack$. Hence, we get
			\begin{align*}
				\NoChargeProfile(c) &= \NoChargeProfile(\EntryCostDominatedStation) + \NoChargeProfile(c) - \NoChargeProfile(\EntryCostDominatedStation)\\
				&= \NoChargeProfile(\EntryCostDominatedStation) + \PWLSlope[\NoChargeProfile][c](\EntryCostDominatedStation) \cdot (c - \EntryCostDominatedStation)\\
				&\overset{(\star)}{\geq} \NoChargeProfile(\EntryCostDominatedStation) + \PWLSlope[\StationProfile][c](\EntryCostDominatedStation) \cdot (c - \EntryCostDominatedStation)\\
				&= \ToBeDominatedProfile(c),
			\end{align*}
			which contradicts our assumption. Here, $(\star)$ holds as $\PWLSlope[\NoChargeProfile][c](\EntryCostDominatedStation) \geq \PWLSlope[\ShiftedStationProfile][c](\EntryCostDominatedStation)$.
			\item[Case 2.1.2]:  $\CostProfileMinCost[\DominatingLabel] < \AssumedCost \leq \CostProfileMaxCost[\ToBeDominatedLabel]$.\\
			We note that ${\Delta q \coloneqq \DominatingProfile(\EntryCostDominating) - \ToBeDominatedProfile(\EntryCostDominating)\ \geq 0}$ holds by Case 2.1.1, such that it holds that ${\Delta c \coloneqq \InverseToBeDominatedProfile(\EntrySoCDominating) - \EntryCostDominating \geq 0}$. Hence, for $c < \EntryCostDominating + \Delta c$, we have $\DominatingProfile(c) \geq \ToBeDominatedProfile(c)$ since $\InverseDeltaChargingCost[q][\cdot]$ are concave for $q \in [\MinSoC, \MaxSoC]$ (cf. Section~\ref{sec:label-propagation}). Using the same argument for inequality $(\star)$, we get the following for $c \geq \EntryCostDominating + \Delta c$:
			\begin{align*}
				\ToBeDominatedProfile(c) &= \EntrySoCDominatedStation + \InverseDeltaChargingCost[\EntrySoCDominatedStation][c - \EntryCostDominatedStation]\\
				&= \EntrySoCDominating + \InverseDeltaChargingCost[\EntrySoCDominating][c - (\EntryCostDominating + \Delta c)]\\
				&\overset{(\star)}{\leq} \EntrySoCDominating + \InverseDeltaChargingCost[\EntrySoCDominating][c - \EntryCostDominating]\\
				&= \DominatingProfile(c),
			\end{align*}
			which contradicts the assumption. See Figure~\ref{fig:proof-example:old-better} for an illustration.
		\end{paragraphs}
		\item[Case 2.2:] $\PWLSlope[\NoChargeProfile][c](\EntryCostDominatedStation) < \PWLSlope[\ShiftedStationProfile][c](\EntryCostDominatedStation)$. \\
		Let $\DominatingLabel \coloneqq \PropagateAndReplace[\OriginalLabel][c'']$, with $c'' = \max\{c \mid c \in \SetOfBreakpoints[{\CostProfile[\OriginalLabel]}] \wedge c \leq c'\}$, and let $\EntryCostDominating \coloneqq c'' + \ArcCost$ such that $\DominatingProfile(c) = \NoChargeProfile(\EntryCostDominating) + \InverseDeltaChargingCost[\EntrySoCDominating][c - \EntryCostDominating]$.
		\begin{paragraphs}
			\item[Case 2.2.1:] $\ToBeDominatedProfile(\AssumedCost) \leq \CostProfileMaxSoC[\DominatingLabel]$.\\
			It follows from the definition of $\DominatingLabel$ that $\PWLSlope[\NoChargeProfile][c]$ is constant on $[\EntryCostDominatedStation, \EntryCostDominating]$. Then, analogous to Case 2.1.1, we have
			\begin{align*}
				\DominatingProfile(\EntryCostDominatedStation) &= \DominatingProfile(\EntryCostDominating) + \DominatingProfile(\EntryCostDominatedStation) - \DominatingProfile(\EntryCostDominating)\\
				&\overset{(\star)}{\geq} \DominatingProfile(\EntryCostDominating) + \PWLSlope[\ShiftedStationProfile][c](\EntryCostDominatedStation) \cdot (\EntryCostDominatedStation - \EntryCostDominating)\\
				&\overset{(\star\star)}{\geq} \DominatingProfile(\EntryCostDominating) + \PWLSlope[\NoChargeProfile][c](\EntryCostDominatedStation) \cdot (\EntryCostDominatedStation - \EntryCostDominating)\\
				&\overset{(\star\star\star)}{=} \ToBeDominatedProfile(\EntryCostDominating) + \PWLSlope[\NoChargeProfile][c](\EntryCostDominatedStation) \cdot (\EntryCostDominatedStation - \EntryCostDominating)\\
				&= \ToBeDominatedProfile(\EntryCostDominatedStation).
			\end{align*}
			Here, $(\star)$ follows from the concavity of $\ShiftedStationProfile$, i.e., $\PWLSlope[\ShiftedStationProfile][c](\EntryCostDominating) \geq \PWLSlope[\ShiftedStationProfile][c](\EntryCostDominatedStation)$, $(\star\star)$ follows from $\PWLSlope[\NoChargeProfile][c](\EntryCostDominatedStation) < \PWLSlope[\StationProfile][c](\EntryCostDominatedStation)$, and $(\star\star\star)$ holds since $\PWLSlope[\NoChargeProfile][c]$ is constant on $[\EntryCostDominating, \EntryCostDominatedStation]$ by the definition of $\EntryCostDominatedStation$ and $\EntryCostDominating$.
			
			Concluding, ${\Delta q \coloneqq \DominatingProfile(\EntryCostDominatedStation) - \EntrySoCDominatedStation \geq 0}$ holds, which implies ${\Delta c \coloneqq \InverseToBeDominatedProfile(\DominatingProfile(\EntryCostDominatedStation))} - {\InverseToBeDominatedProfile(\EntrySoCDominatedStation) \geq 0}$, such that, analogous to Case 2.1.2, we get $\DominatingProfile(c) \geq \ToBeDominatedProfile(c)$ for $c + \Delta c < \EntryCostDominatedStation$ due to the concavity of $\InverseDeltaChargingCost[q][\cdot]$ (cf. Section~\ref{sec:label-propagation}). Furthermore, for $c + \Delta c \geq \EntryCostDominatedStation$, it holds that
			\begin{align*}
				\DominatingProfile(c) &= \EntrySoCDominating + \InverseDeltaChargingCost[\EntrySoCDominating][c - \EntryCostDominatedStation]\\
				&= \EntrySoCDominatedStation + \InverseDeltaChargingCost[\EntrySoCDominatedStation][c - \EntryCostDominatedStation + \Delta c]\\
				&\geq \EntrySoCDominatedStation + \InverseDeltaChargingCost[\EntrySoCDominatedStation][c - \EntryCostDominatedStation]\\
				&= \ToBeDominatedProfile(c),
			\end{align*}
			which contradicts the assumption. See Figure~\ref{fig:proof-example:new-better} for an illustration.
			\item[Case 2.2.2:] $\ToBeDominatedProfile(\AssumedCost) > \CostProfileMaxSoC[\DominatingLabel]$.\\
			Let $\TauLabel \coloneqq \PropagateAndCharge[\OriginalLabel][\PeriodDuration]$ and $\AssumedSoC \coloneqq \ToBeDominatedProfile(\AssumedCost)$.
			Further let ${\DominatingReplacementLabel \coloneqq \PropagateAndReplace[\OriginalLabel][c''']}$ such that $\CostProfileMaxSoC[\DominatingReplacementLabel] = \AssumedSoC$. The existence of $\DominatingReplacementLabel$ follows straightforwardly from ${\AssumedCost \leq \CostProfileMaxCost[\ToBeDominatedLabel]}$ and the existence of $\ToBeDominatedLabel$. Let $\EntryCostDominatingReplacement \coloneqq c''' + \ArcCost$ and $\EntrySoCDominatingReplacement \coloneqq \DominatingReplacementProfile(\EntryCostDominatingReplacement)$.
			Note that ${c'' \leq c''' \leq c'}$, and thus ${\EntryCostDominating \leq \EntryCostDominatingReplacement \leq \EntryCostDominatedStation}$, since the definition of $c''$ implies that $\PWLSlope[\OriginalProfile][c]$ is constant on $\lbrack c'', c' \rbrack$. Hence, analogous to Case 2.2.1, we get $\DominatingReplacementProfile(\AssumedCost) \geq \ToBeDominatedProfile(\AssumedCost)$.
			With this in mind, we recall from Equation~\ref{eq:cost-profile-replacement} that
			\newcommand{\CalcFinalSoCReplacement}{\ChargeFunction[{\ChargerOfVertex[\OriginVertex]}](\EntrySoCDominatingReplacement, \PeriodDuration) - \EntrySoCDominatingReplacement}
			\newcommand{\CalcFinalSoCIntermediate}{\ChargeFunction[{\ChargerOfVertex[\OriginVertex]}](\ChargeFunction[{\ChargerOfVertex[\OriginVertex]}]^{-1}(\AssumedSoC) - \PeriodDuration)}
			\begin{align*}
				\InverseDominatingReplacementProfile(\AssumedSoC) &= \InverseDominatingReplacementProfile(\CostProfileMaxSoC[{\DominatingReplacementLabel}]) = \EntryCostDominatingReplacement + \DeltaChargingCost[\EntrySoCDominatingReplacement][\CalcFinalSoCReplacement]\\
				&= \InverseNoChargeProfile(\EntrySoCDominatingReplacement) + \DeltaChargingCost[\EntrySoCDominatingReplacement][\CalcFinalSoCReplacement]\\
				&= \InverseOriginalProfile(\EntrySoCDominatingReplacement) + \ArcCost  + \DeltaChargingCost[\EntrySoCDominatingReplacement][\CalcFinalSoCReplacement].\\
			\end{align*}
			Recall that $\AssumedSoC \coloneqq \ToBeDominatedProfile(\AssumedCost) = {\ChargeFunction[{\ChargerOfVertex[\OriginVertex]}](\EntrySoCDominatingReplacement, \PeriodDuration)}$, such that substituting $\AssumedSoC$ yields
			\begin{align*}
				&\InverseOriginalProfile(\EntrySoCDominatingReplacement) + \ArcCost  + \DeltaChargingCost[\EntrySoCDominatingReplacement][\CalcFinalSoCReplacement]\\
				&= \InverseOriginalProfile(\EntrySoCDominatingReplacement) + \ArcCost  + \DeltaChargingCost[\EntrySoCDominatingReplacement][\AssumedSoC - \EntrySoCDominatingReplacement]\\
				&\overset{(\star)}{=} \InverseOriginalProfile(\EntrySoCDominatingReplacement) + \ArcCost  + (\DeltaChargingCost[0][\AssumedSoC] - \DeltaChargingCost[0][\EntrySoCDominatingReplacement])\\
				&= \InverseOriginalProfile(\EntrySoCDominatingReplacement) + \ArcCost  + (\InvStationProfile(\AssumedSoC) - \InvStationProfile(\EntrySoCDominatingReplacement))\\
				&= \InvTauProfile(\AssumedSoC).
			\end{align*}
			Here, $(\star)$ holds since
			\begin{align*}
				\DeltaChargingCost[\EntrySoCDominatingReplacement][\AssumedSoC - \EntrySoCDominatingReplacement]
				&=\PeriodEnergyCost\cdot(\AssumedSoC - \EntrySoCDominatingReplacement) + \WearDensityFunction(\EntrySoCDominatingReplacement, \EntrySoCDominatingReplacement + \AssumedSoC - \EntrySoCDominatingReplacement)\\
				&=\PeriodEnergyCost\cdot(\AssumedSoC - \EntrySoCDominatingReplacement) + \WearDensityFunction(\EntrySoCDominatingReplacement + \AssumedSoC - \EntrySoCDominatingReplacement) - \WearDensityFunction(\EntrySoCDominatingReplacement)\\
				&=\PeriodEnergyCost\cdot(\AssumedSoC - \EntrySoCDominatingReplacement) + \WearDensityFunction(\AssumedSoC) - \WearDensityFunction(\EntrySoCDominatingReplacement)\\
				&=(\PeriodEnergyCost\cdot\AssumedSoC + \WearDensityFunction(\AssumedSoC) - \WearDensityFunction(0)) - (\PeriodEnergyCost\cdot\EntrySoCDominatingReplacement + \WearDensityFunction(\EntrySoCDominatingReplacement) - \WearDensityFunction(0))\\
				&=(\PeriodEnergyCost\cdot\AssumedSoC + \WearDensityFunction(0, \AssumedSoC)) - (\PeriodEnergyCost\cdot\EntrySoCDominatingReplacement + \WearDensityFunction(0, \EntrySoCDominatingReplacement))\\
				&=\DeltaChargingCost[0][\AssumedSoC]-\DeltaChargingCost[0][\EntrySoCDominatingReplacement].
			\end{align*}
			Concluding, we have $\InvTauProfile(\AssumedSoC) = \InverseDominatingReplacementProfile(\AssumedSoC)$ and $\DominatingReplacementProfile(\AssumedCost) \geq \ToBeDominatedProfile(\AssumedCost)$ such that  $\InvTauProfile(\AssumedSoC) = \InverseDominatingReplacementProfile(\AssumedSoC) \leq \InverseToBeDominatedProfile(\AssumedSoC)$, which contradicts the assumption. See Figure~\ref{fig:proof-example:new-better-intermediate} for an illustration.
		\end{paragraphs}
	\end{paragraphs}
	\item[Case 3:] $\AssumedCost \geq \CostProfileMaxCost[\ToBeDominatedLabel]$.\\
	Case 2 implies that ${\exists \Label'' \in \SetOfLabels': \CostProfile[\Label''](\CostProfileMaxCost[\ToBeDominatedLabel]) \geq \CostProfileMaxSoC[\ToBeDominatedLabel]}$. Hence, as ${\CostProfile[\ToBeDominatedLabel](\AssumedCost) = \CostProfileMaxSoC[\ToBeDominatedLabel]}$ by definition, we get ${\CostProfile[\Label''](\CostProfileMaxCost[\ToBeDominatedLabel]) \geq \CostProfileMaxSoC[\ToBeDominatedLabel]}$, which contradicts the assumption.
\end{paragraphs}

\noindent Concluding, for any $c \in \mathbb{R}$, there exists a $\Label'' \in \SetOfLabels'$, such that $\CostProfile[\Label''](c) \geq \CostProfile[\Label'](c)$. Hence, $\max_{\Label'' \in \SetOfLabels'}\{\Label''(\AssumedCost)\} \geq \ToBeDominatedProfile(\AssumedCost)$, and Theorem~\ref{theorem:one} holds.
\endproof

\branching*

\proof Assume the contrary, i.e., that schedules $\Schedule_1, \dots, \Schedule_n$ exist in basic feasible ${\sigma = (\ScheduleUsed[\Vehicle_1][\Schedule_1], \dots, \ScheduleUsed[\Vehicle_n][\Schedule_n])}$ with $0 < \ScheduleUsed < 1$, and $\sum_{\ScheduleUsed \in \sigma} \ScheduleUsed \ScheduleMatrix_{\Period, \Charger} < \ChargerCapacity$. Let $\Vehicle' \in \SetOfVehicles$ such that $\exists \Schedule_i, \Schedule_j \in \SetOfVehicleSchedules[\Vehicle']$ with $i \neq j$ and $0 < \ScheduleUsed[\Vehicle'][\Schedule_i] \leq \ScheduleUsed[\Vehicle'][\Schedule_j] < 1$ in $\sigma$. Here, convexity Constraints~\eqref{MP-mip:oneSchedulePerVehicle} ensure the existence of such $\Schedule_i$ and $\Schedule_j$. As charger capacity Constraints~\eqref{MP-mip:stationCapacityRespected} are non-binding, there exists $\epsilon > 0$ such that ${\sigma' = (\ScheduleUsed[\Vehicle_1][\Schedule_1], \dots, \ScheduleUsed[\Vehicle'][\Schedule_i] + \epsilon, \dots, \ScheduleUsed[\Vehicle'][\Schedule_j] - \epsilon, \dots, \ScheduleUsed[\Vehicle_n][\Schedule_n])}$ is feasible. This implies that $\sigma$ is not basic and thus contradicts the assumption.
\endproof

	\section{Fundamentals}
\label{app:technical-fundamentals}
The following sections detail the methodology behind and the derivation of charging functions $\ChargeFunction$ and \gls{abk:wdf}~$\widetilde{\WearDensityFunction}$: we first provide an overview of the electro-chemical fundamentals of the charging process and discuss a charging scheme that prevents overcharging and thus critically damaging the battery's internals in Section~\ref{app:subsec:modeling-recharging-operations}. Afterwards, we show how to model this charging scheme in functions~$\ChargeFunction$ (Section~\ref{app:subsec:modeling-pwl-charging}). Finally, Section~\ref{app:subsec:battery-degradation} details how we capture battery health considerations in our optimization problem. For an in-depth review of battery modeling for \glspl{abk:ecv} in particular and electro-chemical cells in general, we refer to \cite{PelletierJabaliEtAl2017} and \cite{Franco2015}.
\subsection{The Constant Current-Constant Voltage charging scheme}
\label{app:subsec:modeling-recharging-operations}

\glsreset{abk:cc-cv}
A battery's capacity can be measured in several units: Ampere-hours, Coulombs, and kWh, each proper in different application cases. To avoid confusion arising from handling these technicalities in the remainder of this section, we use the concept of \gls{abk:soc}, which expresses the battery's unit-independent state of charge relative to its nominal capacity, i.e., a \gls{abk:soc} of $0\%$ corresponds to an empty and a \gls{abk:soc} of $100\%$ to a full battery.

Most electric vehicles use \Gls{abk:lhio} batteries for energy storage. These are commonly charged with a \gls{abk:cc-cv} charging scheme to prevent critically damaging the battery's internals by overcharging. This process is best understood using the battery model developed in \cite{TremblayDessaintEtAl2007} and \cite{Zang2019}, which is specifically tailored to \gls{abk:ecv} applications.
\citeauthor{TremblayDessaintEtAl2007} model the battery as a controlled voltage source in series with a resistor (see Figure~\ref{fig:ersatzschaltbild_batterie}), which allows expressing the dynamics of the charging process as:
\begin{figure}[b]
	\centering
	\caption{Circuit diagram of the battery model developed in \cite{TremblayDessaintEtAl2007}}.
	\includegraphics[width=0.4\linewidth]{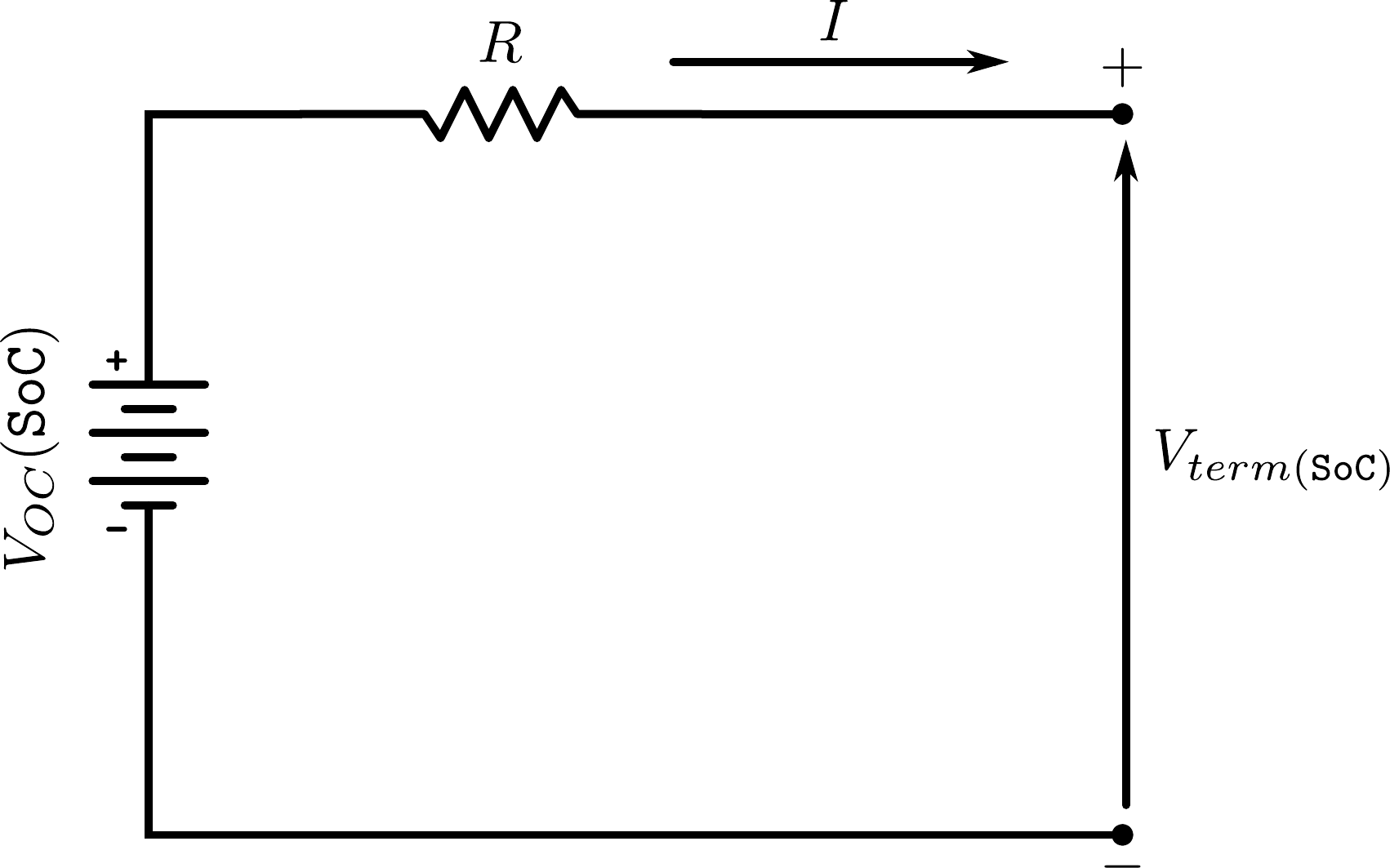}
	\label{fig:ersatzschaltbild_batterie}
\end{figure}
\begin{equation}
	\label{eq:vterm}
	\TerminalVoltage \coloneqq \OpenCircuitVoltage + R \cdot I.
\end{equation}
Equation~\eqref{eq:vterm} states the relationship between terminal voltage ($\TerminalVoltage$), which refers to the voltage measured across the battery's terminals during the charging process, the open-circuit voltage ($\OpenCircuitVoltage$), which corresponds to the terminal voltage in a disconnected state, and the charging current $I$.
The battery's internal resistance is denoted by $R$ and varies with several exogenous factors, e.g., temperature, load, and battery age. For the sake of simplicity, we assume this resistance to be constant (cf. \cite{PelletierJabaliEtAl2017}). \\
The open circuit voltage, $\OpenCircuitVoltage$, increases (non-linearly) with the battery's \gls{abk:soc} and is often used as an indicator of the battery's charge level. In practice, $\OpenCircuitVoltage$ is often approximated using Equation~\eqref{eq:ocv}, parameterized with experimental data \citep[cf.][]{MarraFawzyEtAl2012,PelletierJabaliEtAl2017}:
\begin{equation}
	\label{eq:ocv}
	\OpenCircuitVoltage[\SoC(t)] \coloneqq E_0 - \frac{K}{\SoC(t)} + A\exp(-BQ(1- \SoC(t))).
\end{equation}
Here, $E_0$ denotes the battery's constant voltage, $Q$ corresponds to the battery'sä capacity in ampere-hours, $A$ and $B$ are parameters, and $K$ is the polarization voltage.
To prevent damaging the battery's electrodes, the charging current $I$ and the terminal voltage $\TerminalVoltage$ must remain within charger and battery-dependent bounds $I_{\max}$ and $\MaxTerminalVoltage$, respectively. Usually, the maximum terminal voltage $\MaxTerminalVoltage$ lies well below $\TerminalVoltage[100\%] = \OpenCircuitVoltage[100\%] + R \cdot I_{\max}$ and is hence exceeded before the battery is fully charged. To respect this voltage threshold, the \gls{abk:cc-cv} charging scheme is as follows: first the charging current $I$ is held constant at $I_{\max}$ in the \textit{\acrshort{abk:cc}} charging phase until $\TerminalVoltage$ reaches $\MaxTerminalVoltage$, at which point the \textit{\acrshort{abk:cv}} charging phase begins. Here, the charging current $I$ is steadily reduced such that $\TerminalVoltage$ remains at but does not exceed $\MaxTerminalVoltage$. The \gls{abk:cv} phase ends when the charging current drops below manufacturer recommendations. The \gls{abk:soc} evolution over time during \gls{abk:cc-cv} charging is thus non-linear and concave, as illustrated in Figure~\ref{fig:non-linear-charging-pwl}. 

\subsection{Modeling CC-CV charging}
\label{app:subsec:modeling-pwl-charging}
We capture the charger-specific non-linear behavior of the \gls{abk:cc-cv} charging process in charging functions $\ChargeFunction(\tau)$, which map the time spent charging at charger $\Charger$ to the resulting \gls{abk:soc} when charging with an initially empty battery. In what follows, we describe how to derive $\ChargeFunction(\tau)$ from battery and charger specifications.
Let $I_{\max}^{\Charger}$ be the maximum charging current of charger $\Charger$ and $\tau^{\Charger}_{CV}$ be the point in time where the \gls{abk:cv} phase of charger $\Charger$ begins. We can then express $\ChargeFunction$ using auxiliary functions $\ChargeFunction^{CC}(\tau)$ and $\ChargeFunction^{CV}(\tau)$, which correspond to the \acrshort{abk:cc} and \gls{abk:cv} phases of the charging process, respectively, in Equations \eqref{app:eq:univariate-charge-function-first}-\eqref{app:eq:univariate-charge-function-last}:
\begin{subequations}
	\label{app:eq:univariate-charge-function}
	\begin{equation}
		\ChargeFunction^{CC}(\tau) \coloneqq \ChargeFunction(0) + \frac{I^{\Charger}_{\max} \cdot \tau}{Q},
		\label{app:eq:univariate-charge-function-first}
	\end{equation}
	\begin{equation}
		V_{CC} \coloneqq \OpenCircuitVoltage[\texttt{SoC($\tau$)}] + R \cdot 3600 \cdot Q \cdot \frac{\partial \ChargeFunction^{CV}(\tau)}{\partial \tau},
	\end{equation}
	\begin{equation}
		\frac{\partial \ChargeFunction^{CV}(\tau)}{\partial \tau} \coloneqq \frac{V_{CC} - \OpenCircuitVoltage[\ChargeFunction^{CV}(\tau)]}{R\cdot 3600 \cdot Q},
	\end{equation}
	\begin{equation}
		\label{app:eq:univariate-charge-function-last}
		\widetilde{\ChargeFunction}(\tau) \coloneqq \begin{cases}
			\ChargeFunction^{CC}(\tau) &\text{if } \tau \leq \tau^{\Charger}_{CV}\\
			\ChargeFunction^{CV}(\tau) &\text{otherwise.}\\
		\end{cases}	
	\end{equation}
\end{subequations}
Solving Equations~\eqref{app:eq:univariate-charge-function-first}-\eqref{app:eq:univariate-charge-function-last}, we obtain an accurate model of the charging process, which we can easily incorporate into our planning problem. We note that $\widetilde{\ChargeFunction}$ is concave for realistic charger models and, without loss of generality, extend the definitions of $\widetilde{\ChargeFunction}$ and $\widetilde{\ChargeFunction}^{-1}$ to avoid edge cases in the main body of this work. Equations~\eqref{app:eq:univariate-charge-function-fixed} and \eqref{app:eq:univariate-charge-function-inverse} state the final definitions.
\begin{multicols}{2}
	\noindent
	\begin{equation}
		\label{app:eq:univariate-charge-function-fixed}
		\ChargeFunction(\tau) \coloneqq \begin{cases}
			\widetilde{\ChargeFunction} &\text{if } \tau \leq 0\\
			\widetilde{\ChargeFunction}(\tau_{\max}) &\text{if } \tau \geq \tau_{\max}\\
			\widetilde{\ChargeFunction}(\tau) &\text{otherwise.}\\
		\end{cases}
	\end{equation}
	\begin{equation}
		\label{app:eq:univariate-charge-function-inverse}
		\ChargeFunction^{-1}(\beta) \coloneqq \begin{cases}
			0 &\text{if } \beta \leq \widetilde{\ChargeFunction}(0)\\
			\tau_{\max} &\text{if } \beta \geq \widetilde{\ChargeFunction}(\tau_{\max})\\
			\widetilde{\ChargeFunction}^{-1}(\beta) &\text{otherwise.}\\
		\end{cases}
	\end{equation}
\end{multicols}
\noindent Here, $\tau_{\max}$ represents the point in time at which the charging current falls below manufacturer recommendations. We assume that $\tau_{\max}$ is finite, such that there exists some $\tau_{\max} > 0$ with ${\ChargeFunction(\tau_{\max}) = \MaxSoC}$. Finally, we note that $\ChargeFunction$ are continuous functions and that the main body of this work, in line with \citet{PelletierJabaliEtAl2017}, \citet{MontoyaGueretEtAl2017}, and \citet{FrogerMendozaEtAl2019}, instead uses $\ChargeFunction$ to refer to the piecewise linear approximation of $\ChargeFunction$.
\subsection{Battery degradation}
\label{app:subsec:battery-degradation}
\glsreset{abk:wdf}
The \gls{abk:cc-cv} charging scheme prevents critically damaging a battery through overcharging. However, overcharging is not the only cause of accelerated degradation. More precisely, the magnitude of battery degradation resulting from (dis-)charging depends on (endogenous) factors such as cycle depth, charging current, and residual \gls{abk:soc}. Hence, different charge scheduling decisions have different impacts on battery life. In fact, there exists a trade-off between utilization and battery degradation, that an operator can use to her advantage. Capturing this trade-off requires quantifying the impact of charge scheduling decisions on battery health in an analytical model. Such models take one of two approaches: they either model battery wear based on the underlying electro-chemical processes, or pursue an empirical approach based on experimental data \citep[cf.][]{ReniersMulderEtAl2019}. As data required to parameterize the former is often unavailable and may vary between individual cells, we rely on an empirical approach to quantify battery degradation. To this end, we follow the approach from \citet{PelletierJabaliEtAl2018} and base our model on the work of \cite{HanHanEtAl2014}.

\cite{HanHanEtAl2014} relate battery price to cycle life specifications supplied by manufacturers, specifically to the \gls{abk:dod-acc} curve. Each point on the \gls{abk:dod-acc} curve, $ACC(D)$, corresponds to the number of cycles achievable before the battery becomes unusable when it is cycled at the respective \gls{abk:dod}. For instance, $ACC(20\%) = 2500$ indicates that the battery can be discharged from 100\% to 80\% and then recharged back to 100\% \gls{abk:soc} 2500 times before capacity and power fade render it ineffective. 

The \gls{abk:dod-acc} function establishes a relationship between battery life and price: dividing the battery price by the total energy transferred over $ACC(D)$ cycles gives the average wear cost of (dis-)charging when cycling the battery at a \gls{abk:dod} of $D$, denoted $AWC(D)$.
However, the average wear cost function is only of limited use as it is only valid if the battery is always cycled in the same fashion, i.e., from $100\%$ \gls{abk:soc} to $1-D$ and back to $100\%$. \cite{HanHanEtAl2014} address this issue by combining $n$ (equidistant) points on the \gls{abk:dod-acc} curve according to the following methodology: let ${S \coloneqq [S_1, \dots, S_n]}$ be the \gls{abk:soc} values corresponding to the given \gls{abk:dod-acc} points and let
\begin{equation}
	Q(s) \coloneqq ACC(1 - s) \cdot 2 \cdot (1 - s) \cdot C
\end{equation}
denote the total amount of energy transferred when cycling a battery with capacity $C$ to a \gls{abk:soc} of~$s$.
For the highest \gls{abk:soc} segment $S_n$, the battery price $c_{\text{bat}}$ must equal the wear cost $\widetilde{\WearDensityFunction}(S_n)$ of energy charged on the segment $[S_n, 100\%]$, multiplied by the total amount of energy transferred over the battery's lifespan when cycling at ${D = 1 - S_n}$. Equation~\eqref{app:eq:first-segment-wdf} formalizes this relationship:
\begin{equation}
	\label{app:eq:first-segment-wdf}
	c_{\text{bat}} = Q(S_n) \cdot \widetilde{\WearDensityFunction}(S_n).
\end{equation}
Each charging cycle at a \gls{abk:dod} of $1 - S_{n-1}$ also cycles the battery at $1 - S_{n}$. Hence, the average wear cost must be at least $\widetilde{\WearDensityFunction}(S_n)$. Formally, we have $AWC(1 - S_{n-1}) = \widetilde{\WearDensityFunction}(S_n) + \epsilon$ for some $\epsilon > 0$. We thus have $AWC(1 - S_{i}) = \sum_{j=i}^{n} \widetilde{\WearDensityFunction}(S_j)$ by induction, i.e., the average wear cost corresponds to the sum of wear costs incurred on all utilized segments. Hence, we can generalize Equation~\eqref{app:eq:first-segment-wdf} to Equation~\eqref{app:eq:extended-segment-wdf}:
\begin{equation}
	\label{app:eq:extended-segment-wdf}
	c_{\text{bat}} = Q(S_i) \cdot (\sum_{j=i}^n \widetilde{\WearDensityFunction}(S_j)).
\end{equation}
Equation~\eqref{app:eq:extended-segment-wdf} yields a linear equation system of size $n$, which can be solved for $\widetilde{\WearDensityFunction}(s)$ at discrete points $s \in \{S_1, \dots, S_n\}$ \citep[cf.][]{HanHanEtAl2014}. Setting $\widetilde{\WearDensityFunction}(s) \coloneqq \widetilde{\WearDensityFunction}(S_i)$ for ${s \in \lbrack S_i, S_{i+1}\rbrack, 0 \leq i < n}$ finally yields a piecewise constant function, called the \emph{wear density function}, which gives the unit cost of charging at a certain \gls{abk:soc}.
We utilize this function to compute the \emph{\acrlong{abk:wdf}}, denoted by ${\WearDensityFunction: \SoC \mapsto \texttt{cost}}$. Specifically, we obtain $\WearDensityFunction$ by integrating $\widetilde{\WearDensityFunction}$, which yields a convex piecewise linear function stating the total cost of charging an initially empty battery up to a certain \gls{abk:soc}.
	\section{Implementation details}
\label{app:algorithm-details}

\begin{algorithm}[H]
	\caption{Label setting search}
	\label{alg:detailed-pseudocode-subproblem}
	\textbf{Initialization:}\\
	$\SetOfUnsettledLabels[\SourceNode] \coloneqq \{\Label_{\SourceNode}\}$\label{code:initialize-root-label}\;
	$\NodeQueue \coloneqq \{\SourceNode\}$\label{code:initialize-nodequeue}\;
	\While{$\NodeQueue$.\texttt{notEmpty()}}{
		\label{code:body-begin}
		$\OriginVertex \coloneqq \NodeQueue$.\texttt{pop()}\label{code:extract-label-begin}\;
		$\Label \coloneqq$ \texttt{extract\_min}($\SetOfUnsettledLabels[\OriginVertex]$)\label{code:extract-label-end}\;
		\If{$\OriginVertex = \SinkNode$}{
			\Return $\CostProfileMinCost$\label{code:extracted-at-sink}\;
		}
		\If{$\exists \Label' \in \SetOfSettledLabels[\OriginVertex], \Label' \LabelDominatesLabel \Label$}{
			\label{code:pairwise-dominance-check}
			\Continue\;
		}
		\For{$(\OriginVertex, \TargetVertex) \in \SetOfArcs$}{\label{code:label-propagation-begin}
			\If{$(\OriginVertex, \TargetVertex) \notin \SetOfChargingEdges$}{
				$\SetOfGeneratedLabels \coloneqq \{\Label \underset{(\OriginVertex, \TargetVertex)}{\gets} /\}$\;
			}
			\Else {
				\label{code:spawn-charging-decisions-begin}
				\tcp{Track charging decisions in period $\PeriodOfVertex[\OriginVertex]$}
				$\SetOfGeneratedLabels \coloneqq \{\Label \overunderset{\ChargerOfVertex[\OriginVertex]}{(\OriginVertex, \TargetVertex)}{\gets} c \mid c \in \SetOfBreakpoints[{\CostProfile[\Label]}]\}$\;
				\label{code:replace-at-u}
				
				\tcp{Charge for $\tau = \PeriodDuration$ in $\PeriodOfVertex[\OriginVertex]$}
				$\SetOfGeneratedLabels \coloneqq \SetOfGeneratedLabels \cup \{\Label \underset{(\OriginVertex, \TargetVertex)}{\gets} \PeriodDuration\}$\;\label{code:commit-at-u}
				\label{code:spawn-charging-decisions-end}
			}
			\label{code:label-propagation-end}
			\texttt{remove\_set\_dominated($\SetOfGeneratedLabels$)}\;\label{code:prune-generated-labels}
			\tcp{Enqueue feasible labels}
			\For{$\Label' \in \mathcal{L}_{new}$}{
				\label{code:insert-feasible-begin}
				\If{\texttt{feasible}$(\Label')$}{
					\texttt{insert}$(\SetOfUnsettledLabels[\TargetVertex], \Label')$\;
				}
			}
			\label{code:insert-feasible-end}
		}
		\texttt{insert}$(\SetOfSettledLabels[\OriginVertex], \Label)$\;\label{code:body-end}
	}
	\Return \textit{infeasible}\;
\end{algorithm}
\newcommand{\HashFunction}{\bar{h}}

Algorithm~\ref{alg:detailed-pseudocode-subproblem} details our label-setting search procedure.
The algorithm relies on several data structures: first, it maintains a set of settled ($\SetOfSettledLabels[\Vertex]$) and unsettled ($\SetOfUnsettledLabels[\Vertex]$) labels for each vertex ${\Vertex \in \SetOfVertices}$. These keep track of already developed paths and collect candidates for expansion, respectively. To establish the label-setting property of our algorithm, we store vertices $\Vertex$ with unsettled labels, i.e., potential candidates for expansion, in a priority queue $\NodeQueue$. This queue orders vertices according to the cost of the cheapest unsettled label at the respective vertex, formally, ${\min_{\Label \in \SetOfUnsettledLabels[\Vertex]}\CostProfileMinCost \geq \min_{\Label \in \SetOfUnsettledLabels[\Vertex']}\CostProfileMinCost \Rightarrow v \succeq_{\NodeQueue} v'}$ for $\Vertex, \Vertex' \in \NodeQueue$. We break ties according to the vertex's period index in descending order such that we prefer vertices closer to the sink node.

We initialize $\NodeQueue$ with the source vertex and root label (Lines~\ref{code:initialize-root-label}-\ref{code:initialize-nodequeue}). The main body of the algorithm (Lines~\ref{code:body-begin}-\ref{code:body-end}) iteratively extracts labels from $\NodeQueue$ (Lines~\ref{code:extract-label-begin}-\ref{code:extract-label-end}) and propagates these along all adjacent arcs (Lines~\ref{code:label-propagation-begin}-\ref{code:label-propagation-end}). 
We propagate extracted labels $\Label \in \SetOfLabels_{\OriginVertex}$ along idle and service arcs~$(\Arc)$ according to $\ArcCost$, $\ArcConsumption$, and $\ArcOperation$.
Charging arcs require special treatment: here, it is possible to either commit to a charging decision at a previously visited station vertex $\OriginVertex'$ to then track charging trade-offs at station vertex~$\OriginVertex$, or to commit to charging at $\OriginVertex$, continuing to track decisions at $\OriginVertex'$. 
Line~\ref{code:replace-at-u} handles the former case and creates a label for each non-dominated charging decision at the station vertex~$\OriginVertex'$ tracked by $\Label$ so far, i.e., fixes the amount of charge replenished at~$\OriginVertex'$, and thus the arrival \gls{abk:soc} at $\OriginVertex$ to some value.
Line~\ref{code:commit-at-u} handles the latter case, i.e., spawns a label that charges at $\OriginVertex$ \emph{without} forcing a decision at the tracked station $i'$. This fixes the amount of time spent charging at $\OriginVertex$ to some value.
Note that ignoring the charging opportunity at $\OriginVertex$ corresponds to visiting the respective garage vertex and is thus not explicitly considered.
We prune the set of generated labels according to our set-based dominance criterion (cf. Definition~\ref{def:set-dominance}) in Line~\ref{code:prune-generated-labels}. 

Lines~\ref{code:insert-feasible-begin}-\ref{code:insert-feasible-end} insert feasible labels into $\SetOfUnsettledLabels[\TargetVertex]$, i.e., track potential candidates for expansion at vertex~$\TargetVertex$, updating the vertex queue accordingly. Finally, we settle the original label at vertex~$\OriginVertex$ for future dominance checks.
The algorithm terminates when a label is extracted at the sink (Line~\ref{code:extracted-at-sink}), or no unsettled labels remain.

We implement $\SetOfUnsettledLabels$ as a min-heap and key labels by minimum cost. We postpone pairwise dominance checks against already settled labels to label extraction (Line~\ref{code:pairwise-dominance-check}). This lazy approach serves two purposes: first, it avoids superfluous dominance checks for labels never considered during the search; second, it delays dominance checks as much as possible to maximize the number of candidates for domination.

We further rely on two techniques to speed up dominance checks:
first, we keep $\SetOfSettledLabels[\OriginVertex]$ sorted by maximum reachable \gls{abk:soc}. This allows skipping superfluous dominance checks against settled labels with a lower maximum \gls{abk:soc}. Second, we maintain a hash table of settled and unsettled labels at each vertex. We probe this hash table and abort the dominance check if an equivalent label is found. These strategies minimize the number of (explicit) dominance checks required to maintain our dominance invariant.

	\section{Compact formulation}
\label{sec:compact-formulation}

In the following, we model our planning problem as a mixed integer program, which we state as a shortest path problem on the time-expanded network presented in Section~\ref{sec:network-design}.
\mipref{mips:dyn-tour} comprises the following decision variables: 
binary variable $\TravelsEdge$ indicates that vehicle $k$ traverses arc $(i, j)$. 
Variables $\ConvexMultEntrySoC$ and $\ConvexMultExitSoC$ model charging operations as piecewise linear functions. To this end, continuous variables $\ConvexMultEntrySoC$ and $\ConvexMultExitSoC$ are convex multipliers associated with the breakpoints of the respective piecewise-linear charging functions, i.e., give the contribution of each breakpoint to the function value. We establish a \gls{abk:sos2} relationship between variables associated with the same vertex and vehicle. Variables $\ConvexMultEntrySoCDeg$ and $\ConvexMultExitSoCDeg$ model the \acrshort{abk:wdf} analogously. We use continuous variables $\VehicleArrivalSoC$ and $\VehicleReplenishedCharge$ to track the arrival \gls{abk:soc} and total amount of energy replenished by vehicle $\Vehicle$ at vertex $v \in \SetOfVertices$. With the notation summarized in Table~\ref{table:dyn-tour-mip-param}, our \gls{abk:mip} is as follows.
\begin{table}
	\centering
	\caption{\centering Parameters and variables of the compact formulation.}
	\begin{threeparttable}
		{\footnotesize{\begin{tabular}{@{}ll@{}}
					\toprule
					$\Vertices$ & set of vertices\\
					$\ChargerVertices$ & set of charger vertices\\
					$\VerticesOfCharger$ & set of vertices associated with charger $\Charger$\\
					$\IncomingArcs[\Vertex]$ & set of incoming arcs at vertex $\Vertex$\\
					$\OutgoingArcs[\Vertex]$ & set of outgoing arcs at vertex $\Vertex$\\
					$\SetOfChargingArcs$ & set of charging arcs\\
					$\SetOfServiceArcs$ & set of service arcs\\
					$\PeriodEnergyCost$ & energy cost in period $\Period$\\
					$\ArcConsumption$ & charge consumption of arc $(\Arc)$\\
					$\ArcCost$ & cost of arc $(\Arc)$\\
					\midrule
					$\SetOfChargers$ & set of chargers\\
					$\ChargerCapacity$ & charger capacity for $\AllChargers$\\
					$\SetOfChargerSegments$ & breakpoints of the linearized charging function\\
					$\ChargerSoCBreakpoint$ & SoC associated with breakpoint $\ChargerSegment \in \SetOfChargerSegments$\\
					$\ChargerTimeBreakpoint$ & Time associated with breakpoint $\ChargerSegment \in \SetOfChargerSegments$\\
					\midrule
					$\SetOfDegradationSegments$ & breakpoints of the \gls{abk:wdf}\\
					$\DegradationSoCBreakpoint$ & SoC associated with breakpoint $\Breakpoint \in \SetOfDegradationSegments$\\
					$\DegradationCostBreakpoint$ & Costs per kWh associated with breakpoint $\Breakpoint \in \SetOfDegradationSegments$\\
					\midrule
					$\SetOfPeriods$ & set of periods in the planning horizon\\
					\midrule
					$\SetOfVehicles$ & set of vehicles\\
					$\MaxSoC$ & maximum battery charge level (\gls{abk:soc})\\
					$\MinSoC$ & minimum battery charge level (\gls{abk:soc})\\
					\midrule
					$\VehicleArrivalSoC$ & \gls{abk:soc} with which vehicle $\Vehicle$ arrives at vertex $\Vertex$\\
					$\VehicleReplenishedCharge$ & \gls{abk:soc} charged/discharged at vertex $\Vertex$\\
					$\DegradationOfChargingArc$ & battery wear cost incurred by charging at $\Vertex$\\
					$\TravelsEdge$ & binary variable, indicating whether vehicle $\Vehicle$ traverses arc $(\Edge)$ (${\TravelsEdge = 1}$)\\
					{} & or not (${\TravelsEdge = 0}$)\\
					$\ConvexMultEntrySoC$ & Convex multipliers binding entry \gls{abk:soc} to $\ChargeFunction[{\ChargerOfVertex[\Vertex]}]$\\
					$\ConvexMultExitSoC$ & Convex multipliers binding exit \gls{abk:soc} to $\ChargeFunction[{\ChargerOfVertex[\Vertex]}]$\\
					$\ConvexMultEntrySoCDeg$ & Convex multipliers binding entry \gls{abk:soc} to $\WearDensityFunction$\\
					$\ConvexMultExitSoCDeg$ & Convex multipliers binding exit \gls{abk:soc} to $\WearDensityFunction$\\
					\bottomrule
		\end{tabular}}}
		{}
	\end{threeparttable}
	\label{table:dyn-tour-mip-param}
\end{table}
\setcounter{equation}{0}
\begin{subequations}
\renewcommand{\theequation}{\theparentequation.\arabic{equation}}
\label{mips:dyn-tour}
\setlength{\abovedisplayskip}{0pt}
\setlength{\belowdisplayskip}{0pt}
\setlength{\abovedisplayshortskip}{0pt}
\setlength{\belowdisplayshortskip}{0pt}
\setlength{\parskip}{0pt}
\begin{multline}
	\hfill \min \sum_{\AllVehicles} \sum_{\Vertex \in \ChargerVertices} \VehicleReplenishedCharge \cdot \PeriodEnergyCost[{\PeriodOfVertex[\Vertex]}] + \DegradationOfChargingArc \hfill
	\label{dyn-tour-mip:objective}
\end{multline}
\begin{multline}
	\quad \quad \sum_{(\Arc) \in \OutgoingArcs[\SourceNode]} \TravelsEdge[\Arc] = 1
	\hfill \AllVehicles
	\label{dyn-tour-mip:leave-source}
\end{multline}
\begin{multline}
	\quad \quad \sum_{(\Arc) \in \SetOfServiceArcs[\Tour]} \TravelsEdge[\ServiceArc] \geq 1
	\hfill \Tour \in \SetOfTours, \AllVehicles
	\label{dyn-tour-mip:enter-sink}
\end{multline}
\begin{multline}
	\quad \quad \sum_{(\Arc) \in \IncomingArcs[\Vertex]} \TravelsEdge[\Arc] - \sum_{(\Arc) \in \OutgoingArcs[\Vertex]} \TravelsEdge[\Arc] = 0
	\hfill \Vertex \in \SetOfVertices \setminus \{\SourceNode, \SinkNode\}, \AllVehicles
	\label{dyn-tour-mip:flow-propagation}
\end{multline}
\begin{multline}
	\quad \quad \VehicleArrivalSoC[\OriginVertex] + \ArcConsumption + \VehicleReplenishedCharge[\OriginVertex] \geq \VehicleArrivalSoC[\TargetVertex] - (1 - \TravelsEdge) \cdot \MaxSoC
	\hfill \forall (\Arc) \in \SetOfArcs, \AllVehicles
	\label{dyn-tour-mip:propagate-charge-lower}
\end{multline}
\begin{multline}
	\quad \quad \VehicleArrivalSoC[\OriginVertex] + \ArcConsumption + \VehicleReplenishedCharge[\OriginVertex] \leq \VehicleArrivalSoC[\TargetVertex] + (1 - \TravelsEdge) \cdot \MaxSoC
	\hfill \forall (\Arc) \in \SetOfArcs, \AllVehicles
	\label{dyn-tour-mip:propagate-charge-upper}
\end{multline}
\begin{multline}
	\quad \quad \MinSoC \leq \VehicleArrivalSoC[\Vertex] \leq \MaxSoC
	\hfill \AllVertices, \AllVehicles
	\label{dyn-tour-mip:soc-bounds}
\end{multline}
\begin{multline}
	\quad \quad \VehicleArrivalSoC[\SourceNode] = \MinSoC
	\hfill \AllVehicles
	\label{dyn-tour-mip:initialize-soc}
\end{multline}
\begin{multline}
	\label{dyn-tour-mip:bind-entry-soc-to-pwl}
	\quad\quad \sum_{\Breakpoint \in \SetOfBreakpoints[{\ChargerOfVertex[\Vertex]}]} \ConvexMultEntrySoC \cdot \ChargerSoCBreakpoint[{\ChargerOfVertex[\Vertex]}, \Breakpoint] = \VehicleArrivalSoC[\Vertex] \hfill \Vertex \in \SetOfStationNodes, \AllVehicles
\end{multline}
\begin{multline}
	\label{dyn-tour-mip:entry-soc-pwl-convexity}
	\quad\quad \sum_{\Breakpoint \in \SetOfBreakpoints[{\ChargerOfVertex[\Vertex]}]} \ConvexMultEntrySoC = 1
	\hfill \Vertex \in \SetOfStationNodes, \AllVehicles
\end{multline}
\begin{multline}
	\label{dyn-tour-mip:bind-exit-soc-to-pwl}
	\quad\quad \sum_{\Breakpoint \in \SetOfBreakpoints[{\ChargerOfVertex[\Vertex]}]} \ConvexMultExitSoC \cdot \ChargerSoCBreakpoint[{\ChargerOfVertex[\Vertex]}, \Breakpoint] - \sum_{\Breakpoint \in \SetOfBreakpoints[{\ChargerOfVertex[\Vertex]}]} \ConvexMultEntrySoC \cdot \ChargerSoCBreakpoint[{\ChargerOfVertex[\Vertex]}, \Breakpoint] = \VehicleReplenishedCharge
	\hfill \Vertex \in \SetOfStationNodes, \AllVehicles
\end{multline}
\begin{multline}
	\label{dyn-tour-mip:exit-soc-pwl-convexity}
	\quad\quad \sum_{\Breakpoint \in \SetOfBreakpoints[{\ChargerOfVertex[\Vertex]}]} \ConvexMultExitSoC = 1
	\hfill \Vertex \in \SetOfStationNodes, \AllVehicles
\end{multline}
\begin{multline}
	\quad\quad \sum_{\Breakpoint \in \SetOfBreakpoints[{\ChargerOfVertex[\Vertex]}]} \ConvexMultExitSoC \cdot \ChargerTimeBreakpoint[{\ChargerOfVertex[\Vertex]}, \Breakpoint] - \sum_{\Breakpoint \in \SetOfBreakpoints[{\ChargerOfVertex[\Vertex]}]} \ConvexMultEntrySoC \cdot \ChargerTimeBreakpoint[{\ChargerOfVertex[\Vertex]}, \Breakpoint] \leq \PeriodDuration \cdot \sum_{(\Arc) \in \OutgoingArcs[\Vertex]} \TravelsEdge[\Arc] 
	\hfill \Vertex \in \VerticesOfCharger, \AllVehicles
	\label{dyn-tour-mip:limit-gamma-by-period-length}
\end{multline}
\begin{multline}
	\label{dyn-tour-mip:bind-entry-soc-to-deg-pwl}
	\quad\quad \sum_{\Breakpoint \in \SetOfDegradationSegments} \ConvexMultEntrySoCDeg \cdot \DegradationSoCBreakpoint = \VehicleArrivalSoC[\Vertex] \hfill \Vertex \in \SetOfStationNodes, \AllVehicles
\end{multline}
\begin{multline}
	\label{dyn-tour-mip:entry-soc-deg-pwl-convexity}
	\quad\quad \sum_{\Breakpoint \in \SetOfDegradationSegments} \ConvexMultEntrySoCDeg = 1
	\hfill \Vertex \in \SetOfStationNodes, \AllVehicles
\end{multline}
\begin{multline}
	\label{dyn-tour-mip:bind-exit-soc-to-deg-pwl}
	\quad\quad \sum_{\Breakpoint \in \SetOfDegradationSegments} \ConvexMultExitSoCDeg \cdot \DegradationSoCBreakpoint - \sum_{\Breakpoint \in \SetOfBreakpoints[\ChargerOfVertex]} \ConvexMultEntrySoCDeg \cdot \DegradationSoCBreakpoint = \VehicleReplenishedCharge
	\hfill \Vertex \in \SetOfStationNodes, \AllVehicles
\end{multline}
\begin{multline}
	\label{dyn-tour-mip:exit-soc-deg-pwl-convexity}
	\quad\quad \sum_{\Breakpoint \in \SetOfDegradationSegments} \ConvexMultExitSoCDeg = 1
	\hfill \Vertex \in \SetOfStationNodes, \AllVehicles
\end{multline}
\begin{multline}
	\quad\quad \sum_{\Breakpoint \in \SetOfDegradationSegments} \ConvexMultExitSoCDeg \cdot \DegradationCostBreakpoint - \sum_{\Breakpoint \in \SetOfDegradationSegments} \ConvexMultEntrySoCDeg \cdot \DegradationCostBreakpoint = \DegradationOfChargingArc
	\hfill \Vertex \in \VerticesOfCharger, \AllVehicles
	\label{dyn-tour-mip:limit-rho-by-period-length}
\end{multline}
\begin{multline}
	\quad\quad \sum_{\Vehicle \in \SetOfVehicles} \sum_{(\Arc) \in \OutgoingArcs[\Vertex]} \TravelsEdge[\Arc] \leq \ChargerCapacity[{\ChargerOfVertex[\Vertex]}]
	\hfill \forall \Vertex \in \bigcup_{\Vehicle \in \SetOfVehicles}\SetOfStationNodes
	\label{dyn-tour-mip:capacity}
\end{multline}
\begin{multline}
	\quad\quad \TravelsEdge \in \{0, 1\}
	\hfill \AllEdges, \AllVehicles
	\label{dyn-tour-mip:bound1}
\end{multline}
\begin{multline}
	\quad\quad \VehicleArrivalSoC, \VehicleReplenishedCharge, \DegradationOfChargingArc \geq 0
	\hfill \AllVertices, \AllVehicles
	\label{dyn-tour-mip:bound2}
\end{multline}
\begin{multline}
	\quad\quad \forall \Breakpoint \in \SetOfBreakpoints[{\ChargerOfVertex[\Vertex]}]: \ConvexMultEntrySoC, \ConvexMultExitSoC \in SOS2 \hfill
	\hfill \Vertex \in \ChargerVertices, \AllVehicles
	\label{dyn-tour-mip:bound3}
\end{multline}
\begin{multline}
	\quad\quad \forall \Breakpoint \in \SetOfDegradationSegments: \ConvexMultEntrySoCDeg, \ConvexMultExitSoCDeg \in SOS2 \hfill
	\hfill \Vertex \in \ChargerVertices, \AllVehicles
	\label{dyn-tour-mip:bound4}
\end{multline}
\end{subequations}

\vspace{\baselineskip}
\noindent The objective function \eqref{dyn-tour-mip:objective} minimizes the total cost of the charging schedule, i.e., the sum of energy costs and battery wear incurred. Constraints~\eqref{dyn-tour-mip:leave-source} and \eqref{dyn-tour-mip:enter-sink} enforce an outgoing and incoming arc at the source and sink nodes, respectively. Constraints~\eqref{dyn-tour-mip:flow-propagation} set up flow conservation on all other vertices.
Constraints~\eqref{dyn-tour-mip:propagate-charge-lower}-\eqref{dyn-tour-mip:propagate-charge-upper} propagate the \gls{abk:soc}. The concaveness of the \gls{abk:wdf} requires strict equality. Consumption and charging operations are captured with~$\ArcConsumption$ and~$\VehicleReplenishedCharge[i]$, respectively. Constraints~\eqref{dyn-tour-mip:soc-bounds} ensure that manufacturer \gls{abk:soc} bounds are respected. Constraints~\eqref{dyn-tour-mip:initialize-soc} initialize~$\VehicleArrivalSoC[\SourceNode]$. Constraints~(\ref{dyn-tour-mip:bind-entry-soc-to-pwl}-\ref{dyn-tour-mip:exit-soc-pwl-convexity}) model charging operations as piecewise linear functions. Constraints~\eqref{dyn-tour-mip:limit-gamma-by-period-length} limit the maximum \gls{abk:soc} rechargeable at station nodes, i.e., ensure that the charging rate is respected, and establish a link between~$\TravelsEdge$ and~$\VehicleReplenishedCharge$, such that charging can occur only if the station node is visited. Constraints~\eqref{dyn-tour-mip:bind-entry-soc-to-deg-pwl}-\eqref{dyn-tour-mip:limit-rho-by-period-length} model battery degradation similarly. Constraints~\eqref{dyn-tour-mip:capacity} limit the number of simultaneous charging operations at each charger. Finally, Constraints~\eqref{dyn-tour-mip:bound1}-\eqref{dyn-tour-mip:bound4} state the domain of the decision variables and establish \gls{abk:sos2} sets.

	\section{Benchmark instance generation}
\label{app:benchmark-instance-generation}

\newcommand{\PWLMaxVal}{\nu}
\newcommand{\PWLMinSlope}{\chi_{\min}}
\newcommand{\PWLMaxSlope}{\chi_{\max}}
\newcommand{\NumberOfSegments}{n}

We choose a discretization step size of $30$ minutes and draw period energy prices uniformly from the interval $\lbrack0.5, 1.0\rbrack$ (\euro). We further assume a battery capacity of $80$ kWh with ${\MinSoC \coloneqq 0\%}$ and ${\MaxSoC \coloneqq 100\%}$.
To avoid biases in our numerical study, we generate charging functions $\ChargingFunction$ and the \gls{abk:wdf} randomly according to the following procedure:
Given parameters $\NumberOfSegments$, $\PWLMinSlope, \PWLMaxSlope$, and $\PWLMaxVal$, which correspond to the number of segments, minimum slope, maximum slope, and upper bound respectively, we assign a random weight $w_i$, and a slope drawn from interval $[\PWLMinSlope, \PWLMaxSlope]$ to each segment $i$. We then sort the segments by slope in descending (ascending) order such that the resulting function is convex (concave) for \gls{abk:wdf} and charging functions, respectively.
We transform the piecewise linear functions generated in this fashion to valid \gls{abk:wdf} and charging functions by scaling each segment $i$ to span $\PWLMaxVal \cdot \frac{w_i}{\sum_{j \in [1, n]} w_j}$ on the \gls{abk:soc} and time axes.
We generate the \gls{abk:wdf} with $\PWLMinSlope \coloneqq 0.1$, $\PWLMaxSlope \coloneqq 0.8$, and $\PWLMaxVal \coloneqq 80$. We use durations $\PWLMaxVal \coloneqq \lbrack150, 60, 90, 120, 75, 135\rbrack$ to generate chargers, such that fully charging the battery using the $i^{\text{th}}$ charger takes $\PWLMaxVal_i$ minutes. Unless otherwise specified, we distribute charger capacity evenly across all available chargers.
Finally, we generate a set of three operations for each day and vehicle. Each service operation consumes $50\%$ of the battery capacity, leading to a total discharge of $120$ kWh per day and vehicle. We distribute operation departure times randomly such that vehicles spend a minimum of one hour before each operation at the depot, and center the departure time windows according to the static case.
To ensure that the generated instances are comparable, we use independently seeded random engines for each parameter, such that the set of service operations of a one day instance is a subset of the service plan of the two and three-day instances generated from the same seed.
	\section{Data used in the example}
\label{app:example-details}

\begin{minipage}[b]{0.3\textwidth}
\begin{table}[H]
	\centering
	\caption{WDF.}
	\begin{threeparttable}
		{\footnotesize{
				\begin{tabular}{ccc}
					\toprule
					SoC & Cost & Unit cost \\
					\midrule
					$0$&$0$&-\\
					$2$&$1$&$0.5$\\
					$7$&$7$&$1.2$\\
					\bottomrule
				\end{tabular}	
		}}
		{}
	\end{threeparttable}
\end{table}
\end{minipage}
\begin{minipage}[b]{0.68\textwidth}
	\begin{table}[H]
		\centering
		\caption{Arcs of the time expanded network.}
		\begin{threeparttable}
			{\footnotesize{
					\begin{tabular}{lccc}
						\toprule
						Arc & Consumption ($\ArcConsumption$) & Fixed cost ($\ArcCost$) & Energy cost at origin \\
						\midrule
						$(\SourceNode, v_f)$&$0$&$2$&$0$\\
						$(v_f, v_2)$&$0$&$0$&$2.5$\\
						$(v_2, v_3)$&$1.5$&$0$&$0$\\
						$(v_3, v_g)$&$0$&$0$&$0$\\
						$(v_g, v_5)$&$0$&$0$&$0.75$\\
						$(v_5, \SinkNode) (1)$&$0$&$3.5$&$0$\\
						$(v_5, \SinkNode) (2)$&$0$&$4.25$&$0$\\
						\bottomrule
					\end{tabular}	
			}}
			{}
		\end{threeparttable}
	\end{table}
\end{minipage}

\begin{minipage}{0.45\textwidth}
	\vspace{0pt}
	\begin{table}[H]
		\begin{threeparttable}
			{\footnotesize{
					\begin{tabular}{ccc|ccc}
						\toprule
						\multicolumn{3}{c|}{f} & \multicolumn{3}{c}{g}\\
						\midrule
						Time & SoC & Rate & Time & SoC & Rate \\
						\midrule
						$0$&$0$&-&$0$&$0$&-\\
						$6$&$2$&$0.3333$&$2.5$&$2$&$0.8$\\
						$24.5$&$7$&$0.2707$&$12.25$&$7$&$0.5128$\\
						\bottomrule
					\end{tabular}	
			}}
			{}
		\end{threeparttable}
		\caption{The charging functions used.}
	\end{table}
\end{minipage}
\begin{minipage}{0.45\textwidth}
	\begin{table}[H]
		\begin{threeparttable}
			{\footnotesize{
					\begin{tabular}{ccc|ccc}
						\toprule
						\multicolumn{3}{c|}{f} & \multicolumn{3}{c}{g}\\
						\midrule
						Cost & SoC & Rate & Cost & SoC & Rate \\
						\midrule
						$0$&$0$&-&$0$&$0$&-\\
						$6$&$2$&$0.3333$&$2.5$&$2$&$0.8$\\
						$24.5$&$7$&$0.2702$&$12.25$&$7$&$0.5128$\\
						\bottomrule
					\end{tabular}	
			}}
			{}
		\end{threeparttable}
		\caption{The station cost profiles.}
	\end{table}
\end{minipage}

\begin{table}[H]
	\centering
	\caption{\centering Cost profiles created in the example.}
	\begin{threeparttable}
		{\tiny{
				\begin{tabular}{cllll}
					\toprule
					Cost profile & \multicolumn{4}{c}{Segments}\\
					\midrule
					$\CostProfile[{\Label_{\SourceNode}}]$ & $[-\infty, 0.0) \rightarrow [-\infty, -\infty]$ & $[0.0, \infty) \rightarrow [0.0, 0.0]$ & & \\
					$\CostProfile[{\Label_{v_f}}]$ & $[-\infty, 2.0) \rightarrow [-\infty, -\infty]$ & $[2.0, \infty) \rightarrow [0.0, 0.0]$ & & \\
					$\CostProfile[{\Label_{v_2}}]$ & $[-\infty, 2.0) \rightarrow [-\infty, -\infty]$ & $[2.0, 8.0) \rightarrow [0.0, 2.0]$ & $[8.0, 11.7) \rightarrow [2.0, 3.0)$ & $[11.7, \infty] \rightarrow [3.0, 3.0]$\\
					$\CostProfile[{\Label_{v_3}}]$ & $[-\infty, 6.5) \rightarrow [-\infty, -\infty]$ & $[6.5, 8.0) \rightarrow [0.0, 0.5]$ & $[8.0, 11.7) \rightarrow [0.5, 1.5)$ & $[11.7, \infty] \rightarrow [1.5, 1.5]$\\
					$\CostProfile[{\Label_{v_g}}]$ & $[-\infty, 6.5) \rightarrow [-\infty, -\infty]$ & $[6.5, 8.0) \rightarrow [0.0, 0.5]$ & $[8.0, 11.7) \rightarrow [0.5, 1.5)$ & $[11.7, \infty] \rightarrow [1.5, 1.5]$\\
					$\CostProfile[{\Label^1_{v_5}}]$ & $[-\infty, 6.5) \rightarrow [-\infty, -\infty]$ & $[6.5, 9.0) \rightarrow [0.0, 2.0]$ & $[9.0, 12.9) \rightarrow [2.0, 4.0)$ & $[12.9, \infty] \rightarrow [4.0, 4.0]$\\
					$\CostProfile[{\Label^2_{v_5}}]$ & $[-\infty, 12.9) \rightarrow [-\infty, -\infty]$ & $[12.9, 14.75) \rightarrow [4.0, 4.5]$ & $[14.75, 19.15) \rightarrow [4.5, 5.5)$ & $[14.75, \infty] \rightarrow [5.5, 5.5]$\\
					$\CostProfile[{\Label^3_{v_5}}]$ & $[-\infty, 8.0) \rightarrow [-\infty, -\infty]$ & $[8.0, 9.875) \rightarrow [0.5, 2.0]$ & $[9.875, 14.75) \rightarrow [2.0, 4.5)$ & $[14.75, \infty] \rightarrow [4.5, 4.5]$\\
					$\CostProfile[{\Label^4_{v_5}}]$ & $[-\infty, 11.7) \rightarrow [-\infty, -\infty]$ & $[11.7, 12.325) \rightarrow [1.5, 2.0]$ & $[12.325, 19.15) \rightarrow [2.0, 5.5)$ & $[12.325, \infty] \rightarrow [5.5, 5.5]$\\
					\midrule
					\multicolumn{5}{c}{$q_{(v_5, \SinkNode)} = 4.25$}\\
					\midrule
					$\CostProfile[{\Label^1_{\SinkNode}}]$ & $[-\infty, 11.925) \rightarrow [-\infty, -\infty]$ & $[11.925, 12.9) \rightarrow [0.0, 0.5]$ & $[12.9, \infty) \rightarrow [0.5, 0.5)$ & \\
					$\CostProfile[{\Label^2_{\SinkNode}}]$ & $[-\infty, 12.9) \rightarrow [-\infty, -\infty]$ & $[12.9, 14.75) \rightarrow [0.5, 1.0]$ & $[14.75, 19.15) \rightarrow [1.0, 2.0)$ & $[14.75, \infty] \rightarrow [2.0, 2.0]$\\
					\midrule
					\multicolumn{5}{c}{$q_{(v_5, \SinkNode)} = 3.5$}\\
					\midrule
					$\CostProfile[{\Label^2_{\SinkNode}}]$ & $[-\infty, 13.825) \rightarrow [-\infty, -\infty]$ & $[13.825, 14.75) \rightarrow [0.0, 0.25]$ & $[14.75, 19.15) \rightarrow [0.25, 1.25)$ & $[14.75, \infty] \rightarrow [1.25, 1.25]$\\
					$\CostProfile[{\Label^3_{\SinkNode}}]$ & $[-\infty, 14.2625) \rightarrow [-\infty, -\infty]$ & $[14.2625, 14.75) \rightarrow [0.0, 0.25]$ & & $[14.75, \infty] \rightarrow [0.25, 0.25]$\\
					\bottomrule
				\end{tabular}	
		}}
		{}
	\end{threeparttable}
\end{table}

	\section{Online supplement}
\label{app:online}

Instances, detailed results, and code can be found at \url{https://research.libklein.com/fscp}.
\end{appendices}
\end{document}